\documentclass[11pt]{article}

\pdfoutput=1

\usepackage[T1]{fontenc}
\usepackage[latin9]{inputenc}
\usepackage[a4paper]{geometry}
\usepackage[active]{srcltx}
\usepackage{amsmath}
\usepackage{amssymb}
\usepackage{esint}
\usepackage{ulem}

\makeatletter
\usepackage{textcomp}

\pdfoutput=1 % if your are submitting a pdflatex (i.e. if you have
\usepackage{jheppub}% for details on the use of the package, please
\newcommand*\xbar[1]{%
  \hbox{%
    \vbox{%
      \hrule height 0.5pt % The actual bar
      \kern0.3ex%         % Distance between bar and symbol
      \hbox{%
        \kern-0.0em%      % Shortening on the left side
        \ensuremath{#1}%
        \kern-0.0em%      % Shortening on the right side
      }%
    }%
  }%
}

\usepackage{amsfonts}

\usepackage{bbm}

\newcommand{\be}{\begin{equation}}
\newcommand{\ee}{\end{equation}}
\newcommand{\bea}{\begin{eqnarray}}
\newcommand{\eea}{\end{eqnarray}}

\title{\boldmath Logarithmic supertranslations and supertranslation-invariant Lorentz charges}

\author[a]{Oscar Fuentealba,}
\author[a,b]{Marc Henneaux,}
\author[c]{and C\'{e}dric Troessaert}

\affiliation[a]{Universit\'e Libre de Bruxelles and International Solvay Institutes, ULB-Campus Plaine CP231, B-1050 Brussels, Belgium}
\affiliation[b]{Coll\`ege de France, 11 place Marcelin Berthelot, 75005 Paris, France}
\affiliation[c]{Haute-Ecole Robert Schuman, Rue Fontaine aux M\^{u}res, 13b, B-6800, Belgium}

\emailAdd{oscar.fuentealba@ulb.be}
\emailAdd{marc.henneaux@ulb.be}
\emailAdd{cedric.troessaert@hers.be}

\preprint{}

\abstract{We extend the BMS(4) group by adding logarithmic supertranslations.  This is done by relaxing the boundary conditions on the metric and its conjugate momentum at spatial infinity in order to allow logarithmic terms of carefully designed form in the asymptotic expansion, while still preserving finiteness of the action. Standard theorems of the Hamiltonian formalism are used to derive the (finite) generators of the logarithmic supertranslations.  As the ordinary supertranslations, these depend on a function of the angles.    Ordinary  and logarithmic supertranslations are then shown to form an abelian subalgebra with non-vanishing central  extension. Because of this central term, one can make nonlinear redefinitions of the generators of the algebra so that  the pure supertranslations ($\ell >1$ in a spherical harmonic expansion) and the logarithmic supertranslations  have vanishing brackets with all the Poincar\'e generators, and, in particular, transform in the trivial representation of the Lorentz group.  The symmetry algebra is then the direct sum of the Poincar\'e algebra and the infinite-dimensional abelian algebra formed by the pure supertranslations and the logarithmic supertranslations (with central extension).  The pure supertranslations are thus completely decoupled from the standard Poincar\'e algebra in the asymptotic symmetry algebra.  This implies in particular that one can provide a definition of the angular momentum which is manifestly free from supertranslation ambiguities. An intermediate redefinition providing a partial decoupling of the pure and logarithmic supertranslations is also given. }

\makeatother

\begin{document}
\maketitle \flushbottom

 %\newpage{}

\section{Introduction}

The possibility to perform ``logarithmic translations'' 
\be
x^\mu \rightarrow x^\mu + C^\mu \ln r \label{eq:LogTran}
\ee
without changing the asymptotic form $g_{\mu \nu} - \eta_{\mu \nu} = \mathcal O\left(\frac{1}{r}\right)$ of the metric was already pointed out in the early 1960's \cite{Bergmann:1961zz}.  The physical implications of these coordinate transformations were shown in \cite{AshtekarLog85} not to affect  the definition of energy-momentum and angular momentum, implying that these were somehow pure gauge (see also \cite{AARP95}).  Related studies include  \cite{BeigSchmidt82,Beig:1983sw,Chrusciel89,Ashtekar:1990gc,Compere:2011ve,Troessaert:2017jcm}.

Besides logarithmic translations, logarithmic  {\it supertranslations}  where the coefficients $C^\mu$, instead of being constants, depend on the angles, have also been contemplated  \cite{BeigSchmidt82,Ashtekar:1990gc,Compere:2011ve,Troessaert:2017jcm}.   But because they induce logarithmic terms in the metric and hence conflict with its $1/r$ asymptotic decay, they have not been studied much. However, one might wonder whether this is due to a choice of boundary conditions that is too strict.  The logarithmic terms induced in the metric by logarithmic supertranslations can perhaps be tamed.  In other words, can one extend the formalism so as to accommodate logarithmic supertranslations as symmetries? Exhibiting all the symmetries of a theory usually sheds useful insight on its structure.  

We show in this paper that it is possible to consistently include logarithmic supertranslations provided the coefficients $C^\mu(\theta, \varphi)$ are submitted to parity conditions that we explicitly spell out.  These conditions imply that the logarithmic supertranslations that are incorporated in our formalism are parametrized by a single function of the angles, like the supertranslations. This function is naturally decomposed into an even and an odd part (under the antipodal map) in our $3+1$ presentation, associated with spacelike and normal logarithmic supertranslations, respectively. 

A hint that logarithmic supertranslations could be included, and the way to proceed,  are given by the study of gravity in five dimensions \cite{Fuentealba:2021yvo,Fuentealba:2022yqt}.  The ``Coulomb part'' of the metric decays in $D=5$ as $1/r^2$, and in order to exhibit the full symmetry, it is necessary to include a term that decays slowlier ($1/r$) but takes the explicit form of a diffeomorphism transformation (``improper gauge transformation'' \cite{Benguria:1976in}).  As we shall show, the procedure for including logarithmic supertranslations  in $D=4$ parallels these steps, with the Coulomb part behaving now as $1/r$ and the improper gauge part involving the slowlier decaying logarithmic term $\log r/r$. 

The logarithmic supertranslations provide an infinite-dimensional extension of the standard BMS group exhibited first at null infinity \cite{Bondi:1962px,Sachs:1962wk,Sachs:1962zza,Penrose:1962ij} (for recent reviews, see 
  \cite{Madler:2016xju,Alessio:2017lps,Ashtekar:2018lor}) and later at spatial infinity \cite{Henneaux:2018hdj,Henneaux:2019yax}.  This extension will be called the ``logarithmic BMS algebra''.  With the definition of the Lorentz generators that naturally arises, the logarithmic supertranslations turn out to transform in the same infinite-dimensional irreducible representation of the Lorentz group as the supertranslations (quotientized by the four-dimensional  invariant subspace of ordinary spacetime translations), even though  they are characterized by  functions with  opposite parities under the antipodal map.   Logarithmic and BMS supertranslation charges form furthermore a centrally extended abelian algebra such that the logarithmic supertranslation charges are canonically conjugate to the pure supertranslation charges (corresponding to $\ell >1$ in the expansion in terms of spherical harmonics, i.e., beyond the ordinary energy and momentum). One can take advantage of this property to redefine the homogeneous Lorentz generators so as to eliminate some features of the BMS algebra that have puzzled the relativity community ever since the BMS symmetry was discovered.  These  features arise because pure supertranslations do not commute with the homogeneous Lorentz transformations.  

Three of these intriguing properties  are:  
\begin{itemize}
\item (Angular momentum ambiguity) It follows from the non-vanishing of the bracket of the  pure supertranslations with the homogeneous Lorentz transformations that the angular momentum transforms under pure supertranslations.   This non-invariance comes on top of the familiar non-invariance of the angular momentum under ordinary translations, but there one knows how to define an intrinsic angular momentum in terms of Casimirs of the Poincar\'e algebra, which amounts to compute the angular momentum with respect to the center of mass worldline.  A similar construction for supertranslations appears to be more intricate for the full BMS algebra.
\item (Soft dressing by boosting) The Poisson brackets of the homogeneous Lorentz generators with the pure supertranslations  involve not only the pure supertranslations, but also the standard $4$-momentum $P^\mu$ \cite{Sachs:1962zza}.  Thus if one boosts or rotates a solution with $P^\mu \not=0$ one generates in general a solution with non-vanishing value of some pure supertranslation charges, even if these are zero prior to the action of the Lorentz transformation. This has raised some discussion in the literature.
\item (No invariant Poincar\'e subalgebra) The fact that the Poisson brackets of the homogeneous Lorentz generators with the pure supertranslations involve the pure supertranslations implies that the Poincar\'e subalgebra is not an ideal.  At the same time, because the pure supertranslations do not form an ideal on account of the preceding point, they cannot be meaningfully quotientized out to get the Poincar\'e algebra as a quotient algebra.
\end{itemize}
 All these features can be eliminated by a nonlinear redefinition of the homogeneous Lorentz generators that involves the logarithmic supertranslation charges.  This is because the logarithmic supertranslation charges are canonically conjugate to the pure supertranslation charges. 
 
We explicitly perform the construction of these new Lorentz generators in our paper.  Once this is done, the pure supertranslations and the logarithmic supertranslations have vanishing Poisson brackets with all the Poincar\'e generators. The decoupling of the pure supertranslations agrees with the ideas pursued in \cite{Mirbabayi:2016axw,Bousso:2017dny,Javadinezhad:2018urv,Javadinezhad:2022hhl} along different quantum lines. In particular, there are some similarities between the logarithmic supertranslation charges and the boundary gravitons of \cite{Javadinezhad:2018urv,Javadinezhad:2022hhl} which are conjugate to the supertranslations in that work. Our considerations are, however, purely classical, do not need any form of regularization, and fulfill automatically all necessary Jacobi identity required by consistency since we work with a well-defined Poisson bracket. Furthermore, the new Lorentz generators  manifestly reduce on-shell to surface integrals at infinity.

Our paper is organized as follows.  In the next section, we briefly review the standard treatment of the BMS group at spatial infinity \cite{Henneaux:2018hdj} and recall why logarithmic translations (and a fortiori logarithmic supertranslations) do not appear in that approach.
We then provide in Section \ref{Asymptotia} boundary conditions on the canonical variables that include terms generated by the logarithmic supertranslations.  The boundary conditions involve also a crucial strengthening of the asymptotic behaviour of the constraints which is worked out in Section \ref{constraints}.  The boundary conditions are then shown in Section \ref{Sec:FineAction} to keep the kinetic term in the action finite, a key property that enables one to use standard Hamiltonian methods.  The description of the asymptotic symmetries and the derivation through canonical methods of their charge-generators are then successively achieved in Sections \ref{Sec:AsympSymm1} and \ref{Sec:AsympSymm2}.  The algebra of these canonical generators is computed and analyzed in Section \ref{Sec:AsympSymm3}. We show in particular that the pure supertranslation charges and the logarithmic supertranslation charges are canonically conjugate.  We take advantage of this property in Section \ref{sec:Decoupling} to redefine the symmetry generators  in such a way that the pure supertranslation and logarithmic supertranslation charges have vanishing brackets with all the Poincar\'e generators. We first give a general  algebraic argument to show that this is possible and provide then the explicit redefinitions.  An alternative nonlinear redefinition of the symmetry generators such that the brackets of the pure supertranslations with the Lorentz boosts remain non-trivial but do not involve the ordinary translations (``no soft dressing by boosting'') is also given.  The concluding section (Section \ref{sec:Conclusions}) recapitulates our results and suggests various future developments.  Finally, three appendices of a more technical nature complete our paper.

 \section{The BMS group at spatial infinity \cite{Henneaux:2018hdj}}
 
 We recall in this section how the BMS group (and not just its Poincar\'e subgroup or the bigger Spi group \cite{Ashtekar:1978zz}) emerges as symmetry group of the theory (i.e., of the action and the boundary conditions) at spatial infinity.
 
In the Hamiltonian description of asymptotically flat spacetimes \cite{Dirac:1958sc,Dirac:1958jc,Arnowitt:1962hi}, the possibility to perform logarithmic translations  does not appear.  This is because a consistent Hamiltonian formulation requires additional asymptotic conditions besides the mere requirement that (i) the metric should deviate at infinity from the flat metric by terms of order $1/r$ and (ii) its conjugate momentum should decay as $1/r^2$,
\be 
g_{ij} = \delta_{ij} +  \frac{\xbar h_{ij}}{r} + \mathcal O(r^{-2}), \qquad \pi^{ij} = \frac{\xbar \pi^{ij}}{r^2} + \mathcal O (r^{-3}) , \label{eq:StrictParity}
\ee
where $\xbar h_{ij}$ and $\xbar \pi^{ij}$ depend on the angles only, i.e., $\xbar h_{ij}= \xbar h_{ij}(\mathbf{n})$ and $\xbar \pi^{ij}= \xbar \pi^{ij}(\mathbf{n})$, where $\mathbf{n}$ is the unit normal to the sphere ($(n^i) = (\sin \theta \cos \varphi, \sin \theta \sin \varphi, \cos \theta)$).  For the action to be finite\footnote{By finiteness of the action, we mean finiteness ``on the nose'', without (foliation-dependent) regularization.  This enables one to apply straightforwardly standard theorems of the Hamiltonian  formalism, in particular the connection between symmetries, charges and their algebra.} and the Poincar\'e transformations to be bona fide canonical transformations (with well-defined generators), it is necessary to restrict further the metric (and its conjugate momentum).  

The conditions that have been proposed in the literature are either strict parity conditions on the leading orders of the fields \cite{Regge:1974zd}, 
\be
\xbar h_{ij}(-\mathbf{n}) = \xbar h_{ij}(\mathbf{n}), \qquad \xbar \pi^{ij} (-\mathbf{n}) = - \xbar \pi^{ij}(\mathbf{n}),
\ee
or parity conditions twisted by an improper diffeomorphism \cite{Henneaux:2018hdj,Henneaux:2019yax}\footnote{Strict parity conditions stronger than those of \cite{Regge:1974zd} have been considered later in \cite{Ashtekar:1990gc} from a different perspective.}.  By this, it is meant that one allows an odd (respectively, even) part in $\xbar h_{ij}$ (respectively, $\xbar \pi^{ij}$), but it must take the form of a diffeomorphism at order $\mathcal O (1/r)$ (respectively $\mathcal O (1/r^2)$).   In polar coordinates $(r, x^A)$, ($x^A$ coordinates on the sphere), the diffeomorphism-twisted conditions read explicitly
\begin{align}
g_{rr} & =1+\frac{1}{r}\xbar h_{rr}+\mathcal O(r^{-2})\,,\label{eq:hrr0}\\
g_{rA} & =\mathcal O(r^{-1})\,,\label{eq:hrA0}\\
g_{AB} & =r^{2}\xbar g_{AB}+r \xbar h_{AB} + \mathcal O(1)\,,\label{eq:hAB0}
\end{align}
and
\begin{align}
\pi^{rr} & =\xbar\pi^{rr}+\mathcal O(r^{-1})\,,\label{eq:pirr0}\\
\pi^{rA} & =\frac{1}{r}\xbar\pi^{rA}+\mathcal O(r^{-2})\,,\label{eq:pirA0}\\
\pi^{AB} & =\frac{1}{r^{2}}\xbar\pi^{AB}+\mathcal O(r^{-3})\,,\label{eq:piAB0}
\end{align}
where the leading orders of the metric and the momenta obey the conditions\footnote{In writing the parity conditions, we assume that the coordinates $x^A$ transform as $x^A \rightarrow - x^A$ under the antipodal map. If one uses instead standard polar coordinates for which $\theta$ changes orientation ($\theta \rightarrow \pi - \theta$, $\frac{\partial} {\partial \theta} \rightarrow -\frac{\partial} {\partial \theta}$) but not $\varphi$ ($\varphi \rightarrow \varphi + \pi$, $\frac{\partial} {\partial \varphi} \rightarrow \frac{\partial} {\partial \varphi}$), the necessary adjustements must be made.}:
\begin{align}
\xbar h_{rr} & =(\bar{h}_{rr})^{\text{even}},\label{eq:hbarrr0}\\
\xbar h_{AB} & =(\xbar h_{AB})^{\text{even}}+2(\xbar D_{A}\xbar D_BU+\xbar g_{AB}U)\,,\label{eq:hbarAB0}
\end{align}
and
\begin{align}
\xbar\pi^{rr} & =(\xbar\pi^{rr})^{\text{odd}}-\sqrt{\xbar g}\,\xbar\triangle V\,,\label{eq:pibrr0}\\
\xbar\pi^{rA} & =(\xbar\pi^{rA})^{\text{even}}-\sqrt{\xbar g}\,\xbar D^{A}V\,,\label{eq:pibrA0}\\
\xbar\pi^{AB} & =(\xbar\pi^{AB})^{\text{odd}}+\sqrt{\xbar g}\,(\xbar D^{A}\xbar D^{B}V-\xbar g^{AB}\xbar\triangle V)\,.\label{eq:pibarAB0}
\end{align}
We have imposed the extra condition that the radial-angular components $\xbar h_{rA}$ are zero, which insures that asymptotic Lorentz boosts are canonical transformations \cite{Henneaux:2018hdj,Henneaux:2019yax}. We have also  used the fact that at the leading order relevant to the analysis, the diffeomorphisms linearize so that their finite form is identical to their infinitesimal one. 

The functions $U$ and $V$, which parametrize the improper diffeomorphisms to be included in the asymptotic form of the canonical variables, can be assumed to be odd and even, respectively, since their even/odd parts can be absorbed through redefinitions.  In the canonical description, shifts of the even (respectively, odd) part of $U$ (respectively, $V$) are proper gauge symmetries with zero charges.  Finally $\xbar g_{AB}$ is the metric on the round unit sphere and $\xbar D_A$ the corresponding covariant derivative. The indices $A,B,\dots$ of the barred fields are raised and lowered
using the metric $\xbar g_{AB}$ on the unit 2-sphere at spatial infinity.

The logarithmic translations are eliminated with either the strict parity conditions (\ref{eq:StrictParity}) or the twisted parity conditions (\ref{eq:hrr0})-(\ref{eq:pibarAB0}) because they induce terms that violate these conditions \cite{Fuentealba:2020ghw}\footnote{This argument also eliminates logarithmic translations for the differently twisted parity conditions proposed in \cite{Henneaux:2018cst}, since the radial-radial component  $\xbar h_{rr}$ should also be even in that case, while logarithmic translations induce the odd term $\sim C^i n_i$.}.
The connection between this Hamiltonian result and the insightful work  \cite{AshtekarLog85} on logarithmic translations is that the parity conditions imposed in \cite{Regge:1974zd} or their weakened twisted form \cite{Henneaux:2018hdj,Henneaux:2019yax} imply that the Weyl tensor fulfills  the strict parity conditions that were shown in \cite{AshtekarLog85} to enable one to eliminate the logarithmic translation ambiguities.  Conversely, the strict parity conditions on the Weyl tensor considered in \cite{AshtekarLog85} imply the weakened parity conditions of \cite{Henneaux:2018hdj,Henneaux:2019yax} on the canonical variables assuming that the metric possesses an asymptotic expansion in $1/r$ starting as above at $1/r$ for the metric and $1/r^2$ for the momenta (see appendix of \cite{Henneaux:2018hdj} for precise information). 

With the twisted boundary conditions (\ref{eq:hrr0})-(\ref{eq:pibarAB0}), the asymptotic symmetry group is the infinite-dimensional BMS group  \cite{Henneaux:2018hdj,Henneaux:2019yax}.  The twist is essential since without it,  the remaining asymptotic symmetries are just the Poincar\'e transformations \cite{Regge:1974zd}. The supertranslations generate shifts in $U$ and $V$, which must be therefore be allowed to take arbitrary values.

\section{Asymptotic conditions \label{Asymptotia}}

To include the logarithmic supertranslations, one must impose less stringent boundary conditions on the canonical variables. This enlargement of the asymptotic conditions  must obey two crucial consistency requirements: first, the action must be finite for all allowed  configurations of $g_{ij}(x^k)$ and $\pi^{ij}(x^k)$, even for those which do not obey the original boundary conditions; second, the BMS transformations and the logarithmic supertranslations must be canonical transformations with well-defined (finite) charges.

Deriving a set of boundary conditions that fulfill these consistency conditions follows usually a trial-and-error procedure.  Rather then repeating the somewhat zig-zag way in which the conditions were arrived at, we shall write them first and shall then explain their various properties as we check that all the consistency conditions are met. 

As we indicated, the allowed asymptotic form of the fields must generalize (\ref{eq:hrr0})-(\ref{eq:pibarAB0}) by incorporating, besides the terms involving $U$ and $V$, terms corresponding to changes of the fields under logarithmic supertranslations.  More precisely, one must twist the strict parity conditions of (\ref{eq:StrictParity}) by a diffeomorphism that involves now also logarithmic supertranslations.

By following a trial-and-error procedure, we arrived at the following consistent fall-off of the spatial metric and its conjugate momentum,
\begin{align}
g_{rr} & =1+\frac{1}{r}\xbar h_{rr}+\frac{1}{r^{2}}\left(\ln^{2}r\,h_{rr}^{\log(2)}+\ln r\,h_{rr}^{\log(1)}+h_{rr}^{(2)}\right)+o(r^{-2})\,,\label{eq:hrr}\\
g_{rA} & =\xbar\lambda_{A}+\frac{1}{r}\left(\ln^{2}r\,h_{rA}^{\log(2)}+\ln r\,h_{rA}^{\log(1)}+h_{rA}^{(2)}\right)+o(r^{-1})\,,\label{eq:hrA}\\
g_{AB} & =r^{2}\xbar g_{AB}+r\left(\ln r\,\theta_{AB}+\xbar h_{AB}\right)+\ln^{2}r\,\theta_{AB}^{(2)}+\ln r\,\sigma_{AB}+h_{AB}^{(2)}+o(1)\,,\label{eq:hAB}\\
\pi^{rr} & =\ln r\,\pi_{\text{log}}^{rr}+\xbar\pi^{rr}+\frac{1}{r}\left(\ln^{2}r\pi_{\text{log}(2)}^{rr}+\ln r\pi_{\text{log}(1)}^{rr}+\pi_{(2)}^{rr}\right)+o(r^{-1})\,,\label{eq:pirr}\\
\pi^{rA} & =\frac{\ln r}{r}\pi_{\text{log}}^{rA}+\frac{1}{r}\xbar\pi^{rA}+\frac{1}{r^{2}}\left(\ln^{2}r\pi_{\text{log}(2)}^{rA}+\ln r\pi_{\text{log}(1)}^{rA}+\pi_{(2)}^{rA}\right)+o(r^{-2})\,,\label{eq:pirA}\\
\pi^{AB} & =\frac{\ln r}{r^{2}}\pi_{\text{log}}^{AB}+\frac{1}{r^{2}}\xbar\pi^{AB}+\frac{1}{r^{3}}\left(\ln^{2}r\pi_{\text{log}(2)}^{AB}+\ln r\pi_{\text{log}(1)}^{AB}+\pi_{(2)}^{AB}\right)+o(r^{-3})\,.\label{eq:piAB}
\end{align}
 The twisted parity conditions for the metric coefficients are
 \begin{align}
\xbar h_{rr} & =(\bar{h}_{rr})^{\text{even}}+2\tilde{U}=\text{even}\,,\label{eq:hbarrr}\\
\xbar\lambda_{A} & =D_{A}U^{\text{even}} + \xbar D_{A}\tilde{U}-U_{A}=\text{odd}\,,\label{eq:lambdaA}\\
\theta_{AB} & =2(\xbar D_{A}\xbar D_{B}\tilde{U}+\xbar g_{AB}\tilde{U})=\text{even}\,,\label{eq:thetaAB}\\
\xbar h_{AB} & =(\xbar h_{AB})^{\text{even}}+2(\xbar D_{(A}U_{B)}+\xbar D_A \xbar D_B U^{\text{odd}}+\xbar g_{AB}U)\,,\label{eq:hbarAB}
\end{align}
while those for the momentum components read
\begin{align}
\pi_{\text{log}}^{rr} & =-\sqrt{\xbar g}\,\xbar\triangle\tilde{V}=\text{odd}\,,\label{eq:pirrlog}\\
\xbar\pi^{rr} & =(\xbar\pi^{rr})^{\text{odd}}-2\sqrt{\xbar g}\,\tilde{V}-\sqrt{\xbar g}\,\xbar\triangle V\,,\label{eq:pibrr}\\
\pi_{\text{log}}^{rA} & =-\sqrt{\xbar g}\,\xbar D^{A}\tilde{V}=\text{even}\,,\label{eq:pirAlog}\\
\xbar\pi^{rA} & =(\xbar\pi^{rA})^{\text{even}}+\sqrt{\xbar g}\,\xbar D^{A}\tilde{V}-\sqrt{\xbar g}\,\xbar D^{A}V\,,\label{eq:pibrA}\\
\pi_{\text{log}}^{AB} & =\sqrt{\xbar g}(\xbar D^{A}\xbar D^{B}\tilde{V}-\xbar g^{AB}\xbar\triangle\tilde{V})=\text{odd}\,,\label{eq:piABlog}\\
\xbar\pi^{AB} & =(\xbar\pi^{AB})^{\text{odd}}+\sqrt{\xbar g}\,(\xbar D^{A}\xbar D^{B}V-\xbar g^{AB}\xbar\triangle V)\,,\label{eq:pibarAB}
\end{align}
where $\tilde{U}$ is even and $\tilde{V}$ is odd.  The function $V$ is even (its odd part remains pure gauge), but $U$ has both  odd (as before) and even parts, $U = U^{\text{even}} + U^{\text{odd}}$, the even component being non trivial once the logarithmic terms are included. Finally $U_{A}$ is odd\footnote{We initially allowed also an even part $(U_{A})^{\text{even}}\not=0$, but found that this part must be equal to $U_{A}^{\text{even}}=\xbar D_{A}U^{\text{odd}}$, something we already implemented in the equations in the text.  The ``left-over'' $U_A$ appearing in the text have thus $U_{A}^{\text{even}} = 0$.}.  The reasons behind these parity assignments will be pointed out as we proceed with the various consistency checks.

Since the parity conditions will be used very often in the sequel, we recapitulate them in a table:
\begin{center}
  \begin{tabular}{ | c | c | c |}
    \hline
    Variable & Even component? & Odd component? \\ \hline
   $ \tilde U$ & Yes & No \\ \hline
    $\tilde V$  & No  & Yes \\ \hline
    $U$& Yes & Yes \\ \hline
    $ V$ & Yes & No \\ \hline
    $U_A$  &  No  & Yes \\ \hline
  \end{tabular}
\end{center}
We observe that $V^{\text{odd}} $ is in fact pure gauge, i.e., one may carry it along and observe in the end that it can be gauged away by a proper diffeomorphism, exactly as in  \cite{Henneaux:2018hdj}.  To simplify the computations we set from the outset $V^{\text{odd}} =0$.  The same is actually true for $\tilde U$, which turns out also to be entirely removable by a proper gauge transformation, so that only logarithmic supertranslations with $\tilde V$ are improper.  We keep $\tilde U$ at this stage however, since proving its triviality is one of the essential results of this paper,  which we want to explicitly establish.

Note that most metric and momentum components have mixed parity properties due to the twist, except  $(\xbar h_{rr},\theta_{AB},\pi_{\text{log}}^{rA})$ which are even  and 
$(\xbar\lambda_{A},\pi_{\text{log}}^{rr},\pi_{\text{log}}^{AB})$,  
which are  odd.

A few comments are in order: 
\begin{itemize}
\item Just as $U$ and $V$ parametrize the standard supertranslations, the functions $\tilde{U}$ and $\tilde{V}$ parametrize the logarithmic supertranslations.  Why they are only two of them (where one would expect four) and why they are subject to parity conditions comes from the requirements of consistency of the formalism, as we shall see below.  In fact, there are two ``hidden'' extra functions $\tilde{U}_A$ giving the other two missing logarithmic supertranslations but consistency with the absence of leading logarithmic term in $\lambda_A$, which plays a central role in the integrability of the charges,  forces them to be equal to $\xbar D_A\tilde{U}$, so that they are not independent. This was explicitly taken care of in the above formulas, which explains why $\tilde{U}_A$ does not appear.
\item The parametrization by the functions $\tilde U$, $U$, $U_A$, $\tilde V$ and $V$ of the metric and momentum components  $\xbar h_{rr} $, $\xbar\lambda_{A}$, $\theta_{AB}$, $\xbar h_{AB}$, $\pi_{\text{log}}^{rr}$, $\xbar\pi^{rr}$, $\pi_{\text{log}}^{rA}$, $\xbar\pi^{rA}$, $\pi_{\text{log}}^{AB}$ and $\xbar\pi^{AB}$ involves some  redundancy.  
\begin{itemize}
\item[$\square$] Consider first $\tilde U$, which is even. Given $\theta_{AB}$ (of the required form),  the function $\tilde U$ is determined by (\ref{eq:thetaAB}) up to a solution of $\xbar D_{A}\xbar D_{B} \Upsilon+\xbar g_{AB}\Upsilon = 0$.  But the most general solution of this equation is a linear combination of the $\ell= 1$ spherical harmonics $Y^{\ell}_m$, and so is odd or zero.  Since it must be even, we conclude that $\Upsilon =0$ and that $\tilde U$ is unique (for given $\theta_{AB}$ of the required form).
\item[$\square$] The same argument shows that $U^{\text{odd}}$ is determined by $(\xbar h_{AB})^{\text{odd}}$ from (\ref{eq:hbarAB}) up to $\sum_{m=-1, 0, 1} c_m Y^{1}_m$.  This ambiguity corresponds to spatial translations, which are Minkowski isometries and have no effect on $\xbar h_{AB}$ (but do act effectively on some of the other variables).
\item[$\square$] The functions $U^{\text{even}}$ and $U_A$ (which is odd) are determined from (\ref{eq:lambdaA}) up to a solution of $\xbar D_{A}u^{\text{even}} -u_{A} = 0$.  But the general solution of this equation fulfills $\xbar \Delta u^{\text{even}} -\xbar D^A u_{A}=0$.  We show below that such an ambiguity corresponds to a proper gauge transformation, so that the redundancy in the description of $\xbar \lambda _A$ has a clear interpretation.
\item[$\square$] The equation (\ref{eq:hbarrr}) can then be viewed as defining $(\xbar h_{rr})^{\text{even}}$ given $\xbar h_{rr}$, while the even part of (\ref{eq:hbarAB}) defines $(\xbar h_{AB})^{\text{even}}$, once a choice for $U^{\text{even}}$ and $U_A$ has been made from (\ref{eq:lambdaA}).
\item[$\square$] Similarly, one finds from the momenta relations that $\tilde{V}$ is potentially determined up to a constant, but since it is odd, it is unique and carries no redundancy.    By contrast, the function $V$ is even and so is determined up to a constant.  This ambiguity corresponds to time translations, which are Minkowski isometries.  Time translations are redundancies for the asymptotic fields considered here, but they are not when one takes into account all the other fields.  
\end{itemize}
\item The next terms in the expansions ($\frac{1}{r^{2}}\ln^{2}r\,h_{rr}^{\log(2)}$, etc) must be explictly written because they can potentially contribute to divergences and so one must  check that they are harmless. They are  generated when one performs logarithmic supertranslations (so it would be inconsistent to set them equal to zero), but their explicit expressions are complicated because the non-abelian character of the diffeomorphisms becomes relevant below the leading orders, and the finite form of the transformations is intricate.  We shall only need the equations that these lower order terms fulfill as a result of the Hamiltonian and momentum constraints $\mathcal H \approx 0$, $\mathcal H_i \approx 0$, which in fact determine some of them completely in terms of the leading order variables. 
\item We have written  in the expansion (\ref{eq:hrr}) for $g_{rr}$ the subleading terms as  $o(r^{-2})$ rather then $\mathcal O(r^{-3})$ because there are contributions such as $r^{-3} \ln r $, which are $o(r^{-2})$ but not $\mathcal O(r^{-3})$.  The same feature holds for the other expansions.
\item We will show below when computing the charges of the asymptotic symmetries that $\tilde U$ can be completely gauged away by a proper diffeomorphism.  This justifies the statement made above that it would not be a limitation to set $\tilde U = 0$, but - as also announced above - we shall refrain from doing so since we want to establish its pure gauge character.
\item It is interesting to compare the status of $\xbar \lambda_A$ for the various boundary conditions.   
\begin{itemize}
\item[$\square$] The strict parity conditions of \cite{Regge:1974zd} do not restrict $\xbar \lambda_A$, except that it must be odd.  By a  coordinate transformation generated by a $\mathcal O(1)$ odd vector field, which is an allowed proper gauge transformation in that context, one can set $\xbar \lambda_A$ equal to zero.     
\item[$\square$] With the boundary conditions of \cite{Henneaux:2018hdj} that allows for a twist by an $\mathcal O(1)$ even vector field,  $\xbar \lambda_A$ is not arbitrary any more since this would conflict with the integrability of the asymptotic boost charges.  It is restricted to obey $\xbar D_A \xbar \lambda^A = 0$ (see appendix C of \cite{Henneaux:2019yax} for a detailed discussion). This enables one to set it equal to zero by a proper gauge transformation \cite{Henneaux:2019yax}, a permissible simplification adopted in  \cite{Henneaux:2018hdj}.   \item[$\square$] Finally, with the more general twist involving a logarithmic supertranslation,  $\xbar \lambda_A \not= 0$ because it transforms under such  transformations (the $\xbar D_{A}\tilde{U}$-term in (\ref{eq:lambdaA})).  For that reason, one must carry $\xbar \lambda_A$.  As we shall verify, this does not conflict with integrability of the boost charges even though $\xbar D_A \xbar \lambda^A \not= 0$ in general,  because of the presence of the additional asymptotic fields.  
\end{itemize}
\end{itemize}

\section{Asymptotic form of the constraints\label{constraints}}

In addition to the above asymptotic behaviour of the fields, we must impose that the Hamiltonian and momentum constraint functions 
\begin{align}
\mathcal{H} & =\frac{1}{\sqrt{g}}\left(\pi^{ij}\pi_{ij}-\frac{1}{2}\pi^{2}\right)-\sqrt{g}R\,,\label{eq:HamG}\\
\mathcal{H}_{i} & =-2\nabla^{j}\pi_{ij}\,,\label{eq:MomG}
\end{align}
go asymptotically to zero faster than what the decay of the fields imply.  As we shall see, this is necessary in order to have  finite Lorentz charges and finite kinetic term. This extra requirement is similar to what was found in  \cite{Henneaux:2018hdj}.   Writing out explicitly the consequences of this faster decay  is the purpose of this section.

The first step is to expand the constraint functions in the limit $r \rightarrow \infty$.  We start with the Hamiltonian constraint.

\subsection{Hamiltonian constraint}

In order to compute the asymptotic form of the Hamiltonian constraint
\eqref{eq:HamG}, we need the asymptotic form of the spatial curvature.
Using the useful expressions of the radial $2+1$ decomposition of the spatial metric given in Appendix \ref{app-decomp}, we find the somewhat involved formulas:
\begin{equation}
^{(3)}R=\frac{1}{r^{3}}\left(\ln rR_{\text{log}}^{(3)}+R^{(3)}\right)+\frac{1}{r^{4}}\left(\ln^{2}rR_{\text{log}(2)}^{(4)}+\ln rR_{\text{log}(1)}^{(4)}+R^{(4)}\right)+o\left(r^{-4}\right)\,,
\end{equation}
where
\begin{align}
R_{\text{log}}^{(3)} & =\frac{1}{2}\left(\xbar D_{A}\xbar D_{B}\theta^{AB}-\xbar\triangle\,\xbar\theta\right)\,,\\
R^{(3)} & =-\xbar\triangle\,\xbar h-\xbar\triangle\,\xbar h_{rr}+\xbar D_{A}\xbar D_{B}\xbar h^{AB}+2\xbar D_{A}\xbar\lambda^{A}\,,\\
R_{\log(2)}^{(4)}&=-\theta^{(2)}+\xbar D^{A}\xbar D^{B}\theta_{AB}^{(2)}-\xbar\triangle\,\theta^{(2)}-2h_{rr}^{\log(2)}-\xbar\triangle\,h_{rr}^{\log(2)}\nonumber\\
&\quad+\frac{1}{4}\Big(3\theta_{B}^{A}\theta_{A}^{B}-\theta^{2}-\xbar D^{A}\theta\xbar D_{A}\theta+\xbar D_{C}\theta_{B}^{A}\xbar D^{C}\theta^{B}\Big)\,,\\
R_{\log(1)}^{(4)} & =4\theta^{(2)}+4k_{\log(2)}^{(2)}-2h_{rr}^{\log(1)}-\xbar\triangle\,h_{rr}^{\log(1)}-\sigma+\xbar D^{A}\xbar D^{B}\sigma_{AB}-\xbar\triangle\,\sigma-\frac{1}{2}\big(3\theta_{B}^{A}\theta_{A}^{B}-\theta^{2}\big)\nonumber\\
&\quad+\frac{3}{2}\xbar h_{B}^{A}\theta_{A}^{B}-\frac{1}{2}\xbar h_{rr}\theta-\frac{1}{2}\xbar h\,\theta-\xbar\lambda_{A}\,\xbar D^{A}\theta-\theta\xbar D^{A}\xbar\lambda_{A}+\frac{1}{2}\xbar D_{A}\theta\xbar D^{A}h_{rr}-\xbar D_{A}\xbar h_{B}^{A}\xbar D^{B}\theta\nonumber\\
 & \quad+\theta_{B}^{A}\xbar D_{A}\xbar D^{B}\xbar h-\theta_{B}^{A}\xbar D^{B}\xbar D_{C}\xbar h_{A}^{C}+\frac{1}{2}\xbar D_{A}\theta\xbar D^{A}\xbar h-\theta_{B}^{A}\xbar D^{B}\xbar\lambda_{A}+\theta_{B}^{A}\xbar D_{A}\xbar D^{B}\xbar h_{rr} \nonumber\\
 & \quad-\theta_{B}^{A}\xbar D_{C}\xbar D^{B}\xbar h_{A}^{C}+\theta_{B}^{A}\xbar\triangle\,\xbar h_{A}^{B}+\frac{1}{2}\xbar D_{A}\theta_{C}^{B}\xbar D^{A}\xbar h_{B}^{C}\,,\\
 R^{(4)} & =2\sigma+2k_{\log(1)}^{(2)}-2h_{rr}^{(2)}-\xbar\triangle\,h_{rr}^{(2)}-h^{(2)}+\xbar D^{A}\xbar D_{B}h_{A}^{(2)B}-\xbar\triangle\,h^{(2)}+2\xbar h_{rr}^{2}+\xbar h_{rr}\xbar\triangle\,\xbar h_{rr}\nonumber\\
 &\quad+\frac{1}{2}\xbar D_{A}\xbar h_{rr}\xbar D^{A}\xbar h_{rr}+\frac{3}{4}\xbar h_{B}^{A}\xbar h_{A}^{B}-\frac{1}{4}\xbar h^{2}+\xbar h_{B}^{A}\xbar D^{B}\xbar D_{A}\xbar h-\xbar h_{B}^{A}\xbar D_{A}\xbar D^{C}\xbar h_{C}^{B}-\frac{1}{4}\xbar D_{A}\xbar h\,\xbar D^{A}\xbar h\nonumber\\
 & \quad-\xbar D_{A}\xbar h_{B}^{A}\xbar D^{C}\xbar h_{C}^{B}+\xbar D^{A}\xbar h\,\xbar D_{B}\xbar h_{A}^{B}-\xbar h_{B}^{A}\xbar D_{C}\xbar D^{B}\xbar h_{A}^{C}+\xbar h_{B}^{A}\xbar\triangle\,\xbar h_{A}^{B}-\frac{1}{2}\xbar D_{A}\xbar h_{C}^{B}\xbar D^{C}\xbar h_{B}^{A}\nonumber\\
 & \quad+\frac{3}{4}\xbar D_{C}\xbar h_{A}^{B}\xbar D^{C}\xbar h_{B}^{A}-\frac{1}{2}\xbar h_{rr}\xbar h-\frac{1}{2}\xbar D_{A}\xbar h\,\xbar D^{A}\xbar h_{rr}+\xbar D_{A}\xbar h_{rr}\xbar D^{B}\xbar h_{B}^{A}+\xbar h_{A}^{B}\xbar D^{A}\xbar D_{B}\xbar h_{rr}\nonumber\\
 & \quad-\xbar h_{rr}\xbar D_{A}\xbar\lambda^{A}-2\xbar\lambda_{A}\xbar D^{A}\xbar h_{rr}-\xbar\lambda_{A}\xbar D^{A}\xbar h-\xbar h\xbar D_{A}\xbar\lambda^{A}-\xbar h_{A}^{B}\xbar D^{A}\xbar\lambda_{B}+2\xbar\lambda_{A}\xbar\lambda^{A}-2\xbar\lambda^{A}\xbar D_{A}\xbar D_{B}\xbar\lambda^{A}\nonumber\\
 & \quad-\xbar D_{A}\xbar\lambda^{A}\xbar D_{B}\xbar\lambda^{B}+2\xbar\lambda_{A}\xbar\triangle\,\xbar\lambda^{A}-\frac{1}{2}\xbar D_{A}\xbar\lambda^{B}\xbar D_{B}\xbar\lambda^{A}+\frac{3}{2}\xbar D_{A}\xbar\lambda_{B}\xbar D^{A}\xbar\lambda^{B}+\frac{3}{4}\theta_{A}^{B}\theta_{B}^{A}-\frac{1}{4}\theta^{2}\nonumber\\
 & \quad-\frac{3}{2}\xbar h_{A}^{B}\theta_{B}^{A}+\frac{1}{2}\xbar h\,\theta-\xbar h_{rr}\theta+\xbar\lambda_{A}\xbar D^{A}\theta+\theta\xbar D_{A}\xbar\lambda^{A}-\theta_{A}^{B}\xbar D_{B}\xbar\lambda^{A}\,.
\end{align}

In the above formulas, which involve the sub-subleading terms in the asymptotic expansion of the fields (among which $\sigma_{AB}$), the variables $\{k_{\log(2)}^{(2)},k_{\log(1)}^{(2)},k^{(2)}\}$ are the sub-subleading
coefficients of the fall-off of the 2-dimensional extrinsic curvature:
\begin{align}
K_{AB} & =-r\xbar g_{AB}-\frac{1}{2}\ln r\,\theta_{AB}+\frac{1}{2}\left(-\xbar h_{AB}+\xbar h_{rr}\xbar g_{AB}-\theta_{AB}+\xbar D_{A}\xbar\lambda_{B}+\xbar D_{B}\xbar\lambda_{A}\right)\\
 & \quad+\frac{1}{r}\left(\ln^{2}r\,k_{\log(2)AB}^{(2)}+\ln r\,k_{\log(1)AB}^{(2)}+k_{AB}^{(2)}\right)+o\left(r^{-1}\right)\,.
\end{align}

A direct computation shows that the square root of the determinant of the spatial metric reads
\begin{equation}
\sqrt{g}=\sqrt{\xbar g}\left[r^{2}+\frac{r}{2}\left(\ln r\,\theta+\xbar h_{rr}+\xbar h\right)+o\left(r\right)\right]\,.
\end{equation}
Taking into account the fall-off of the conjugate momentum given above, we then get that
the Hamiltonian constraint \eqref{eq:HamG} behaves asymptotically as
\begin{equation}
\mathcal{H}=-\frac{1}{r}\left(\ln r\sqrt{\xbar g}R_{\text{log}}^{(3)}+\sqrt{\xbar g}R^{(3)}\right)+\frac{1}{r^{2}}\left(\ln^{2}r\mathcal{H}_{\log(2)}+\ln r\mathcal{H}_{\log(1)}+\mathcal{H}^{(2)}\right)+o\left(r^{-2}\right)\,,
\end{equation}
where
\begin{align}
\mathcal{H}_{\log(2)}&=-\sqrt{\xbar g}\Big(R_{\text{log}(2)}^{(4)}+\frac{1}{2}\theta R_{\text{log}}^{(3)}\Big)\nonumber\\
&\quad+\frac{1}{\sqrt{\xbar g}}\Big[\frac{1}{2}(\pi_{\log}^{rr})^{2}+2\pi_{\log}^{rA}\pi_{\log A}^{r}+\pi_{\log}^{AB}\pi_{\log AB}-\pi_{\log}^{rr}\pi_{\log}-\pi_{\log}^{2}\Big]\,,\\
\mathcal{H}_{\log(1)}&=-\sqrt{\xbar g}\Big[R_{\text{log}(1)}^{(4)}+\frac{1}{2}(\xbar h_{rr}+\xbar h)R_{\text{log}}^{(3)}+\frac{1}{2}\theta R^{(3)}\Big]\nonumber\\
&\quad+\frac{1}{\sqrt{\xbar g}}\Big(\xbar\pi^{rr}\pi_{\log}^{rr}-\xbar\pi\,\pi_{\log}^{rr}+4\xbar\pi^{rA}\pi_{\log A}^{r}+2\xbar\pi_{B}^{A}\pi_{\log A}^{B}-\xbar\pi^{rr}\pi_{\log}-\xbar\pi\,\pi_{\log}\Big)\,,\label{eq:Hlog(1)}\\
\mathcal{H}^{(2)}&=-\sqrt{\xbar g}\Big[R^{(4)}+\frac{1}{2}(\xbar h_{rr}+\xbar h)R^{(3)}\Big]+\frac{1}{\sqrt{\xbar g}}\Big[\frac{1}{2}(\xbar\pi^{rr})^{2}+2\xbar\pi^{rA}\xbar\pi_{A}^{r}+\xbar\pi^{AB}\xbar\pi_{AB}-\xbar\pi^{rr}\xbar\pi-\xbar\pi^{2}\Big]\,.\label{eq:H(2)}
\end{align}

It follows from the definition of the field $\theta_{AB}$ that $R_{\log}^{(3)}=0$,
\be
\xbar D_{A}\xbar D_{B}\theta^{AB}-\xbar\triangle\theta_{A}^{A}  =0\,.\label{eq:Cons1}
\ee
  Thus for generic boundary conditions,  $\mathcal{H}=\mathcal{O}\left(r^{-1}\right)$. However, we will impose a faster fall-off, namely $\mathcal{H}= o\left(r^{-2}\right)$. Therefore we add to the asymptotic conditions on the metric and its momentum given in  the previous section the additional requirement:
\be \mathcal{H}= o\left(r^{-2}\right) \qquad \Leftrightarrow \qquad R^{(3)} = 0, \qquad \mathcal{H}_{\log(2)} = 0, \qquad \mathcal{H}_{\log(1)}= 0, \qquad \mathcal{H}^{(2)} = 0.
\ee

\subsection{Momentum constraint}

If one expands the radial component of the momentum constraint asymptotically,  one gets
\begin{equation}
\mathcal{H}^{r}=\frac{\ln r}{r}\mathcal{H}_{\text{log}}^{r}+\frac{1}{r}\mathcal{H}_{(1)}^{r}+\frac{1}{r^{2}}\left(\ln^{2}r\mathcal{H}_{\log(2)}^{r}+\ln r\mathcal{H}_{\log(1)}^{r}+\mathcal{H}_{(2)}^{r}\right)+o\left(r^{-2}\right)\,,
\end{equation}
with
\begin{align}
\mathcal{H}_{\text{log}}^{r} & =-2\left(\xbar D_{A}\pi_{\log}^{rA}-\pi_{\text{log}A}^{A}\right)\,,\label{eq:Gr1Log}\\
\mathcal{H}_{(1)}^{r} & =-2(\xbar D_{A}\xbar\pi^{rA}-\xbar\pi_{A}^{A}+\pi_{\text{\ensuremath{\log}}}^{rr})\,,\label{eq:Gr1}\\
\mathcal{H}_{\text{log}(2)}^{r} & =-2\xbar D_{A}\pi_{\log(2)}^{rA}+2\pi_{\log(2)}^{rr}+2\pi_{\log(2)A}^{A}+\theta_{AB}\pi_{\log}^{AB}\,,\\
\mathcal{H}_{\text{log}(1)}^{r} & =-2\xbar D_{A}\pi_{\log(1)}^{rA}+2\pi_{\log(1)}^{rr}-4\pi_{\log(2)}^{rr}+2\pi_{\log(1)A}^{A}+\xbar h_{rr}\pi_{\log}^{rr}+\theta_{A}^{B}\xbar\pi_{B}^{A}+2\xbar\lambda_{A}\pi_{\log}^{rA}+\xbar h_{AB}\pi_{\log}^{AB}\nonumber \\
 & \quad+\theta_{AB}\pi_{\log}^{AB}-2\pi_{\log}^{rA}\xbar D_{A}\xbar h_{rr}-2\xbar\lambda^{A}\xbar D_{B}\pi_{\log A}^{B}-2\pi_{\log}^{AB}\xbar D_{A}\xbar\lambda_{B}-2\xbar h_{rr}\xbar D_{A}\pi_{\log}^{rA}\,,\\
\mathcal{H}_{(2)}^{r} & =-2\xbar D_{A}\pi_{(2)}^{rA}+2\pi_{(2)}^{rr}-2\pi_{\log(1)}^{rr}+2\pi_{(2)A}^{A}+\xbar h_{rr}\xbar\pi^{rr}-2\xbar h_{rr}\pi_{\log}^{rr}+2\xbar\lambda_{A}\xbar\pi^{rA}+\xbar h_{A}^{B}\xbar\pi_{B}^{A}+\theta_{A}^{B}\xbar\pi_{B}^{A}\nonumber \\
 & \quad-2\xbar\lambda_{A}\pi_{\log}^{rA}-2\xbar\pi^{rA}\xbar D_{A}\xbar h_{rr}-2\xbar\lambda^{A}\xbar D_{B}\xbar\pi_{A}^{B}-2\xbar\pi^{AB}\xbar D_{A}\xbar\lambda_{B}-2\xbar h_{rr}\xbar D_{A}\xbar\pi^{rA}\,.
\end{align}

Similarly, the fall-off of the angular components of the momentum constraint is
given by
\begin{equation}
\mathcal{H}_{A}=\ln r\,\mathcal{H}_{A}^{\text{log}}+\mathcal{H}_{A}^{(0)}+\frac{1}{r}\left(\ln^{2}r\,\mathcal{H}_{A}^{\text{log(2)}}+\ln r\,\mathcal{H}_{A}^{\text{log(1)}}+\mathcal{H}_{A}^{(1)}\right)+o\left(r^{-1}\right)\,,
\end{equation}
where 
\begin{align}
\mathcal{H}_{A}^{\text{log}} & =-2(\xbar D_{B}\pi_{\text{log}A}^{B}+\pi_{\text{log}A}^{r})\,,\\
\mathcal{H}_{A}^{(0)} & =-2(\xbar D_{B}\xbar\pi_{\,\,\,A}^{B}+\xbar\pi_{\,\,\,A}^{r}+\pi_{\text{log}\,A}^{r})\,,\\
\mathcal{H}_{A}^{\log(2)} & =-2\xbar D_{B}\pi_{\log(2)A}^{B}-2\theta_{A}^{B}\xbar D_{C}\pi_{\log B}^{C}-\pi_{\log}^{BC}\xbar D_{A}\theta_{BC}\,,\\
\mathcal{H}_{A}^{\log(1)} & =-4\pi_{\log(2)}^{rA}-2\xbar D_{C}\left(\pi_{\log}^{BC}\xbar h_{B}^{A}\right)+\pi_{\log}^{BC}\xbar D^{A}\xbar h_{BC}-2\xbar D_{C}\left(\xbar\pi^{BC}\theta_{B}^{A}\right)+\xbar\pi^{BC}\xbar D^{A}\theta_{BC}\nonumber \\
 & \quad-4\theta_{B}^{A}\pi_{\log}^{rB}-2\xbar D_{B}\left(\pi_{\log}^{rB}\xbar\lambda^{A}\right)-2\xbar D_{B}\pi_{\log(1)}^{AB}+\pi_{\log}^{rr}\xbar D^{A}\xbar h_{rr}+2\pi_{\log}^{rB}\xbar D^{A}\xbar\lambda_{B}\,,\\
\mathcal{H}_{A}^{(1)} & =-2\pi_{\log(1)}^{rA}-2\pi_{\log}^{rB}\xbar h_{B}^{A}-2\xbar\pi^{rB}\theta_{B}^{A}-2\xbar D_{B}(\xbar\pi^{rB}\xbar\lambda^{A})+2\xbar\pi^{rB}\xbar D^{A}\xbar\lambda_{B}\nonumber \\
 & \quad-2\xbar D_{B}(\xbar\pi^{BC}\xbar h_{C}^{A})+\xbar\pi^{rr}\xbar D^{A}\xbar h_{rr}+\xbar\pi^{BC}\xbar D^{A}\xbar h_{BC}\,.
\end{align}
It follows from the expressions 
of the fields $\pi_{\text{log}}^{ij}$ that the logarithmic leading terms
indentically vanish,
$\mathcal{H}_{\text{log}}^{r}=\mathcal{H}_{A}^{\text{log}}=0$, 
\be
\xbar D_{A}\pi_{\log}^{rA}-\pi_{\text{log}A}^{A}=0\,, \qquad \xbar D_{B}\pi_{\text{log}A}^{B}+\pi_{\text{log}A}^{r} = 0 \, .
\ee
 Accordingly, the components of the momentum constraint go generically as 
$\mathcal{H}_{r}=\mathcal{O}\left(r^{-1}\right)$ and $\mathcal{H}_{A}=\mathcal{O}\left(r^{0}\right)$.
However, a faster fall-off will
be necessary for finiteness of the formalism  and we thus impose
\begin{equation}
\mathcal{H}_{r}=o\left(r^{-2}\right)\,,\quad\text{and}\quad\mathcal{H}_{A}=o\left(r^{-1}\right)\,.
\end{equation}
This is equivalent to 
\be
\mathcal{H}_{(1)}^{r}= 0, \qquad \mathcal{H}_{\log(2)}^{r}= 0, \qquad \mathcal{H}_{\log(1)}^{r}= 0, \qquad \mathcal{H}_{(2)}^{r}= 0,
\ee
and
\be
\mathcal{H}_{A}^{(0)}= 0, \qquad \mathcal{H}_{A}^{\text{log(2)}}= 0, \qquad \mathcal{H}_{A}^{\text{log(1)}}=0, \qquad \mathcal{H}_{A}^{(1)}= 0.
\ee

Imposing that the constraints decay faster than what follows from the asymptotic conditions on the fields is consistent in that it does not eliminate classical solutions, for which the constraints hold to all orders. Furthermore, this requirement is compatible with the asymptotic diffeomorphism symmetries given below because (i) the constraints transform among themselves under diffeomorphisms (they are first class); and (ii) in each case we are imposing the same extra condition that the first four non-trivial terms in the expansion vanish.

\section{Action and Finiteness of the symplectic structure}
\label{Sec:FineAction}

The action of General Relativity in four spacetime dimensions in Hamiltonian
form reads
\begin{equation}
I_{H}[g_{ij}, \pi^{ij}, N,  N^k] =\int dtd^{3}x\left(\pi^{ij}\dot{g}_{ij}-N\mathcal{H}-N^{i}\mathcal{H}_{i}\right) + \mathcal B\,,\label{eq:Action}
\end{equation}
where $\mathcal B$ is the integral over time of a boundary term at spatial infinity, which depends on the asymptotic form of the lapse and the shift and will be discussed below. We assume that the metric is fixed at the two time boundaries as appropriate for the ``$p \dot{q}$''-form - here $\int d^{3}x\left(\pi^{ij}\dot{g}_{ij}\right)$ -  of the kinetic term. This is the form of the action needed for computing, say, the transition amplitude $<g^{(2)}_{ij}(\mathbf{x}), t_2 \vert g^{(1)}_{ij}(\mathbf{x}), t_1>$. If instead of fixing the $q$'s at the time boundaries,  one would fix the $p$'s, one would need to make an integration by parts in time to convert $p \dot{q}$ to $- q \dot{p}$, i.e., here, $-\int d^{3}x\left(g_{ij}\dot{\pi}^{ij}\right)$. This is well known and not peculiar to gravity.  Such changes of representations are not the object of our paper, which focuses instead on the boundary terms at spatial infinity. 

The dynamical equations in Hamiltonian form follow from extremizing the action with respect to the dynamical fields $g_{ij}$ and $\pi^{ij}$. 
Variation of the Hamiltonian action with respect to $N$ (lapse function)
and $N^{i}$ (shift vector) yields the Hamiltonian and momentum constraints $\mathcal H \approx 0$, $\mathcal H_i \approx 0$. 

We now prove  the finiteness of the kinetic term with our boundary conditions, something which is not manifest since there are for instance terms like $\int d r  \frac{\log r}{r} $. 
For definiteness, we work  in spherical
coordinates.  Recall that the momenta carry a density weight, so the integration in spherical coordinates is just $\int dr  d \theta d \varphi \pi^{ij} \dot{g}_{ij}$, without $\sqrt{g}$.   We write this integral as $\int dr \oint d^2x$, where the $2$-sphere integral $\oint \equiv \oint_{S_2}$ involves functions depending only on the angles., i.e., we integrate first over the angles and then over $r$.  

The ingredients for proving finiteness of the kinetic term are:
\begin{itemize}
\item The (strict) parities of $\tilde U$,  $\tilde V$ and $U_A$, which are
\be
\tilde{U} = \tilde{U}^{\text{even}}, \qquad \tilde{V} = \tilde{V}^{\text{odd}}, \qquad U_A = U_A^{\text{odd}}.
\ee
\item The faster decrease of the constraints, but for this step, it is sufficient to have $\mathcal{H}= o\left(r^{-1}\right)$, $\mathcal{H}_{r}=o\left(r^{-1}\right)$ and $\mathcal{H}_{A}=o\left(1\right)$,  i.e.,
\begin{align}
R^{(3)} & = -\xbar\triangle\,\xbar h-\xbar\triangle\,\xbar h_{rr}+\xbar D_{A}\xbar D_{B}\xbar h^{AB}+2\xbar D_{A}\xbar\lambda^{A} = 0\,, \label{eq:Cons2}\\
\mathcal{H}_{(1)}^{r} & =-2(\xbar D_{A}\xbar\pi^{rA}-\xbar\pi_{A}^{A}+\pi_{\text{\ensuremath{\log}}}^{rr}) = 0 \, ,\label{eq:Cons4}\\
\mathcal{H}_{A}^{(0)} & =-2(\xbar D_{B}\xbar\pi_{\,\,\,A}^{B}+\xbar\pi_{\,\,\,A}^{r}+\pi_{\text{log}\,A}^{r}) = 0\,, \label{eq:Cons5}
\end{align}
from which one easily derives
\be
\xbar D_{A}\xbar D_{B}\xbar\pi^{AB}+\xbar\pi_{A}^{A}  =0\,,\label{eq:Cons3}\, \\
\ee
using the expressions of $\pi_{\text{\ensuremath{\log}}}^{rr}$ and $\pi_{\text{log}\,A}^{r}$ in terms of $\tilde V$.
\end{itemize}

Finiteness of the kinetic term is demonstrated as follows.  Inserting the asymptotic expansion of the fields inside the kinetic term of the action, one finds
\begin{align}
\int d^{3}x\pi^{ij}\dot{h}_{ij} & =\int drd^{2}x\Big[\frac{\ln^{2}r}{r}\pi_{\text{log}}^{AB}\dot{\theta}_{AB}+\frac{\ln r}{r}\left(\pi_{\text{log}}^{rr}\dot{\xbar h}_{rr}+\pi_{\text{log}}^{AB}\dot{\xbar h}_{AB}+\xbar\pi^{AB}\dot{\theta}_{AB}+2\pi_{\text{log}}^{rA}\dot{\xbar\lambda}_{A}\right)\label{eq:SymStr} \\
 & \qquad\qquad\qquad+\frac{1}{r}\left(\xbar\pi^{rr}\dot{\xbar h}_{rr}+\xbar\pi^{AB}\dot{\xbar h}_{AB}+2\xbar\pi^{rA}\dot{\xbar\lambda}_{A}\right)+o(r^{-1})\Big]\,.\label{eq:SymStr2}
\end{align}
The $o(r^{-1})$ are harmless since the integral $\int o(r^{-1}) dr$ does not diverge at infinity.  The other written terms are potentially harmful and one must verify that they are zero.  We examine in turn the coefficients of  $\frac{\ln^{2}r}{r}$,  $\frac{\ln r}{r}$ and $\frac{1}{r}$.

\begin{itemize}
\item 
 For the first term in the r.h.s. of \eqref{eq:SymStr}, we express $\pi_{\text{log}}^{AB}$ in terms of $\tilde{V}$
using \eqref{eq:piABlog}
and  get that
\begin{align}
\oint d^{2}x\pi_{\text{log}}^{AB}\dot{\theta}_{AB} & =-\oint d^{2}x\sqrt{\xbar g}(\xbar g^{AB}\xbar\triangle\tilde{V}-\xbar D^{A}\xbar D^{B}\tilde{V})\dot{\theta}_{AB} \nonumber \\
 & =-\oint d^{2}x\sqrt{\xbar g}(\xbar\triangle\dot{\theta}_{A}^{A}-\xbar D^{A}\xbar D^{B}\dot{\theta}_{AB})\tilde{V} \nonumber \\
 & \approx0\,.
\end{align}
This  integral over the sphere vanishes by virtue of \eqref{eq:Cons1}.  Hence, the coefficient of $\frac{\ln^{2}r}{r}$ is zero.
\item For the second term in the r.h.s. of \eqref{eq:SymStr}, we express the ``log'' momenta in terms of $\tilde V$ 
from \eqref{eq:pirrlog} and \eqref{eq:piABlog} to get
\begin{align}
& \oint d^{2}x\left(\pi_{\text{log}}^{rr}\dot{\xbar h}_{rr}+\pi_{\text{log}}^{AB}\dot{\xbar h}_{AB}+\xbar\pi^{AB}\dot{\theta}_{AB}+2\pi_{\text{log}}^{rA}\dot{\xbar\lambda}_{A}\right)   \\
&  \qquad = \oint d^{2}x\Big[\bar{\pi}^{AB}\dot{\theta}_{AB}-\sqrt{\xbar g}(\xbar\triangle\,\dot{\xbar h}+\xbar\triangle\,\dot{\xbar h}_{rr}-\xbar D_{A}\xbar D_{B}\dot{\xbar h}^{AB}-2\xbar D_{A}\dot{\xbar\lambda}^{A})\tilde{V}\Big]\\
 & \qquad \approx\oint d^{2}x\,\bar{\pi}^{AB}\dot{\theta}_{AB}\,,
\end{align}
where the constraint equation \eqref{eq:Cons2} was used. By expressing now
 $\theta_{AB}$ 
in terms of $\tilde{U}$, one finds that this remaining term becomes
\begin{equation}
\oint d^{2}x\,\bar{\pi}^{AB}\dot{\theta}_{AB}=2\oint d^{2}x\sqrt{\xbar g}\left(\xbar D_{A}\xbar D_{B}\xbar\pi^{AB}+\xbar\pi_{A}^{A}\right)\dot{\tilde{U}}\approx0\,.
\end{equation}
It vanishes by virtue of \eqref{eq:Cons3}. Hence, the coefficient of $\frac{\ln r}{r}$ is also zero.

\item  Finally, for the $1/r$ term given in \eqref{eq:SymStr2}, we
use the expressions \eqref{eq:pibrr}, \eqref{eq:pibrA} and \eqref{eq:pibarAB} for the momenta.
After some integrations by parts we get: 
\begin{align}
&\oint d^{2}x\left(\bar{\pi}^{rr}\dot{\bar{h}}_{rr}+\bar{\pi}^{AB}\dot{\bar{h}}_{AB}+2\bar{\pi}^{rA}\dot{\bar{\lambda}}_{A}\right)  \\
&=\oint d^{2}x\Big\{\big[(\xbar\pi^{rr})^{\text{odd}}-2\sqrt{\xbar g}\tilde{V}\big]\dot{\xbar h}_{rr}+(\xbar\pi^{AB})^{\text{odd}}\dot{\xbar h}_{AB}+2\big[(\xbar\pi^{rA})^{\text{even}}+\sqrt{\xbar g}\,\xbar D^{A}\tilde{V}\big]\dot{\xbar\lambda}_{A}\Big\}\\
 & \quad-\oint d^{2}x\sqrt{\xbar g}\big(\xbar\triangle\,\dot{\xbar h}+\xbar\triangle\,\dot{\xbar h}_{rr}-\xbar D_{A}\xbar D_{B}\dot{\xbar h}^{AB}-2\xbar D_{A}\dot{\xbar\lambda}^{A}\Big)V\,.
\end{align}
The last sphere integral vanishes by virtue of \eqref{eq:Cons2}.  Furthermore, since $\xbar h_{rr}$ is even and its coefficient is odd, while $\xbar \lambda_A$ is odd while its coefficient is even, the only term that remains in the integral of the first line is $(\xbar\pi^{AB})^{\text{odd}}\dot{\bar{h}}_{AB}$.  We transform this term by inserting the expression \eqref{eq:hbarAB} and integrating by part, which yields (recalling that $\xbar U_A$ is odd), 
\begin{align}
&\oint d^{2}x\left(\xbar\pi^{rr}\dot{\xbar h}_{rr}+\xbar\pi^{AB}\dot{\xbar h}_{AB}+2\xbar\pi^{rA}\dot{\xbar\lambda}_{A}\right) \\
& =2\oint d^{2}x\sqrt{\xbar g}\left(\xbar D_{A}\xbar D_{B}(\xbar\pi^{AB})^{\text{odd}}+(\xbar\pi_{A}^{A})^{\text{odd}}\right)\dot{{U}}^{\text{odd}}\approx0\,.
\end{align}
This integral vanishes again by virtue of \eqref{eq:Cons3}. Hence, the coefficient of $\frac{1}{r}$ is also zero.
\end{itemize}

The vanishing of the $\frac{1}{r}$ term, which is necessary for the finiteness of the kinetic term, actually forces the parity of the functions $\tilde{U}$, $\tilde{V}$ and the form of $U_A^{\text{even}}$,  if we assume that $\xbar \pi^{rr}_{\text{odd}}$, $\xbar \pi_{\text{odd}}$ and $\xbar h^{\text{even}}_{rr}$ are independent fields.  Indeed, the $1/r$ term reads, using the expressions of the momenta and the constraint \eqref{eq:Cons2},
\begin{equation}
\oint d^{2}x\Big\{\big[(\xbar\pi^{rr})^{\text{odd}}-2\sqrt{\xbar g}\tilde{V}\big]\dot{\xbar h}_{rr}+(\xbar\pi^{AB})^{\text{odd}}\dot{\xbar h}_{AB}+2\big[(\xbar\pi^{rA})^{\text{even}}+\sqrt{\xbar g}\,\xbar D^{A}\tilde{V}\big]\dot{\xbar\lambda}_{A}\Big\}\,.
\end{equation}
Inserting the explicit form of the fields, i.e.,
\begin{align}
\xbar h_{rr} & =(\bar{h}_{rr})^{\text{even}}+2\tilde{U}\,,\\
\xbar\lambda_{A} & =D_{A}U + \xbar D_{A}\tilde{U}-U_{A}\,,\\
\xbar h_{AB} & =(\xbar h_{AB})^{\text{even}}+2(\xbar D_{(A}U_{B)}+\xbar g_{AB}U)\,,
\end{align}
(where we allow a priori an independent $U_A^{\text{even}}$-term, not necessarily fulfilling $U_A^{\text{even}}=\xbar D_A U^{\text{odd}}$),
we find that the above integral becomes
\begin{align}
\oint d^{2}x\Big\{&-2\big[ \xbar D_A (\xbar\pi^{AB})^{\text{odd}}+(\xbar\pi^{rB})^{\text{even}}\big]\dot{U}_B-2\big[\xbar D_A(\xbar\pi^{rA})^{\text{even}}-\xbar\pi^{\text{odd}}\big]\dot{U}\label{eq:constraints} \\
&-2\sqrt{\xbar g}\,\big(\xbar \triangle\,\tilde{V}+\tilde{V}\big)\dot{\tilde{U}}-2\sqrt{\xbar g}\,\xbar D_A\tilde{V}\big(\dot{U}_A-\xbar D_A \dot{U}\big)\\
&+2\big[(\xbar\pi^{rr})^{\text{odd}}-\xbar D_A(\xbar\pi^{rA})^{\text{even}}\big]\dot{\tilde{U}}-2\sqrt{\xbar g}\,\tilde{V}(\dot{\xbar h}_{rr})^{\text{even}}\Big\}\,.\label{eq:3rdline}
\end{align}
The line \eqref{eq:constraints} vanishes by virtue of the asymptotic constraints (\ref{eq:Cons4}), (\ref{eq:Cons5}), which imply 
\begin{align}
\xbar D_A(\xbar\pi^{rA})^{\text{even}}-\xbar\pi^{\text{odd}}&=0\,,\\
\xbar D_A (\xbar\pi^{AB})^{\text{odd}}+(\xbar\pi^{rB})^{\text{even}}&=0\,.
\end{align}
It follows from line \eqref{eq:3rdline} that $\tilde{U}$ must be even and $\tilde{V}$ must be odd. The above integral reduces then to
\begin{align}
-2\oint d^{2}x\sqrt{\xbar g}\,\xbar D_A\tilde{V}\big(\dot{U}_A-\xbar D_A \dot{U}\big)\,,
\end{align}
which vanishes if and only if
\begin{equation}
U_A^{\text{even}}=\xbar D_A U^{\text{odd}}\,,
\end{equation}
as we wanted to prove.

We have thus established the finiteness of the kinetic term in the action and hence of the symplectic form.  Finiteness of the symplectic structure is a cornerstone of our approach because it enables one to straightforwardly apply standard  Hamiltonian theorems and, in particular, momentum map considerations.  Note that finiteness holds exactly, without having to perform any regularization procedure.

\section{Asymptotic symmetries}
\label{Sec:AsympSymm1}

\subsection{Form of vector fields}
Since our boundary conditions involve explicitly terms generated by logarithmic supertranslations and ordinary supertranslations, we expect them to be invariant under diffeomorphisms parametrized by vector fields $\left(\xi^{\perp}\equiv\xi,\xi^{i}\right)$  combining both kind of transformations. Their exact form will be shown to be
\begin{align}
\text{\ensuremath{\xi}} & =br+\ln r \left( \tilde{T} +\tilde{T}_{(b)} \right)+T+ T_{(b)} + o(1)\,, \label{eq:AsympDiff1}\\
\xi^{r} & =\ln r\,\tilde{W}+W+o(1)\,, \label{eq:AsympDiff2}\\
\xi^{A} & =Y^{A}+\frac{\ln r}{r}\left(\frac{2b}{\sqrt{\xbar g}}\pi_{\log}^{rA} + \bar{D}^{A}\tilde{W}\right)+\frac{1}{r}\left(\frac{2b}{\sqrt{\xbar g}}\xbar\pi^{rA} + \bar{D}^{A}(W)^{\text{odd}}\right)+ \frac{1}{r}I^{A} +o(r^{-1})\,, \label{eq:AsympDiff3}
\end{align}
where
\begin{align}
\tilde{T}_{(b)} & =\partial_{A}b\xbar\lambda^{A}-\left(\xbar\triangle+2\right)^{-1}\left(\xbar D_{A}\xbar D_{B}+\xbar g_{AB}\right)\left[b\,\left(\xbar D^{A}\xbar\lambda^{B}-\frac{1}{2}\theta^{AB}\right)\right]\,,\\
T_{(b)} & =-\frac{1}{2}b\xbar h\,.
\end{align}
The function $\tilde{T}_{(b)}$ is well defined because the operator  $\left(\xbar\triangle+2\right)^{-1}$ acts on a function with no $\ell=1$ spherical harmonic component.  The ambiguity in $\tilde{T}_{(b)}$, spanned by the linear combinations of the $Y^\ell_m$ with $\ell = 1$, will be shown to be a proper gauge transformation. 

The independent parameters in (\ref{eq:AsympDiff1})-(\ref{eq:AsympDiff3}) are:
\begin{itemize}
\item the function $b=b_i n^i$, which parametrizes the Lorentz boosts; 
\item the vector $Y^{A}$, which parametrizes the spatial rotations;
\item the functions $\tilde{T}$ (odd) and $\tilde{W}$ (even), which parametrize logarithmic supertranslations; however, as we shall see, the transformations generated by $\tilde{W}$ are proper (zero charge);
\item the function $T$ (even) and the odd part of $W$, which parametrize the familiar supertranslations; 
\item the even part of $W$ and the vector $I^A$ (odd), which generate a new type of supertranslations; however, only the combination $\xbar \triangle W^{\text{even}} - \xbar D_A I^A$ is physically relevant in the sense that  if $\xbar \triangle W^{\text{even}} - \xbar D_A I^A = 0$, these new supertranslations are proper (see below).
\end{itemize}

We summarize the properties of these independent parameters in the following table:
\begin{center}
  \begin{tabular}{ | c | c | c | c|}
    \hline
    Variable & Parity & Nature of diffeomorphism  & Proper or improper?\\ \hline
   $ \tilde W$ & Even & Logarithmic supertranslations & Proper \\ \hline
    $\tilde T$  & Odd  & Logarithmic supertranslations & Improper (except $\ell=1$ component)\\ \hline
    $W^{\text{odd}}$& Odd & Supertranslations & Improper \\ \hline
    $ T$ & Even & Supertranslations & Improper \\ \hline
      $W^{\text{even}}$& Even & New supertranslations & Improper if $\xbar \triangle W^{\text{even}} - \xbar D_A I^A  \not=0$\\ \hline
    $I^A$  &  Odd  & New supertranslations & Improper if $\xbar \triangle W^{\text{even}} - \xbar D_A I^A \not= 0$\\ \hline
  \end{tabular}
\end{center}

The independent piece of (\ref{eq:AsympDiff1}) is thus $br+\ln r  \tilde{T} +T$, which can be chosen freely (within the class of boosts for $b$, odd functions on the sphere for $\tilde{T}$ and even functions on the sphere for $T$).  The extra term 
\begin{equation}
\xi_{(b)}=\ln r\tilde{T}_{(b)}+T_{(b)}\,,
\end{equation}
must be added when $b \not=0$ in order to make the boost charges integrable (or, what is the same, in order for the boosts to leave the action invariant, without surface terms at spatial infinity).   Similarly, the written  terms  in $\xi^r$ can be chosen freely, while the  independent piece of (\ref{eq:AsympDiff3}) is $\xi^{A}  =Y^{A}+ \frac{1}{r}I^{A}$. The other terms involving $b$, $\tilde W$ and $W$ are included to preserve the boundary conditions on $\lambda_A$ under boosts ($b$), logarithmic supertranslations ($\tilde W$) and supertranslations ($W$).

The need to add extra correcting terms to the ``naked'' parameters is not a surprise since it just generalizes to the logarithmic supertranslation case what was found to be  already necessary in the Hamiltonian formulation of the standard BMS symmetry \cite{Henneaux:2018hdj}.  While the need for the correcting terms associated with the preservation of the boundary conditions will be made obvious in the next subsections, the justification for those associated with integrability will have to wait until Section \ref{Sec:AsympSymm2}.

\subsection{Transformations  of the canonical variables}

The transformations generated by the diffeomorphisms (\ref{eq:AsympDiff1})-(\ref{eq:AsympDiff3}) define asymptotic symmetries if and only if they preserve the asymptotic conditions and leave the action invariant (up to possible terms at the time boundaries).

If the asymptotic conditions are preserved, the diffeomorphisms induce well-defined transformations of the asymptotic fields appearing in the asymptotic expansion of the canonical variables.  We deal with this point in this subsection.

The action of diffeomorphisms $\left(\xi,\xi^{i}\right)$  on the canonical variables is given by
 \begin{align}
\delta g_{ij} & =\frac{2\xi}{\sqrt{g}}\left(\pi_{ij}-\frac{1}{2}g_{ij}\pi\right)+\mathcal{L}_{\xi}g_{ij}\,,\label{eq:dh-diff}\\
\delta\pi^{ij} & =-\xi\sqrt{g}\left(R^{ij}-\frac{1}{2}g^{ij}R\right)+\frac{\xi g^{ij}}{2\sqrt{g}}\left(\pi^{mn}\pi_{mn}-\frac{1}{2}\pi^{2}\right)\nonumber \\
 & \quad-\frac{2\xi}{\sqrt{g}}\left(\pi^{im}\pi_{m}^{j}-\frac{1}{2}\pi^{ij}\pi\right)+\sqrt{g}\left(\nabla^{i}\nabla^{j}\xi-g^{ij}\triangle\xi\right)+\mathcal{L}_{\xi}\pi^{ij}\,,\label{eq:dp-diff}
\end{align}
where the spatial Lie derivatives are given by
\begin{align}
\mathcal{L}_{\xi}g_{ij} & =2g_{k(i}\partial_{j)}\xi^{k}+\xi^{k}\partial_{k}g_{ij}\,,\\
\mathcal{L}_{\xi}\pi^{ij} & =-2\partial_{k}\xi^{(i}\pi^{j)k}+\partial_{k}(\xi^{k}\pi^{ij})\,.
\end{align}

One quick way to arrive at these formulas is to observe that they are certainly valid for diffeomorphisms that decrease sufficiently fast at infinity that their canonical generator $G_{\xi,\xi^{i}}\big[g_{ij},\pi^{ij}\big]=\int d^{3}x (\xi\mathcal{H}+\xi^{i}\mathcal{H}_{i})$ without surface term is well defined (proper gauge transformations, $G_{\xi,\xi^{i}} \approx 0$).  The variations of the canonical variables are then just obtained by taking their brackets with  $\int d^{3}x (\xi\mathcal{H}+\xi^{i}\mathcal{H}_{i})$, which yields (\ref{eq:dh-diff})- (\ref{eq:dp-diff}).
Since these formulas  are local in space, however, they hold true independently of the asymptotic behaviour of $(\xi,\xi^{i})$  and can be used even if a surface term must be added to $\int d^{3}x (\xi\mathcal{H}+\xi^{i}\mathcal{H}_{i})$ to get a well-defined generator (improper gauge transformation).

There might be extra terms proportional to the constraints in (\ref{eq:dh-diff})- (\ref{eq:dp-diff}) when  the parameters $\left(\xi^{\perp}\equiv\xi,\xi^{i}\right)$ involve the fields -- as here, through the dependent correcting terms added to preserve the boundary conditions and make the boost charges integrable --  but since the constraints decrease very fast at infinity, these terms play no role in our considerations.

\subsubsection{Transformations under spatial diffeomorphisms}

We  first consider diffeomorphisms acting on the equal-time hypersurfaces ($(\xi,\xi^{i}) = (0, \xi^{i})$).  One directly gets from (\ref{eq:dh-diff}) and (\ref{eq:AsympDiff2}), (\ref{eq:AsympDiff3}) that the boundary conditions are preserved and
that the successive terms in the asymptotic expansion transform as
\begin{itemize}
\item Leading order:
\begin{align}
\delta_{\xi^{i}}\theta_{AB} & =\mathcal{L}_{Y}\theta_{AB}+2\left(\xbar D_{(A}\xbar D_{B)} \tilde W+\xbar g_{AB}\tilde{W}\right)\,.
\end{align}
\item First subleading order:
\begin{align}
\delta_{\xi^{i}}\xbar h_{rr} & =Y^{A}\partial_{A}\xbar h_{rr}+2\tilde{W}\,,\\
\delta_{\xi^{i}}\xbar\lambda_{A} & =\mathcal{L}_{Y}\lambda_{A}+\xbar D_{A}W^{\textrm{even}}-I_{A}+\xbar D_{A} \tilde{W}\,,\\
\delta_{\xi^{i}}\xbar h_{AB} & =\mathcal{L}_{Y}\xbar h_{AB}+2\left(\xbar D_{(A}I_{B)}+\xbar D_A  \xbar D_B W^{\textrm{odd}} + \xbar g_{AB}W\right)\,.
\end{align}
\end{itemize}
Note that the term $\frac{\ln r}{r} \bar{D}^{A}\tilde{W}+\frac{1}{r} \bar{D}^{A}(W)^{\text{odd}}$ must be included in $\xi^A$ in order to maintain the asymptotic form of the mixed component $g_{rA}$.   If we had allowed in $\xi^{A}$  the term $\frac{\ln r}{r}\tilde{I}^{A}+\frac{1}{r}I^{A}+o(r^{-1})$ with arbitrary $\tilde{I}^{A}$ and with $I^{A}$ having an arbitrary even piece, we would have found the extra terms
\begin{equation}
(\delta_{\xi^{i}}g_{rA})^{\textrm{extra}}=\ln r\left(\xbar D_{A}\tilde{W}-\tilde{I}_{A}\right)+ \xbar D_{A}W^{\textrm{odd}}-I_{A}^{\textrm{even}}\,,
\end{equation}
which violate the asymptotic decay of $g_{rA}$ unless $\tilde{I}_{A} = \xbar D_{A}\tilde{W}$ (to eliminate the $\ln r$-piece) and $I_{A}^{\textrm{even}} = \xbar D_{A}W^{\textrm{odd}}$ (so that $\xbar \lambda_A$ is purely odd). 

Turn now to the conjugate momentum.  Under the action of $\xi^{i}$, the successive terms in the asymptotic expansion transform as
\begin{itemize}
\item Leading order:
\begin{align}
\delta_{\xi^{i}}\pi_{\log}^{rr} & =\partial_{A}\left(Y^{A}\pi_{\log}^{rr}\right)\,,\\
\delta_{\xi^{i}}\pi_{\log}^{rA} & =\mathcal{L}_{Y}\pi_{\log}^{rA}\,,\\
\delta_{\xi^{i}}\pi_{\log}^{AB} & =\mathcal{L}_{Y}\pi_{\log}^{AB}\,,\,.
\end{align}
\item First subleading order:
\begin{align}
\delta_{\xi^{i}}\xbar\pi^{rr} & =\partial_{A}\left(Y^{A}\xbar\pi^{rr}\right)\,,\\
\delta_{\xi^{i}}\xbar\pi^{rA} & =\mathcal{L}_{Y}\xbar\pi^{rA}\,,\\
\delta_{\xi^{i}}\xbar\pi^{AB} & =\mathcal{L}_{Y}\xbar\pi^{AB}\,.
\end{align}
\end{itemize}
This follows directly from (\ref{eq:dp-diff}) and (\ref{eq:AsympDiff2}), (\ref{eq:AsympDiff3}).

\subsubsection{Transformations under normal  diffeomorphisms}

The action of the ``naked'' boosts $\text{\ensuremath{\xi}}  =br $ on $g_{rA}$ violate the boundary conditions.  Indeed, it follows from (\ref{eq:dh-diff}) that $g_{rA}$ transforms as
\begin{equation}
\delta_{\xi}g_{rA}=\ln r\frac{2b}{\sqrt{\xbar g}}\pi_{\log A}^{r}+\frac{2b}{\sqrt{\xbar g}}\xbar\pi_{A}^{r}+o(1)\,.
\end{equation}
For this reason, as announced above, 
one must perform a corrective gauge transformation $\tilde{I}_{A}^{(b)}=\frac{2b}{\sqrt{\xbar g}}\pi_{\log A}^{r}$ in order to maintain the condition $g_{rA}=\mathcal{O}(1)$.
However,  the resulting transformation is non-canonical and even generates divergences in the charge (see section
\ref{time-like charge} below). To cure
this problem, the leading corrective transformation $\tilde{I}_{A}^{(b)}$ must be accompanied by a subleading
corrective gauge transformation $I_{A}^{(b)}=\frac{2b}{\sqrt{\xbar g}}\xbar\pi_{A}^{r}$.   Similarly, integrability dictate the need to include the corrective terms $\xi_{(b)}=\ln r\tilde{T}_{(b)}+T_{(b)}$ (see same section
\ref{time-like charge} below).

The normal diffeomorphisms $\xi$ with all the corrections included, 
\begin{align}
\text{\ensuremath{\xi}} & =br+\ln r \left( \tilde{T} +\tilde{T}_{(b)} \right)+T+ T_{(b)} \,, \\
\xi^{r} & =0, \qquad 
\xi^{A}  =\frac{\ln r}{r}\frac{2b}{\sqrt{\xbar g}}\pi_{\log}^{rA} +\frac{1}{r}\frac{2b}{\sqrt{\xbar g}}\xbar\pi^{rA} \,,
\end{align}
preserve the asymptotic decay of the fields and their action on the  successive terms in the asymptotic expansion can be worked out to be:
\begin{itemize}
\item Leading order:
\begin{align}
\delta_{b}\theta_{AB} & =\frac{2b}{\sqrt{\xbar g}}\left(\pi_{AB}^{\log}-\xbar g_{AB}\pi_{\log}\right)+\frac{4}{\sqrt{\xbar g}}\xbar D_{(A}\left[b\pi_{\log B)}^{r}\right] \, .
\end{align}
\item First subleading order:
\begin{align}
\delta_{b}\xbar h_{rr} & =\frac{b}{\sqrt{\xbar g}}\left(\xbar\pi^{rr}-\xbar\pi\right)\,,\\
\delta_{b}\xbar\lambda_{A} & =\frac{2b}{\sqrt{\xbar g}}\pi_{\log A}^{r}\,,\\
\delta_{b}\xbar h_{AB} & =\frac{b}{\sqrt{\xbar g}}\left[2\xbar\pi_{AB}-\xbar g_{AB}\left(\xbar\pi^{rr}+\xbar\pi\right)\right]+\frac{4}{\sqrt{\xbar g}}\xbar D_{(A}\left[b\xbar\pi_{B)}^{r}\right]\,.
\end{align}
\end{itemize}
These transformations preserve the parity conditions.

For the conjugate
momentum, one finds:
\begin{itemize}
\item Leading order:
\begin{align}
\delta_{\xi}\pi_{\log}^{rr} & =\sqrt{\xbar g}\left(-\frac{1}{2}b\theta-\frac{1}{2}\partial_{A}b\xbar D^{A}\theta+\partial_{A}b\xbar D_{B}\theta^{AB}\right)-\sqrt{\xbar g}\,\xbar\triangle\,\tilde{T}'\,,\\
\delta_{\xi}\pi_{\log}^{rA} & =\frac{\sqrt{\xbar g}}{2}\left[b\left(\xbar D_{B}\theta^{AB}-\xbar D^{A}\theta\right)+\partial_{B}b\theta^{AB}\right]-\sqrt{\xbar g}\xbar D^{A}\tilde{T}'\,,\\
\delta_{\xi}\pi_{\log}^{AB} & =\frac{\sqrt{\xbar g}}{2}\Big[b\Big(\theta^{AB}+\xbar\triangle\theta-\xbar D_{C}\xbar D^{A}\theta^{BC}\Big)+\partial_{C}b\Big(\xbar D^{C}\theta^{AB}-2\xbar D^{A}\theta^{BC}+\xbar g^{AB}\xbar D^{C}\theta\Big)\Big]\nonumber\\
 & \quad+\sqrt{\xbar g}\left(\xbar D^{A}\xbar D^{B}\tilde{T}'-\xbar g^{AB}\xbar\triangle\,\tilde{T}'\right)\,.
\end{align}
\item First subleading order:
\begin{align}
\delta_{\xi}\xbar\pi^{rr} & =\frac{\sqrt{\xbar g}}{2}\Big[b\left(6\xbar h_{rr}-\xbar h-2\theta+4\xbar D_{A}\xbar\lambda^{A}+\xbar D_{A}\xbar D_{B}\xbar h^{AB}\right) \nonumber \\
& \qquad+\partial_{A}b\left(4\xbar\lambda^{A}-\xbar D^{A}\xbar h+2\xbar D^{B}\xbar h_{B}^{A}\right)\Big]  -\sqrt{\xbar g}(\xbar\triangle\,T'+2\tilde{T}')\,,\\
\delta_{\xi}\xbar\pi^{rA} & =\frac{\sqrt{\xbar g}}{2}\Big[b\big(2\xbar\lambda^{A}+\xbar D^{B}\xbar h_{B}^{A}-\xbar D^{B}\theta_{B}^{A}+\xbar\triangle\,\xbar\lambda^{A}+\xbar D_{B}\xbar D^{A}\xbar\lambda^{B}-2\xbar D^{A}\xbar h_{rr}-\xbar D^{A}\xbar h\nonumber\\
 & \quad+\xbar D^{A}\theta-2\xbar D^{A}\xbar D_{B}\lambda^{B}\big)+\partial_{B}b\left(\xbar h^{AB}-\theta^{AB}+\xbar D^{B}\xbar\lambda^{A}-\xbar D^{A}\xbar\lambda^{B}\right)\Big]\nonumber\\
& \qquad +\sqrt{\xbar g}\left(-\xbar D^{A}T'+\xbar D^{A}\tilde{T}'\right)\,,\\
\delta_{\xi}\xbar\pi^{AB} & =\frac{\sqrt{\xbar g}}{2}\Big\{ b\Big[3\xbar h^{AB}-\xbar\triangle\,\xbar h^{AB}-2\xbar D^{C}\xbar D^{(A}\xbar h_{C}^{B)}-2\xbar D^{(A}\xbar\lambda^{B)}+\xbar D^{A}\xbar D^{B}\xbar h_{rr}+\xbar D^{A}\xbar D^{B}\xbar h\nonumber\\
 & \quad+\xbar g^{AB}\Big(-2\xbar h+2\xbar D_{A}\xbar\lambda^{A}-\xbar\triangle\,\xbar h_{rr}+\xbar D^{A}\xbar D_{B}\xbar h_{A}^{B}\Big)\Big]+\partial_{C}b\Big[\xbar D^{C}\xbar h^{AB}-2\xbar D^{(A}h^{B)C}\nonumber\\
 & \quad+\xbar g^{AB}\Big(-2\xbar\lambda^{C}-\xbar D^{C}\xbar h_{rr}-\xbar D^{C}\xbar h_{rr}+2\xbar D^{D}\xbar h_{D}^{C}\Big)\Big]\Big\}+\sqrt{\xbar g}\Big(\xbar D^{A}\xbar D^{B}T'-\xbar g^{AB}\xbar\triangle\,T'\Big)\,,
\end{align}
\end{itemize}
with
\begin{align}
\tilde{T}' &= \tilde{T}+\tilde{T}_{(b)}\,,\\
T' &= T+T_{(b)}\,.
\end{align}

It follows in particular from the above formulas that the transformation laws of the fields $\tilde{U}$ and
$\tilde{V}$ under diffeomorphisms with both normal and spatial components are
\begin{align}
\delta_{\xi,\xi^{i}}\tilde{U} & =\mathcal{L}_{Y}\tilde{U}-b\tilde{V}+\tilde{W}\,,\\
\delta_{\xi,\xi^{i}}\tilde{V} & =\mathcal{L}_{Y}\tilde{V}-\frac{1}{2}b\tilde{U}-\frac{1}{2}\partial_{A}b\xbar D^{A}\tilde{U}+\tilde{T}+\tilde{T}_{(b)}\,.
\end{align}

 \section{Canonical generator}
 \label{Sec:AsympSymm2}
 We now turn to the computation of the canonical generator $G_{\xi,\xi^{i}}\big[g_{ij},\pi^{ij}\big]$, which takes the form
\begin{equation}
G_{\xi,\xi^{i}}\big[g_{ij},\pi^{ij}\big]=\int d^{3}x\Big(\xi\mathcal{H}+\xi^{i}\mathcal{H}_{i}\Big)+Q_{\xi,\xi^{i}}\big[g_{ij},\pi^{ij}\big]\,,
\end{equation}
where the surface term must be determined, following standard Hamiltonian methods,  in such a way that $\iota_{\xi,\xi^{i}} \Omega = - d_V G$.  Here, $\iota_{\xi,\xi^{i}} \Omega$ is the internal contraction of the symplectic form with the phase space vector field $X_{\xi,\xi^{i}}$ associated with $(\xi,\xi^{i})$ and $d_V$  is the exterior derivative in phase space.  The equation guarantees $d_V (\iota_{\xi,\xi^{i}} \Omega) = 0$, i.e, $\mathcal L_{X_{\xi,\xi^{i}}} \Omega= 0$, which expresses the invariance of the symplectic structure under the phase space transformation (exactly and not up to surface terms at spatial infinity).

Because the symplectic form takes the standard bulk form $\Omega = \int d^3x \, d_V \pi^{ij} \wedge d_V g_{ij}$ (without surface contribution), this amounts to applying the method of \cite{Regge:1974zd} (see \cite{Henneaux:2018gfi} for more information).  Namely, one computes $d_V G_{\xi,\xi^{i}}\big[g_{ij},\pi^{ij}\big]$ and gets an equation for $d_V Q_{\xi,\xi^{i}}\big[g_{ij},\pi^{ij}\big]$ by requesting that the surface terms obtained through the integrations by parts necessary to bring $d_V G_{\xi,\xi^{i}}\big[g_{ij},\pi^{ij}\big]$ to the canonical form $\int d^{3}x (A^{ij} d_V g_{ij} + B_{ij} d_V \pi^{ij})$ be cancelled by $d_V Q_{\xi,\xi^{i}}\big[g_{ij},\pi^{ij}\big]$.  Of course, for this to work, one must get a finite and $d_V$-closed (``integrable'') expression for $d_V Q_{\xi,\xi^{i}}\big[g_{ij},\pi^{ij}\big]$.  In order  to check these properties and derive the surface term, one can of course use the boundary conditions.

The surface term that one gets from the variation of the bulk part of $G_{\xi,\xi^{i}}\big[g_{ij},\pi^{ij}\big]$ reads
$$-\oint d^{2}x\Big[2\sqrt{g}\xi\delta K+\sqrt{g}g^{BC}\delta g_{CA}\Big(\xi K_{B}^{A}+\frac{1}{\lambda}(\partial_{r}\xi-\lambda^{D}\partial_{D}\xi)\delta_{B}^{A}\Big)+2\xi^{i}\delta\pi_{i}^{r}-\xi^{r}\delta g_{jk}\pi^{jk}\Big]\,,$$
where we have made the change of notations $d_V \rightarrow \delta$ to follow the tradition in the field. 
It follows that the compensation condition that determines  $Q_{\xi,\xi^{i}}\big[g_{ij},\pi^{ij}\big]$ is
\begin{equation}
\delta Q_{\xi,\xi^{i}}\big[g_{ij},\pi^{ij}\big]=\oint d^{2}x\Big[2\sqrt{g}\xi\delta K+\sqrt{g}g^{BC}\delta g_{CA}\Big(\xi K_{B}^{A}+\frac{1}{\lambda}(\partial_{r}\xi-\lambda^{D}\partial_{D}\xi)\delta_{B}^{A}\Big)+2\xi^{i}\delta\pi_{i}^{r}-\xi^{r}\delta g_{jk}\pi^{jk}\Big]\,, \label{eq:VarQ}
\end{equation}
where $K_{B}^{A}$ is the extrinsic curvature of the $2$-spheres in a $2+1$ decomposition, see Appendix \ref{app-decomp}.

We shall now verify that with the above boundary conditions and improved diffeomorphism parameters, not only is the right-hand side of this equation finite but also integrable, yielding a finite, well-defined $Q_{\xi,\xi^{i}}\big[g_{ij},\pi^{ij}\big]$.

\subsection{Generators of spatial diffeomorphisms \label{space-like charge}}

We shall compute in great detail the generators of spatial diffeomorphisms to illustrate both the importance of the boundary conditions and the inner consistency of the formalism.
Inserting the variations of the fields under spatial diffeomorphisms in (\ref{eq:VarQ}) yields
\begin{align}
\delta Q_{\xi^{i}} & =2r\ln r\oint d^{2}xY_{A}\delta\pi_{\text{log}}^{rA}\\
& \quad +2r\oint d^{2}xY_{A}\delta\xbar\pi^{rA}\\
 & \quad+2\ln^{2}r\oint d^{2}x\left[Y_{A}\delta\left(\pi_{\text{log}(2)}^{rA}+\pi_{\log}^{rB}\theta_{B}^{A}\right)+\tilde{W}\delta\left(\pi_{\log}^{rr}-\xbar D_{A}\pi_{\log}^{rA}\right)\right]\\
 & \quad+2\ln r\oint d^{2}x\Big[Y_{A}\delta\left(\pi_{\text{log}(1)}^{rA}+\pi_{\log}^{rr}\xbar\lambda^{A}+\pi_{\log}^{rB}\xbar h_{B}^{A}+\xbar\pi^{rB}\theta_{B}^{A}\right) \nonumber \\
 & \qquad \qquad \qquad \qquad \qquad +\tilde{W}\delta\left(\xbar\pi^{rr}-\xbar D_{A}\xbar\pi^{rA}\right)+W\delta\pi_{\log}^{rr}+I_{A}\delta\pi_{\log}^{rA}\Big]\\
 & \quad+\oint d^{2}x\left[2Y_{A}\delta\xbar\pi_{(2)}^{rA}+2W\delta\xbar\pi^{rr}+2I_{A}\delta\xbar\pi^{rA}\right]\,,
\end{align}
where
\begin{equation}
\xbar\pi_{(2)}^{rA}=\pi_{(2)}^{rA}+\xbar\pi^{rB}\xbar h_{B}^{A}+\xbar\pi^{rr}\xbar\lambda^{A}\,.
\end{equation}
As it is manifest from this expression, the variation of the charge is plagued with four different types of potential divergences, which we must show are actually zero.  It is here that we shall need the faster decay of the constraints.

 We examine the potential divergences in turn, starting with the most ``dangerous'' ones.
\begin{itemize}
\item The coefficient of $r\ln r$ is zero by parity,  since $\pi_{\log}^{rA}$
is even while $Y_A$ is odd.
\item For the term proportional to $r$, we use the asymptotic expression $\xbar\pi^{rA}=(\xbar\pi^{rA})^{\text{even}}+\sqrt{\xbar g}\,\xbar D^{A}\tilde{V}-\sqrt{\xbar g}\,\xbar D^{A}V$.
Taking into account the parities of the fields, the integral reduces then to
\begin{equation}
\oint d^{2}x\sqrt{\xbar g}\,Y_{A}\xbar D^{A}V\,,
\end{equation}
which vanishes under integration by parts by using the Killing equation
$\xbar D_{A}Y^{A}=0$.\newline

\item For the term proportional to $\ln^{2}r$, we first observe that
the integral that involves $\tilde{W}$ vanishes by parity ($\tilde{W}$
is even). Using also that $Y_{A}\pi_{\log}^{rB}\theta_{B}^{A}$
is odd, we therefore find that this term reduces  to
\begin{equation}
\oint d^{2}x\,Y_{A}\pi_{\text{log}(2)}^{rA}\,.\label{eq:Int3}
\end{equation}
In order to show that this expression is zero,  we use the assumption that  the angular component $\mathcal H_A$
of the momentum constraint goes to zero as $o(r^{-1})$. In particular,
from $\mathcal{H}_{A}^{\text{log(1)}}=0$ (coefficient of $r^{-1}\ln r$ in $\mathcal{H}_{A}$),
we can relate the sub-subleading coefficient $\pi_{\text{log}(2)}^{rA}$
with the leading and subleading coefficients in the asymptotic expansion of the
fields. Indeed, the equation $\mathcal{H}_{A}^{\text{log(1)}}=0$ reads
\begin{align}
-4\pi_{\log(2)}^{rA}-2\xbar D_{C}\left(\pi_{\log}^{BC}\xbar h_{B}^{A}\right)+\pi_{\log}^{BC}\xbar D^{A}\xbar h_{BC}-2\xbar D_{C}\left(\xbar\pi^{BC}\theta_{B}^{A}\right)+\xbar\pi^{BC}\xbar D^{A}\theta_{BC}\\
-4\theta_{B}^{A}\pi_{\log}^{rB}-2\xbar D_{B}\left(\pi_{\log}^{rB}\xbar\lambda^{A}\right)-2\xbar D_{B}\pi_{\log(1)}^{AB}+\pi_{\log}^{rr}\xbar D^{A}\xbar h_{rr}+2\pi_{\log}^{rB}\xbar D^{A}\xbar\lambda_{B} & =0\,.\label{eq:2nd-line}
\end{align}
Now,  all the terms in the second line \eqref{eq:2nd-line} are even,
thus once they are inserted in the integrand $Y_{A}\pi_{\log(2)}^{rA}$,
they will vanish (under the integral). Hence, the integral \eqref{eq:Int3}
becomes
\begin{gather}
\frac{1}{4}\oint d^{2}xY_{A}\left[-2\xbar D_{C}\left(\pi_{\log}^{BC}\xbar h_{B}^{A}\right)+\pi_{\log}^{BC}\xbar D^{A}\xbar h_{BC}\right]+\frac{1}{4}\oint d^{2}xY_{A}\left[-2\xbar D_{C}\left(\xbar\pi^{BC}\theta_{B}^{A}\right)+\xbar\pi^{BC}\xbar D^{A}\theta_{BC}\right]\,.\label{eq:Int31}
\end{gather}

\begin{itemize}
\item[$\square$] Let us focus on the first integral
\begin{equation}
\frac{1}{4}\oint d^{2}xY_{A}\left[-2\xbar D_{C}\left(\pi_{\log}^{BC}\xbar h_{B}^{A}\right)+\pi_{\log}^{BC}\xbar D^{A}\xbar h_{BC}\right]\,.\label{eq:Int32}
\end{equation}
Using again parity arguments, we see that only the odd part of $\xbar h_{AB}$ contributes,
which is ``pure gauge'', i.e.,
\begin{equation}
\xbar h_{AB}^{\text{odd}}=2\left(\xbar D_{A}\xbar D_{B}U_{\text{odd}}+\xbar g_{AB}U_{\text{odd}}\right)\,.
\end{equation}
Recall that $U_{A}^{\text{even}}=\xbar D_{A}U_{\text{odd}}$. Thus,
by using that $Y_{A}$ is a Killing vector of the 2-sphere,
\begin{equation}
\xbar D_{(A}Y_{B)}=0\,,\qquad\xbar D_{A}\xbar D_{B}Y_{C}=-\xbar g_{AB}Y_{C}+\xbar g_{AC}Y_{B}\,,
\end{equation}
that $\pi_{\text{log}}^{AB}=\sqrt{\xbar g}(\xbar D^{A}\xbar D^{B}\tilde{V}-\xbar g^{AB}\xbar\triangle\tilde{V})$,
and integrating by parts, we get that the first integral in \eqref{eq:Int32}
becomes
\begin{equation}
\frac{1}{4}\oint d^{2}xY_{A}\left[-2\xbar D_{C}\left(\pi_{\log}^{BC}\xbar h_{B}^{A}\right)\right]=\oint d^{2}x\sqrt{\xbar g}Y_{A}\left(\xbar D^{A}\xbar\triangle\,\tilde{V}+2\xbar D^{A}\tilde{V}\right)U_{\text{odd}}\,.
\end{equation}
We also used the property that the commutator of two covariant derivatives on the
round 2-sphere implies $[\xbar D_{A},\xbar D_{B}]\xbar D_{C}\tilde{V}=-\xbar g_{BC}\xbar D_{A}\tilde{V}+\xbar g_{AC}\xbar D_{B}\tilde{V}$.
For the second integral in \eqref{eq:Int32}, besides of making use
of exactly the same ingredients as before, we must also use the expression for the commutator of
two covariant derivatives of a (symmetric) tensor field, which reads
\begin{equation}
[\xbar D_{B},\xbar D^{A}]\pi_{\log}^{BC}=2\pi_{\log}^{AC}-\pi_{\log}\xbar g^{AC}\,.
\end{equation}
We obtain then the equality
\begin{equation}
\frac{1}{4}\oint d^{2}xY_{A}\pi_{\log}^{BC}\xbar D^{A}\xbar h_{BC}=-\oint d^{2}x\sqrt{\xbar g}\,Y_{A}\left(\xbar D^{A}\xbar\triangle\,\tilde{V}+2\xbar D^{A}\tilde{V}\right)U_{\text{odd}}\,.
\end{equation}
Adding the two terms in  \eqref{eq:Int32}, we get that the first integral in \eqref{eq:Int31} vanishes,
\begin{equation}
\frac{1}{4}\oint d^{2}xY_{A}\left[-2\xbar D_{C}\left(\pi_{\log}^{BC}\xbar h_{B}^{A}\right)+\pi_{\log}^{BC}\xbar D^{A}\xbar h_{BC}\right]=0\,.
\end{equation}
\item[$\square$] Now, for the second integral in \eqref{eq:Int31},
\begin{equation}
\frac{1}{4}\oint d^{2}xY_{A}\left[-2\xbar D_{C}\left(\xbar\pi^{BC}\theta_{B}^{A}\right)+\xbar\pi^{BC}\xbar D^{A}\theta_{BC}\right]\,,\label{eq:Int33}
\end{equation}
the steps are very similar. We first note that (by parity) only the
odd part of $\xbar\pi^{rA}$ and the even part of $\xbar\pi^{AB}$
contribute. These read
\begin{align}
\xbar\pi_{\text{odd}}^{rA} & =-\sqrt{\xbar g}\,\xbar D^{A}V_{\text{even}}\,,\\
\xbar\pi_{\text{even}}^{AB} & =\sqrt{\xbar g}\,(\xbar D^{A}\xbar D^{B}V_{\text{even}}-\xbar g^{AB}\xbar\triangle V_{\text{even}})\,.
\end{align}

Making use of the identities $\xbar D_{C}\xbar\pi_{\text{even}}^{BC}=-\xbar\pi_{\text{odd}}^{rB}$
and $\xbar D_{A}\theta_{BC}=\xbar D_{B}\theta_{AC}$, the integral
\eqref{eq:Int33} can be re-written as
\begin{equation}
\frac{1}{4}\oint d^{2}x\left(2Y_{A}\xbar\pi_{\text{odd}}^{rB}\theta_{B}^{A}-Y_{A}\xbar\pi_{\text{even}}^{BC}\xbar D_{C}\theta_{B}^{A}\right)\,.
\end{equation}
It is easy to see that the first term of the above integral becomes
(after integration by parts and using the Killing equation)
\begin{equation}
\frac{1}{4}\oint d^{2}x\left(2Y_{A}\xbar\pi_{\text{odd}}^{rB}\theta_{B}^{A}\right)=\frac{1}{2}\oint d^{2}x\sqrt{\xbar g}Y_{A}\xbar D_{B}\theta^{AB}V_{\text{even}}\,.
\end{equation}
The second term, on the other hand, can be written as (after integration
by parts)
\begin{equation}
\frac{1}{4}\oint d^{2}x\left(-Y_{A}\xbar\pi_{\text{even}}^{BC}\xbar D_{C}\theta_{B}^{A}\right)=\frac{1}{4}\oint d^{2}x\sqrt{\xbar g}\left[\xbar D^{C}\xbar D^{B}\left(Y_{A}\xbar D_{C}\theta_{B}^{A}\right)+\xbar\triangle\left(Y_{A}\xbar D^{A}\theta\right)\right]V_{\text{even}}\,.
\end{equation}
Expanding the derivatives in the above integral, using the identities
of the Killing vector $Y^{A}$ and the commutator of two covariant
derivatives, we get that the above term reduces to
\begin{equation}
\frac{1}{4}\oint d^{2}x\left(-Y_{A}\xbar\pi_{\text{even}}^{BC}\xbar D_{C}\theta_{B}^{A}\right)=-\frac{1}{2}\oint d^{2}x\sqrt{\xbar g}\,Y_{A}\xbar D_{B}\theta^{AB}V_{\text{even}}\,.
\end{equation}

Thus, the integral \eqref{eq:Int33} also vanishes
\begin{equation}
\frac{1}{4}\oint d^{2}xY_{A}\left[-2\xbar D_{C}\left(\xbar\pi_{\text{even}}^{BC}\theta_{B}^{A}\right)+\xbar\pi_{\text{even}}^{BC}\xbar D^{A}\theta_{BC}\right]=0\,.
\end{equation}
\end{itemize}
Consequently, we have proven that
\begin{equation}
\oint d^{2}xY_{A}\pi_{\text{log}(2)}^{rA}=0\,,
\end{equation}
so that the  divergent term involving $\ln^{2}r$ is actually absent.

\item For the divergent term proportional to $\ln r$, we proceed along lines that adopt similar arguments. For the term proportional to $\tilde{W}$, we use the constraint $\mathcal{H}_{(1)}^{r}\approx0$ in \eqref{eq:Gr1}, to get
\be
\oint d^{2}x\tilde{W\delta}\big(\xbar\pi^{rr}-\xbar D_{A}\xbar\pi^{rA}\big)  =\oint d^{2}x\tilde{W}\delta\left(\xbar\pi^{rr}-\xbar\pi_{A}^{A}+\pi_{\text{\ensuremath{\log}}}^{rr}\right)\,,
\ee
Note that by virtue of the parity conditions of the leading coefficients of the momenta, the quantity
$\left(\xbar\pi^{rr}-\xbar\pi_{A}^{A}+\pi_{\text{\ensuremath{\log}}}^{rr}\right)$
is strictly odd. Therefore, since $\tilde{W}$ is even,  the integral vanishes. 
Consider now the terms involving $W$ and $I_{A}$.  These become
\begin{equation}
\oint d^{2}x\Big(W^{\text{odd}}\delta\pi_{\text{log}}^{rr}+I_{A}^{\text{even}}\delta\pi_{\text{log}}^{rA}\Big)=\oint d^{2}xW^{\text{\text{odd}}}\delta\Big(\pi_{\text{log}}^{rr}-\xbar D_{A}\pi_{\text{log}}^{rA}\Big)=0\,,
\end{equation}
 where the relation $I_{A}^{\text{even}}=\xbar D_{A}W^{\text{\text{odd}}}$ was used.
Thus, the total integral reduces to the term coming from the spatial
rotations
\begin{equation}
\oint d^{2}xY_{A}\left(\pi_{\text{log}(1)}^{rA}+\pi_{\log}^{rr}\xbar\lambda^{A}+\pi_{\log}^{rB}\xbar h_{B}^{A}+\xbar\pi^{rB}\theta_{B}^{A}\right)\,.
\end{equation}

Since the second term $Y_A(\pi_{\log}^{rr}\xbar\lambda^{A})$
is odd,  it does not contribute. Thus, our remaining duty is
to show that
\begin{equation}
\oint d^{2}xY_{A}\left(\pi_{\text{log}(1)}^{rA}+\pi_{\log}^{rB}\xbar h_{B}^{A}+\xbar\pi^{rB}\theta_{B}^{A}\right)=0\,.\label{eq:Int4}
\end{equation}

As for the divergence proportional to $\ln^{2}r$,  we will make use of the fast fall-off of
the angular components of the momentum constraint. In this case, we
need the equation $\mathcal{H}_{A}^{(1)} = 0$ expressing that the $r^{-1}$ term in $\mathcal{H}_{A}$ vanishes,
\begin{align}
-2\pi_{\log(1)}^{rA}-2\pi_{\log}^{rB}\xbar h_{B}^{A}-2\xbar\pi^{rB}\theta_{B}^{A}-2\xbar D_{B}(\xbar\pi^{rB}\xbar\lambda^{A})+2\xbar\pi^{rB}\xbar D^{A}\xbar\lambda_{B}\\
-2\xbar D_{B}(\xbar\pi^{BC}\xbar h_{C}^{A})+\xbar\pi^{rr}\xbar D^{A}\xbar h_{rr}+\xbar\pi^{BC}\xbar D^{A}\xbar h_{BC} & =0\,.
\end{align}

Replacing $\pi_{\log(1)}^{rA}$ in \eqref{eq:Int4}, we find that this
integral becomes
\begin{align}
&\oint d^{2}xY_{A}\left[-\xbar D_{B}(\xbar\pi^{rB}\xbar\lambda^{A})+\xbar\pi^{rB}\xbar D^{A}\xbar\lambda_{B}\right]\nonumber\\
&+\frac{1}{2}\oint d^{2}xY_{A}\left[-2\xbar D_{B}(\xbar\pi^{BC}\xbar h_{C}^{A})+\xbar\pi^{rr}\xbar D^{A}\xbar h_{rr}+\xbar\pi^{BC}\xbar D^{A}\xbar h_{BC}\right]\,.
\end{align}
In the first integral
\begin{equation}
\oint d^{2}xY_{A}\left[-\xbar D_{B}(\xbar\pi^{rB}\xbar\lambda^{A})+\xbar\pi^{rB}\xbar D^{A}\xbar\lambda_{B}\right]\,,
\end{equation}
we see that only the odd part of $\xbar\pi^{rB}$ contributes, which
is ``pure gauge'': $\xbar\pi_{\text{odd}}^{rA}=-\sqrt{\xbar g}\,\xbar D^{A}V_{\text{even}}$.
Thus using the identities for $Y^{A}$ and integrating by parts, we get that the
above integral becomes
\begin{equation}
\oint d^{2}xY_{A}\left[-\xbar D_{B}(\xbar\pi^{rB}\xbar\lambda^{A})+\xbar\pi^{rB}\xbar D^{A}\xbar\lambda_{B}\right]=\oint d^{2}xY_{A}\xbar D^{A}\xbar D^{B}\xbar\lambda_{B}V_{\text{even}}\,. \label{eq:Firstlambda}
\end{equation}

The second integral
\begin{equation}
\frac{1}{2}\oint d^{2}xY_{A}\left[-2\xbar D_{B}(\xbar\pi^{BC}\xbar h_{C}^{A})+\xbar\pi^{rr}\xbar D^{A}\xbar h_{rr}+\xbar\pi^{BC}\xbar D^{A}\xbar h_{BC}\right]\,,
\end{equation}
can be split in two integrals as (by making use of the asymptotic form of the variables)
\begin{gather}
\frac{1}{2}\oint d^{2}xY_{A}\left[-2\xbar D_{B}(\xbar\pi_{\text{odd}}^{BC}\xbar h_{C}^{\text{odd}A})+\xbar\pi_{\text{odd}}^{BC}\xbar D^{A}\xbar h_{BC}^{\text{odd}}\right]\label{eq:Int41}\\
+\frac{1}{2}\oint d^{2}xY_{A}\left[-2\xbar D_{B}(\xbar\pi_{\text{even}}^{BC}\xbar h_{C}^{\text{even}A})+\xbar\pi_{\text{even}}^{rr}\xbar D^{A}\xbar h_{rr}+\xbar\pi_{\text{even}}^{BC}\xbar D^{A}\xbar h_{BC}^{\text{even}}\right]\,.\label{eq:Int42}
\end{gather}
For the terms in \eqref{eq:Int41}, we use the identities
for the Killing vector $Y^{A}$,  the relation $\xbar D_{C}\xbar\pi_{\text{odd}}^{BC}=-\xbar\pi_{\text{even}}^{rB}-\pi_{\log}^{rB}$,
as well as
\begin{equation}
\xbar h_{AB}^{\text{odd}}=2\left(\xbar D_{A}\xbar D_{B}U_{\text{odd}}+\xbar g_{AB}U_{\text{odd}}\right)\,,
\end{equation}
to show that
\begin{equation}
\frac{1}{2}\oint d^{2}xY_{A}\left[-2\xbar D_{B}(\xbar\pi_{\text{odd}}^{BC}\xbar h_{C}^{\text{odd}A})+\xbar\pi_{\text{odd}}^{BC}\xbar D^{A}\xbar h_{BC}^{\text{odd}}\right]=0\,,
\end{equation}
after integration by parts and some straightforward algebra.

For the terms in \eqref{eq:Int42}, we use the asymptotic conditions
\begin{align}
\xbar\pi_{\text{even}}^{rr} & =-\sqrt{\xbar g}\,\xbar\triangle\,V_{\text{even}\,,}\\
\xbar\pi_{\text{odd}}^{rA} & =-\sqrt{\xbar g}\,\xbar D^{A}V_{\text{even}}\,,\\
\xbar\pi_{\text{even}}^{AB} & =\sqrt{\xbar g}\,(\xbar D^{A}\xbar D^{B}V_{\text{even}}-\xbar g^{AB}\xbar\triangle V_{\text{even}})\,,
\end{align}
to derive
\begin{multline}
\frac{1}{2}\oint d^{2}xY_{A}\left[-2\xbar D_{B}(\xbar\pi_{\text{even}}^{BC}\xbar h_{C}^{\text{even}A})+\xbar\pi_{\text{even}}^{rr}\xbar D^{A}\xbar h_{rr}^{\text{}}+\xbar\pi_{\text{even}}^{BC}\xbar D^{A}\xbar h_{BC}^{\text{even}}\right]\\
=\frac{1}{2}\oint d^{2}xY_{A}\xbar D^{A}\left(\xbar D^{B}\xbar D^{C}\xbar h_{BC}^{\text{even}}-\xbar\triangle\,\xbar h^{\text{even}}-\xbar\triangle\,\xbar h_{rr}\right)V_{\text{even}}\,,
\end{multline}
after some direct algebra and integrations by parts.

Putting  together this result and (\ref{eq:Firstlambda}), we obtain the required result that the integral 
\begin{multline}
\frac{1}{2}\oint d^{2}xY_{A}\left[-2\xbar D_{B}(\xbar\pi^{BC}\xbar h_{C}^{A})+\xbar\pi^{rr}\xbar D^{A}\xbar h_{rr}+\xbar\pi^{BC}\xbar D^{A}\xbar h_{BC}\right]\\
=\frac{1}{2}\oint d^{2}xY_{A}\xbar D^{A}\left(\xbar D^{B}\xbar D^{C}\xbar h_{BC}^{\text{even}}-\xbar\triangle\,\xbar h^{\text{even}}-\xbar\triangle\,\xbar h_{rr}+2\xbar D^{B}\xbar\lambda_{B}\right)V_{\text{even}}=0\,.
\end{multline}
vanishes due to the asymptotic form of the Hamiltonian constraint. Thus, we have established \eqref{eq:Int4} and there is no divergence proportional to $\ln r$.
\end{itemize}

This completes the proof that the surface term in the generator of spatial asymptotic symmetries is free of divergences.  It is manifestly integrable and 
given by
\begin{align}
Q_{\xi^{i}} & =\oint d^{2}x\left(2Y_{A}\xbar\pi_{(2)}^{rA}+2W\xbar\pi^{rr}+2I_{A}\xbar\pi^{rA}\right)\,.
\end{align}
We note that $\tilde W$ does not appear in this expression.  This indicates that the corresponding diffeomorphism is a proper gauge transformation.  One could use this proper gauge freedom to set $\theta_{AB} = 0$ since $\tilde U$ transforms as $\tilde U \rightarrow \tilde U + \tilde W$ under the logarithmic supertranslations generated by $\tilde W$.

\subsection{Generators of normal  diffeomorphisms\label{time-like charge}}

The equation for $\delta Q_{\xi}$ reads, in the case of normal diffeomorphisms 
\begin{align}
\delta Q_{\xi} & =r\ln r\oint d^{2}x\sqrt{\xbar g}\,b\delta\theta \\
& \quad +r\oint d^{2}x\sqrt{\xbar g}\,b\delta\left(2\xbar h_{rr}+\xbar h+2\xbar D_{A}\xbar\lambda^{A}-\theta\right)\\
 & \quad+\ln^{2}r\oint d^{2}x\sqrt{\xbar g}\,b\delta\left(2\theta^{(2)}+2k_{\log(2)}^{(2)}+\frac{1}{4}\theta^{2}-\frac{3}{4}\theta_{A}^{B}\theta_{B}^{A}\right)\\
 & \quad +\ln r \delta Q_{\log}
   +\delta Q_{(0)}
  + \delta Q_{\xi_{(b)}^{i}}
\end{align}
where
\begin{align}
\delta Q_{\log} & =\oint d^{2}x\sqrt{\xbar g}\Big[2\tilde{T}\delta\left(\xbar h_{rr}+\xbar D_{A}\xbar\lambda^{A}\right)+b\Big(2\delta\sigma+2\delta k_{\log(1)}^{(2)}+\frac{1}{2}\delta(-3\xbar h_{A}^{B}\theta_{B}^{A}+h\theta)\\
 & \quad-\xbar h_{rr}\delta\theta+\frac{1}{4}\delta(3\theta_{A}^{B}\theta_{B}^{A}-\theta^{2})+\theta\xbar D_{A}\delta\xbar\lambda^{A}-2\theta_{AB}\xbar D^{A}\delta\xbar\lambda^{B}-\delta\theta_{AB}\xbar D^{A}\xbar\lambda^{B}\Big)-\partial_{A}b\xbar\lambda^{A}\delta\theta\Big]\,,
\end{align}
and
\begin{align}
\delta Q_{(0)} & =\oint d^{2}x\sqrt{\xbar g}\,\Big[\tilde{T}\delta\xbar h+2T\delta\big(\xbar h_{rr}+\xbar D_{A}\xbar\lambda^{A}-\frac{1}{2}\theta\big)+b\Big(\frac{1}{2}\xbar h\delta\xbar h-\frac{3}{2}\xbar h_{A}^{B}\delta\xbar h_{B}^{A}-\xbar h_{rr}\delta\xbar h+2\delta h^{(2)}\label{eq:Qlog}\\
 & \quad+2\delta k^{(2)}+\xbar h_{A}^{B}\delta\theta_{B}^{A}-\frac{1}{2}\xbar h\delta\theta+\frac{1}{2}\theta_{A}^{B}\delta\xbar h_{B}^{A}+\xbar h\,\xbar D_{A}\delta\xbar\lambda^{A}-2\xbar h_{A}^{B}\xbar D_{B}\delta\xbar\lambda^{A}-\delta\xbar h_{A}^{B}\xbar D_{B}\xbar\lambda^{A}\Big)-\partial_{A}b\xbar\lambda^{A}\delta\xbar h\Big]\,,
\end{align}
and where $\delta Q_{\xi_{(b)}^{i}}$ is the contribution due to the  correcting gauge transformation
that maintains the asymptotic form of $g_{rA}$.  Using the formulas of the previous section,with 
\begin{equation}
\xi_{(b)}^{A}=\frac{\ln r}{r}\frac{2b}{\sqrt{\xbar g}}\pi_{\log}^{rA}+\frac{1}{r}\frac{2b}{\sqrt{\xbar g}}\xbar\pi^{rA}\,,
\end{equation}
one finds
\begin{equation}
Q_{\xi_{(b)}^{i}}=\ln r\oint d^{2}x\sqrt{\xbar g}\left(4b\xbar D_{A}\tilde{V}\xbar D^{A}V\right)+\oint d^{2}x\frac{2b}{\sqrt{\xbar g}}\xbar\pi^{rA}\xbar\pi_{A}^{r}\,.
\end{equation}
The corrective term $\xi_{(b)}=\ln r\tilde{T}_{(b)}+T_{(b)}$, necessary for integrability, has not been included yet since we want to show  why it must be added.

The charge of normal diffeomorphisms possesses thus a priori the same different types of divergences as for spatial ones.
However, it is easy to see that the coefficients of $r\ln r$
and $r$ are zero by using the parity of the asymptotic fields and  the equation
$\xbar D_{A}\xbar D_{B}b+\xbar g_{AB}b=0$  for the boost parameter. 

We show in Appendix \ref{app-van} that the remaining divergences also cancel.  The computation is very similar to the one which we explicitly carried out in the previous section for the spatial diffeomorphism charges. Key in establishing the absence of divergences in $\delta Q_{\xi}$ is the faster  fall-off of the Hamiltonian constraint, which is part of our boundary conditions. 

Once all dangerous terms have been shown to vanish, one finds that $\delta Q_{\xi}$ reduces to the finite term:
\begin{align}
\delta Q_{\xi} & =\oint d^{2}x\sqrt{\xbar g}\,\Big[\tilde{T}\delta\xbar h+2T\delta\big(\xbar h_{rr}+\xbar D_{A}\xbar\lambda^{A}-\frac{1}{2}\theta\big)+b\Big(\frac{1}{2}\xbar h\delta\xbar h-\frac{3}{2}\xbar h_{A}^{B}\delta\xbar h_{B}^{A}-\xbar h_{rr}\delta\xbar h+2\delta h^{(2)}\\
 & \quad+2\delta k^{(2)}+\xbar h_{A}^{B}\delta\theta_{B}^{A}-\frac{1}{2}\xbar h\delta\theta+\frac{1}{2}\theta_{A}^{B}\delta\xbar h_{B}^{A}+\xbar h\,\xbar D_{A}\delta\xbar\lambda^{A}-2\xbar h_{A}^{B}\xbar D_{B}\delta\xbar\lambda^{A}-\delta\xbar h_{A}^{B}\xbar D_{B}\xbar\lambda^{A}\Big)-\partial_{A}b\xbar\lambda^{A}\delta\xbar h\Big]\,,
\end{align}
While the terms proportional to $\tilde T$ and $T$ are integrable, those involving the boosts, which can be rewitten as
\begin{align}
\delta Q_{b} & =\oint d^{2}x\sqrt{\xbar g}\,\Big\{ b\,\delta\Big[2k^{(2)}+2h^{(2)}-\xbar h\,\xbar h_{rr}+\frac{1}{4}\xbar h^{2}-\frac{3}{4}\xbar h_{A}^{B}\xbar h_{B}^{A}-2\xbar h_{AB}\Big(\xbar D^{A}\xbar\lambda^{B}-\frac{1}{2}\theta^{AB}\Big)+\frac{2}{\xbar g}\xbar\pi^{rA}\xbar\pi_{A}^{r}\Big]\\
 & \quad\qquad\qquad\quad+b\xbar h\delta\Big(\xbar h_{rr}+\xbar D_{A}\xbar\lambda^{A}-\frac{1}{2}\theta\Big)-\partial_{A}b\xbar\lambda^{A}\delta\xbar h+b\Big(\xbar D_{A}\xbar\lambda_{B}-\frac{1}{2}\theta_{AB}\Big)\delta\xbar h^{AB}\Big\}\,.
\end{align}
are not integrable due to the second line.

In order to get rid of these non-integrable terms,
we perform the correcting diffeomorphism
$\xi_{(b)}=\ln r\tilde{T}_{(b)}+T_{(b)}$ 
announced above, to get that the boost charge is
\begin{equation}
Q_{b}=\oint d^{2}x\sqrt{\xbar g}\,b\,\Big[2k^{(2)}+2h^{(2)}-\xbar h\,\xbar h_{rr}+\frac{1}{4}\xbar h^{2}-\frac{3}{4}\xbar h_{A}^{B}\xbar h_{B}^{A}-2\xbar h_{AB}\Big(\xbar D^{A}\xbar\lambda^{B}-\frac{1}{2}\theta^{AB}\Big)+\frac{2}{\xbar g}\xbar\pi^{rA}\xbar\pi_{A}^{r}\Big]\,.
\end{equation}

The total surface term for normal diffeomorphisms then reads
\begin{align}
Q_{\xi} & =\oint d^{2}x\sqrt{\xbar g}\,\Big\{\tilde{T}\xbar h+2T\big(\xbar h_{rr}+\xbar D_{A}\xbar\lambda^{A}-\frac{1}{2}\theta\big)+b\,\Big[2k^{(2)}+2h^{(2)}-\xbar h\,\xbar h_{rr}\\
 & \quad\qquad\qquad\quad+\frac{1}{4}\xbar h^{2}-\frac{3}{4}\xbar h_{A}^{B}\xbar h_{B}^{A}-2\xbar h_{AB}\Big(\xbar D^{A}\xbar\lambda^{B}-\frac{1}{2}\theta^{AB}\Big)+\frac{2}{\xbar g}\xbar\pi^{rA}\xbar\pi_{A}^{r}\Big]\Big\}\,.
\end{align}

Collecting the surface integrals for spatial and normal diffeomorphisms, we conclude that the total charge to be added to the bulk term $\int d^{3}x(\xi\mathcal{H}+\xi^{i}\mathcal{H}_{i})$ when the diffeomorhisms take the asymptotic form (\ref{eq:AsympDiff1})-(\ref{eq:AsympDiff3}) is given by
\begin{align}
Q_{\xi,\xi^{i}} & =\oint d^{2}x\,\Big\{\sqrt{\xbar g}\,\tilde{T}\xbar h+2\sqrt{\xbar g}\,T\big(\xbar h_{rr}+\xbar D_{A}\xbar\lambda^{A}-\frac{1}{2}\theta\big)+\sqrt{\xbar g}\,b\,\Big[2k^{(2)}+2h^{(2)}-\xbar h\,\xbar h_{rr}\\
 & \quad\qquad\qquad\quad+\frac{1}{4}\xbar h^{2}-\frac{3}{4}\xbar h_{A}^{B}\xbar h_{B}^{A}-2\xbar h_{AB}\Big(\xbar D^{A}\xbar\lambda^{B}-\frac{1}{2}\theta^{AB}\Big)+\frac{2}{\xbar g}\xbar\pi^{rA}\xbar\pi_{A}^{r}\Big]\\
 & \quad\qquad\qquad\quad+2Y_{A}\xbar\pi_{(2)}^{rA}+2W\xbar\pi^{rr}+2I_{A}\xbar\pi^{rA}\Big\}\,.
\end{align}
We note  - also as announced - that the $\ell = 1$ spherical harmonic component of $\tilde T$ drops out from the charge and defines therefore a proper gauge transformation since the odd part of $\xbar h$ annihilates it (under the integral sign).

\section{Logarithmic BMS Algebra}
\label{Sec:AsympSymm3}

\subsection{Rewriting of charges}
It is useful to rewrite $Q_{\xi,\xi^{i}}$ in a way that makes manifest the proper gauge transformations. 
By direct computation using the asymptotic form of the fields, one finds that the canonical
charge can be recast as 
\begin{align}
Q_{\xi,\xi^{i}} & =\oint d^{2}x\Big\{\sqrt{\xbar g}\,b\,\Big[2k^{(2)}+2h^{(2)}-\xbar h\,\xbar h_{rr}+\frac{1}{4}\xbar h^{2}-\frac{3}{4}\xbar h_{A}^{B}\xbar h_{B}^{A}-2\xbar h_{AB}\Big(\xbar D^{A}\xbar\lambda^{B}-\frac{1}{2}\theta^{AB}\Big)+\frac{2}{\xbar g}\xbar\pi^{rA}\xbar\pi_{A}^{r}\Big]\nonumber \\
 & \quad\qquad\quad\quad+2Y_{A}\xbar\pi_{(2)}^{rA}+2\sqrt{\xbar g}\,T^{\text{even}}\Big(\xbar h_{rr}+\xbar D_{A}\xbar\lambda^{A}-\frac{1}{2}\theta\Big)+2W^{\text{odd}}\Big(\xbar\pi^{rr}-\xbar\pi+\pi_{\log}^{rr}\Big)\nonumber\\
& \quad\qquad\quad\quad+2\sqrt{\xbar g}\,T_{\log}U-2\sqrt{\xbar g}\,W_{\log}V\Big\}\,,
\end{align}
with
\begin{align}
k^{(2)} & =-\frac{1}{2}\sigma+h_{rr}^{(2)}-\frac{3}{4}\xbar h_{rr}^{2}+\frac{1}{4}\xbar h_{rr}\xbar h-\xbar\lambda_{A}\xbar\lambda^{A}+\frac{1}{4}\xbar h_{rr}\theta\\
 & \quad+\xbar D_{A}h_{r}^{(2)A}+\frac{1}{2}\xbar\lambda_{A}\xbar D^{A}\xbar h-\frac{1}{2}\xbar h_{rr}\xbar D_{A}\xbar\lambda^{A}-\xbar\lambda^{A}\xbar D_{B}\xbar h_{A}^{B}\,,
\end{align}
where $T_{\log}=(\xbar\triangle+2)\tilde{T}=\text{odd}$ and $W_{\text{log}}=\xbar\triangle\,W^{\text{even}}-\xbar D^{A}I_{A}^{\text{odd}}=\text{even}$. 
Boosts, spatial rotations and supertranslations parametrized by $T^{\text{even}}$ and $W^{\text{odd}}$ are all improper, as in the case where logarithmic supertranslations are not included \cite{Henneaux:2018hdj}. The new improper gauge symmetries arising from the extension of the formalism are the logarithmic supertranslations in time, parametrized by the odd function $T_{\log}$, which has no $\ell = 1$ spherical harmonic component, and the subleading supertranslations parametrized by the even function $W_{\text{log}}$, which has no $\ell = 0$ spherical harmonic component.  These subleading transformations become improper and can be triggred because the condition $\xbar \lambda_A = 0$ is not any more necessary for integrability when the new asymptotc fields are introduced.

Note also that the ambiguity in $U$ and $V$ mentioned in Section  \ref{Asymptotia} is indeed irrelevant for the  charges, as it should.

\subsection{Transformations of the parameters}

The bracket of any two canonical generators is computed by evaluating the variation of one generator under the canonical transformation generated by the other.
One finds in this manner
\begin{equation}
\{G_{\xi_{1}}\big[g_{ij},\pi^{ij}\big],G_{\xi_{2}}\big[g_{ij},\pi^{ij}\big]\}=G_{\hat{\xi}}+\mathcal{C}_{\{\xi_{1},\xi_{2}\}}\,,
\end{equation}
where the terms $\mathcal{C}_{\{\xi_{1},\xi_{2}\}}$ correspond to
central terms between the usual and the ``logarithmic'' supertranslations, and where 
\begin{align}
\hat{Y}^{A} & =Y_{1}^{B}\partial_{B}Y_{2}^{A}+\xbar g^{AB}b_{1}\partial_{B}b_{2}-(1\leftrightarrow2)\,, \label{eq:Lorentz1}\\
\hat{b} & =Y_{1}^{B}\partial_{B}b_{2}-(1\leftrightarrow2)\,,\label{eq:Lorentz2}\\
\hat{T} & =Y_{1}^{A}\partial_{A}T_{2}-3b_{1}W_{2}-\partial_{A}b_{1}\xbar D^{A}W_{2}-b_{1}\xbar\triangle W_{2}-(1\leftrightarrow2)\,,\label{eq:SuperTrans1}\\
\hat{W} & =Y_{1}^{A}\partial_{A}W_{2}-b_{1}T_{2}-(1\leftrightarrow2)\,,\label{eq:SuperTrans2}\\
\hat{T}^{\log} & =Y_{1}^{A}\partial_{A}T_{2}^{\log}-3b_{1}W_{2}^{\log}-\partial_{A}b_{1}\xbar D^{A}W_{2}^{\log}-b_{1}\xbar\triangle W_{2}^{\log}-(1\leftrightarrow2)\,,\label{eq:SuperLogTrans1}\\
\hat{W}^{\log} & =Y_{1}^{A}\partial_{A}W_{2}^{\log}-b_{1}T_{2}^{\log}-(1\leftrightarrow2)\,.\label{eq:SuperLogTrans2}
\end{align}
where from now on $W \equiv W^{\text{odd}}$.
 
In order to arrive at these formulas, 
the following transformation laws were found useful,
\begin{align}
&\delta_{\xi.\xi^{i}}\Big(\xbar h_{rr}+\xbar D_{A}\xbar\lambda^{A}-\frac{1}{2}\theta_{A}^{A}\Big)  =\mathcal{L}_{Y}\Big(\xbar h_{rr}+\xbar D_{A}\xbar\lambda^{A}-\frac{1}{2}\theta_{A}^{A}\Big)+\frac{b}{\sqrt{\xbar g}}\Big(\xbar\pi^{rr}-\xbar\pi_{A}^{A}+\pi_{\log}^{rr}\Big)+W_{\log}\,,\\
&\delta_{\xi,\xi^{i}}\Big(\xbar\pi^{rr}-\xbar\pi_{A}^{A}+\pi_{\text{\ensuremath{\log}}}^{rr}\Big)  =\mathcal{L}_{Y}\Big(\xbar\pi^{rr}-\xbar\pi_{A}^{A}+\pi_{\text{\ensuremath{\log}}}^{rr}\Big) \nonumber \\
& \qquad \qquad \qquad \qquad +\sqrt{\xbar g}\Big[b\xbar\triangle+\partial_{A}b\xbar D^{A}+3b\Big]\Big(\xbar h_{rr}+\xbar D_{A}\xbar\lambda^{A}-\frac{1}{2}\theta_{A}^{A}\Big)-\sqrt{\xbar g}\,T_{\log}\,,\\
&\delta_{\xi,\xi^{i}}U  =Y^{A}\partial_{A}U-bV+W\,,\\
&\delta_{\xi,\xi^{i}}V  =Y^{A}\partial_{A}V-b\xbar\triangle U-\partial_{A}b\xbar D^{A}U-3bU+T\,,
\end{align}
as well as the transformation rules of the sub-subleading terms given in Appendix \ref{app-transf}.

The relations (\ref{eq:Lorentz1}) and (\ref{eq:Lorentz2}) encode the homogeneous Lorentz algebra. The relations (\ref{eq:SuperTrans1}) and (\ref{eq:SuperTrans2}) indicate how the ordinary supertranslation parameters transform under the homogeneous Lorentz group, described in $3+1$ Hamiltonian terms.  They match the relations found previously in \cite{Troessaert:2017jcm,Henneaux:2018hdj}.  

Finally, the relations (\ref{eq:SuperLogTrans1}) and (\ref{eq:SuperLogTrans2}) define the Lorentz representation of the logarithmic supertranslations.  As we have indicated, ${T}^{\log}$ is odd and starts in a spherical harmonic expansion at $\ell = 3$.  Similarly, ${W}^{\log}$ is even and starts at $\ell = 2$. It is easy to verify that (\ref{eq:SuperLogTrans1}) has no $\ell=1$ component and that  (\ref{eq:SuperLogTrans2}) has no $\ell = 0$ component, so that these properties are preserved under Lorentz transformations.  Since the minimum spin in the representation of $({T}^{\log}, {W}^{\log})$  is $\ell_0 = 2$ and since the transformation of ${T}^{\log}$ (respectively, ${W}^{\log}$) under boosts  does not involve ${T}^{\log}$ (respectively, ${W}^{\log}$), one can conclude that the other parameter $\ell_1$ characterizing the irreducible representation in which  $({T}^{\log}, {W}^{\log})$ transform vanishes, $\ell_1 = 0$ (see \cite{Naimark62,Gel'fand63,HarishChandra47} for more information). Thus, the parameters associated with logarithmic supertranslations transform according to the irreducible representation $(\ell_0=2, \ell_1=0)$ of the homogeneous Lorentz group.  This is the ``tail'' \cite{Gel'fand63} of the finite-dimensional vector representation $(\ell_0 = 0, \ell_1=2)$ of the ordinary translations. As it is known the supertranslations are in the semi-direct sum $(\ell_0 = 0, \ell_1=2) \oplus_\sigma  (\ell_0=2, \ell_1=0)$, the tail being given here by the proper supertranslations (beyond the translations). The tail appears therefore twice when logarithmic translations are switched on, but it is  realized by functions with different parities under the antipodal map. The ordinary translations appear only once, because the transformations in the finite-dimensional vector representation $(0,2)$ are proper gauge transformations on the logarithmic side and are factored out.

\subsection{Central terms}

We close the discussion of the algebra by observing that the non-vanishing components of the central terms appear in the brackets
of the usual and ``logarithmic'' supertranslations. Specifically,
we have that
\begin{align}
\mathcal{C}_{\{T,W_{\log}\}} & =-\,\mathcal{C}_{\{W_{\log},T\}}=2\oint d^{2}x\sqrt{\xbar g}\,T\,W_{\log}\,,\label{eq:CentralCharge1}\\
\mathcal{C}_{\{W,T_{\log}\}} & =-\,\mathcal{C}_{\{T_{\log},W\}}=-2\oint d^{2}x\sqrt{\xbar g}\,W\,T_{\log}\,.\label{eq:CentralCharge2}
\end{align}
This is a centrally extended abelian algebra. It is easy (and instructive) to verify that these relations are compatible with the Jacobi identity. The opposite parities of $T$ (even) and $T_{\log}$ (odd) on the one hand, and of $W$ (odd) and $ W_{\log}$ on the other hand, as well as the fact that these functions transform in (almost) the same representation of the Lorentz group, are essential for guaranteeing a non-trivial central charge and consistency with the Jacobi identity.

We note the important fact that the central charges vanish for ordinary translations in time (zero mode of $T$) and space ($\ell = 1$ spherical harmonics of $W$).  This implies  that the transformation rules of the energy and the linear momentum are unaffected by the extension of the formalism.  In particular, the energy and momentum are unchanged if one performs a logarithmic supertranslation. By contrast, the angular momentum does transform under logarithmic supertranslations since these are in a non-trivial representation of the Lorentz group.  This is of course the logarithmic analog of the fact that the angular momentum transforms under supertranslations, or even, for that matter, under ordinary translations. How to extract an intrinsic angular momentum (analog of the angular momentum ``in the center of mass frame'') will not be studied here because a dramatic simplification of the algebra which bypasses this question can be achieved.

It is clear that the central charge is invertible in the remaining  sector of the pure supertranslations and the logarithmic supertranslations.  The relations (\ref{eq:CentralCharge1})-(\ref{eq:CentralCharge2}) actually express that the generators of pure supertranslations are canonically conjugate to those of logarithmic supertranslations. More precisely, if we respectively denote by $S_1$ the generators of the pure even $T$-supertranslations, by $S_2$ the generators of the pure odd $W$-supertranslations, by $L_1$ the generators of the logarithmic odd $T_{\log}$-supertranslations and by $L_2$ the generators of the logarithmic even $W_{\log}$-supertranslations, the relations (\ref{eq:CentralCharge1})-(\ref{eq:CentralCharge2}) read\footnote{In $S_\alpha$, the index $\alpha$ runs not only over the discrete values $1,2$, but involves also the argument $(x) \equiv (x^A)$ of functions on the sphere, e.g, $S_1 \approx  2\sqrt{\xbar g}\Big(\xbar h_{rr}+\xbar D_{A}\xbar\lambda^{A}-\frac{1}{2}\theta\Big)_{\ell>0}(x)$ etc.}
\be \{L_\alpha , S_\beta\} = \sigma_{\alpha \beta} \, , \qquad \{S_\alpha , S_\beta\} = 0 \, , \qquad \{L_\alpha , L_\beta\} = 0 \ee
with 
\be
(\sigma_{\alpha \beta}) = \begin{pmatrix} 0& - 2I^{\text{even}} \\ 2 I^{\text{odd}} & 0 \end{pmatrix} \, .
\ee
Here $I^{\text{even}}$ is the unit operator in the space of even functions on the sphere,
 \be
 I^{\text{even}}(x,x') = \frac12 \Big(\delta^{(2)}(x-x') + \delta^{(2)}(x+x') \Big)
 \ee
 while $I^{\text{odd}}$ is the unit operator in the space of odd functions on the sphere,
 \be
 I^{\text{odd}}(x,x') = \frac12 \Big(\delta^{(2)}(x-x') - \delta^{(2)}(x+x') \Big) \, .
 \ee

 The Jacobi identity involving one Lorentz generator, one $S_\alpha$ and one $L_\alpha$  is equivalent to the invariance of the bilinear form $\sigma_{\alpha \beta}$ under Lorentz transformations.

\section{Decoupling of the pure supertranslations from the Poincar\'e generators in the logarithmic BMS algebra}
\label{sec:Decoupling}

In this section, we will show how we can take advantage of the central charge to redefine the Lorentz generators such that their action on the pure and logarithmic supertranslations is trivial. This decoupling of the supertranslations from the Poincar\'e algebra  is performed by exploiting the fact just observed that the pure and logarithmic supertranslations are canonically conjugate.

First, we prove that it is possible to achieve this goal on general grounds, using general algebraic considerations.  We then proceed to display the explicit computation of the realization of the decoupling mechanism.

\subsection{General algebraic considerations}
The logarithmic BMS algebra found in the previous section has the following structure
\begin{flalign}
\{M_a, M_b\} & = f_{ab}^c M_c\,,\\
\{M_a, T_i\} & = R_{ai}^{\phantom{ai}j}T_j\,,\\
\{M_a, S_\alpha\} & = G_{a\alpha}^{\phantom{a\alpha}i}T_i + G_{a\alpha}^{\phantom{a\alpha}\beta}S_\beta\,,\label{eq:bracketMS}\\
\{M_a, L^\alpha\} & = - G_{a\beta}^{\phantom{a\alpha}\alpha}L^\beta\,, \label{eq:bracketMS2}\\
\{L^\alpha, S_\beta\} & = \delta^\alpha_\beta\,, \label{eq:bracketMS3}
\end{flalign}
where $M_a$, $T_i$, $S_{\alpha}$ and $L^{\beta}$ are respectively the generators of the Lorentz transformations, of the standard translations, of the pure supertranslations and of the logarithmic supertranslations.  

Before proceeding, a word of explanation concerning (\ref{eq:bracketMS2}) and (\ref{eq:bracketMS3}).  As we have seen, the generators of logarithmic supertranslations $L_\alpha$ transform in the same representation as the $S_\alpha$ (modulo the standard translations), i.e., $\{M_a, L_\alpha\}  =  G_{a\alpha}^{\phantom{a\alpha}\beta}L_\beta$.  They also fulfill  $\{L_\alpha, S_\beta\} = \sigma_{\alpha \beta}$.  We define $L^\alpha = \sigma^{\alpha \beta} L_\beta$,  where $\sigma^{\alpha \beta}\sigma_{\beta \gamma} =   \delta^\alpha_\gamma$ (inverse of $\sigma_{\alpha \beta}$).  Then (\ref{eq:bracketMS3}) follows immediately from the definitions, while (\ref{eq:bracketMS2}) is an immediate consequence of the Lorentz invariance of $\sigma_{\alpha \beta}$, 
\be
\delta_a \sigma_{\alpha \beta} \equiv G_{a\alpha}^{\phantom{a\alpha}\gamma}\sigma_{\gamma \beta} + G_{a\beta}^{\phantom{a\beta}\gamma}\sigma_{\alpha\gamma} =0\, .
\ee

We now redefine the Lorentz generators by adding new terms as follows,
\begin{eqnarray}
\tilde {M}_a &=& M_a - G_{a\beta}^{\phantom{a\alpha}i}L^\beta T_i- G_{a\beta}^{\phantom{a\alpha}\gamma}L^\beta S_\gamma\\
& =& M_a - L^\beta \{M_a,S_\beta\}\,.\label{eq:M-mod}
\end{eqnarray}
The extra terms have been added in order to trivialise the action of the Lorentz algebra on pure supertranslations by leveraging the non-zero bracket between $L^\alpha$ and $S_\beta$. One easily shows that the action of the new Lorentz generators $\tilde M_a$ on both pure and logarithmic supertranslations vanishes
\begin{equation}
\{\tilde {M}_a, S_\alpha\}= \{\tilde {M}_a, L^\alpha\}=0\,,
\end{equation}
while the bracket $\{\tilde {M}_a, T_i\}$ does not suffer any modification.  The redefinition (\ref{eq:M-mod}) of the homogeneous Lorentz generators is bilinear in the generators of the inhomogeneous logarithmic BMS algebra $T_i, S_\alpha,L_\alpha$.  The corresponding transformations differ from the original Lorentz transformations by a field-dependent logarithmic supertranslation, a field-dependent translation and a field-dependent pure supertranslation,
\begin{eqnarray}
&&\hspace{-1.1cm} \tilde{\delta}_a F = \{F , \tilde{M}_a\} \nonumber \\
&& \hspace{-.9cm} = \delta_a F + (- G_{a\beta}^{\phantom{a\alpha}i} T_i- G_{a\beta}^{\phantom{a\alpha}\gamma} S_\gamma) \{F, L^\beta\} + (- G_{a\beta}^{\phantom{a\alpha}i} L^\beta) \{F,T_i\}+ (- G_{a\beta}^{\phantom{a\alpha}\gamma} L^\beta) \{F, S_\gamma\}
\end{eqnarray}
which, together, constitute by construction a canonical transformation (i.e., have an integrable, well-defined, canonical generator).

One also easily verifies that the Poincar\'e subalgebra is invariant under this redefinition of the Lorentz generators
\begin{equation}
\{\tilde {M}_a, \tilde {M}_b\}=f_{ab}^c\tilde{M}_c\quad,\quad \{\tilde{M}_a, T_i\}  = R_{ai}^{\phantom{ai}j}T_j\,.
\end{equation}

The modification of $M_a$ given in Eq. \eqref{eq:M-mod} is the most drastic option. There exists a milder version that also decouples the pure supertranslations from the translations without trivialising the action of the Lorentz algebra on the former, 
\begin{equation}\label{eq:M-mod2}
\mathcal M_a = M_a - G_{a\beta}^{\phantom{a\alpha}i}L^\beta T_i\,.
\end{equation}
With this alternative redefinition, only the contribution from the standard translations $T_i$ is removed from the right hand side of \eqref{eq:bracketMS}:
\begin{equation}
\{\mathcal M_a, S_\alpha\}  = G_{a\alpha}^{\phantom{a\alpha}\beta}S_\beta\,,
\end{equation}
while all other Poisson brackets remain invariant. Both pure and logarithmic supertranslations $S_\alpha$, $L_\alpha$ transform under the non-trivial infinite-dimensional Lorentz representation characterized by $G_{a\alpha}^{\phantom{a\alpha}\beta}$ (or its dual for $L^\alpha$).  Therefore,   with this redefinition, one cannot induce soft hair by boosting a solution with no soft hair and with $P^\mu \not=0$, but the angular momentum ``ambiguity'' remains.  

\subsection{Explicit computations}

\subsubsection{New Lorentz transformations and charges}

We now provide the detailed formulas.  According to the derivations of the previous subsection, the combinations of field-dependent logarithmic supertranslations, field-dependent translations and field-dependent pure supertranslations that compensate the action of the Lorentz transformations on pure supertranslations and logarithmic supertranslations are explicitly described by vector fields asymptotically given by 
\begin{align}
W_{\log}^{(b,Y)}&=-\Big[\frac{b}{\sqrt{\xbar g}}(\xbar \pi^{rr}-\xbar \pi+\pi^{rr}_{\log})\Big]_{\ell\geq 2} - \Big[\mathcal{L}_Y(\xbar h_{rr}+\xbar D_A \xbar \lambda^A-\frac{1}{2}\theta)\Big]_{\ell\geq 2}\,, \\
T_{\log}^{(b,Y)}&=\Big[\left(3b  +\partial_A b \xbar D^A+b\xbar \triangle\right) (\xbar h_{rr}+\xbar D_A \xbar \lambda^A-\frac{1}{2}\theta)\Big]_{\ell\geq 3}+ \Big[\frac{1}{\sqrt{\xbar g}}\mathcal{L}_Y (\xbar \pi^{rr}-\xbar \pi+\pi^{rr}_{\log})\Big]_{\ell \geq 3}\,,\\
W^{(b,Y)}&=b (V)_{\ell \geq 2}- \mathcal{L}_Y (U)_{\ell \geq 3}\,,\\
T^{(b,Y)}&=\left(3b+\partial_A b \xbar D^A+b\xbar \triangle\right)(U)_{\ell \geq 3}-\mathcal{L}_Y (V)_{\ell \geq 2}\,.
\end{align}
Here the notation $\ell\geq n$ means that in the spherical harmonic expansion of the functions, only the modes with $\ell$ greater than or equal to $n$ are considered. 

The charges associated with these vector fields are integrable and well-defined and read
\begin{align}
Q^{\text{extra}}_b&=\oint d^2x \Big[2b(V)_{\ell \geq 2}(\xbar \pi^{rr}-\xbar \pi+\pi^{rr}_{\log})\nonumber\\
&\qquad \qquad \quad+2\sqrt{\xbar g}\Big(\xbar h_{rr}+\xbar D_A \xbar \lambda^A-\frac{1}{2}\theta\Big)\left(3b+\partial_A b \xbar D^A+b\xbar \triangle\right)(U)_{\ell\geq 3}\Big]\,,\\ 
Q^{\text{extra}}_Y&=\oint d^2x \Big[2\sqrt{\xbar g}\mathcal{L}_Y \Big(\xbar h_{rr}+\xbar D_A \xbar \lambda^A-\frac{1}{2}\theta\Big)(V)_{\ell \geq 2}+2\mathcal{L}_Y(\xbar \pi^{rr}-\xbar \pi+\pi^{rr}_{\log})(U)_{\ell\geq 3}\Big]\,.
\end{align}
The new Lorentz generators $\tilde {Q}_b$ and $\tilde Q_Y$ are obtained by adding these extra terms to the original generators $Q_b$ and $Q_Y$:
\begin{align}
\tilde{Q}_{b}&=Q_b+Q^{\text{extra}}_b\,,\\
\tilde{Q}_{Y}&=Q_Y+Q^{\text{extra}}_Y\,.
\end{align}
These boundary terms  must of course be supplemented by weakly vanishing bulk terms with vector field $\xi^\mu_{(b,Y)}=(\xi^{\perp}_{(b,Y)},\xi^i_{(b,Y)})$ having the given asymptotic behavior.

\subsubsection{Poisson brackets of Lorentz charges with all supertranslations}

A direct computation shows that these new Lorentz charges possess indeed the advertised property  of acting non-trivially only on pure translations. The Poisson brackets of Lorentz boosts with supertranslations take the form
\begin{align}
\{\tilde{Q}_b,Q_T\}&=2\oint d^2x \Big[-b (T)_{\ell=0}\Big](\xbar \pi^{rr}-\xbar \pi+\pi^{rr}_{\log})=Q_{\hat{W}}\,,\\
\{\tilde{Q}_b,Q_W\}&=2\oint d^2x \sqrt{\xbar g}\Big[-\left(3b+\partial_A b \xbar D^A+b\xbar \triangle\right)(W)_{\ell=1}\Big]\Big(\xbar h_{rr}+\xbar D_A \xbar \lambda^A-\frac{1}{2}\theta\Big)=Q_{\hat{T}}\,,
\end{align}
while the Poisson brackets of spatial rotations with supertranslations are given by
\begin{align}
\{\tilde{Q}_Y,Q_T\}&=0\,,\\
\{\tilde{Q}_Y,Q_W\}&=2\oint d^2x \Big[Y^A \partial_A (W)_{\ell=1}\Big](\xbar \pi^{rr}-\xbar \pi+\pi^{rr}_{\log})=Q_{\hat{W}}\,.
\end{align}
We can read off the new algebra of the parameters:
\begin{equation}
        \hat W = Y^A \partial_A (W)_{\ell=1}-b(T)_{\ell=0}, \qquad \hat T = \left(3b+\partial_A b \xbar D^A+b\xbar \triangle\right)(W)_{\ell=1}\,,
\end{equation}
where we see that only the contributions from the translations are left.
Similarly, one can show that the Poisson brackets of all Lorentz charges with logarithmic supertranslations identically vanish
\begin{align}
&\{\tilde{Q}_b,Q_{T_{\log}}\}=\{\tilde{Q}_b,Q_{W_{\log}}\}=0\,,\\
&\{\tilde{Q}_Y,Q_{T_{\log}}\}=\{\tilde{Q}_Y,Q_{W_{\log}}\}=0\,.
\end{align}

\subsubsection{Poisson brackets of new Lorentz charges}

We explicitly verify in this subsection the key property of our construction, namely,  that the redefinitions of the Lorentz charges also leave the homogeneous Lorentz subalgebra invariant.

For this computation it will be useful to have at hand the new action of the Lorentz transformations on the fields:
\begin{align}
\tilde{\delta}_{b,Y}\Pi&=\Big[\sqrt{\xbar g}(3b\mathcal{T} +\partial_A b \xbar D^A\mathcal{T}+b\xbar \triangle \mathcal{T})\Big]_{\ell=1}+\Big[\mathcal{L}_Y \Pi\Big]_{\ell=1}\,,\\
\tilde{\delta}_{b,Y}\mathcal{T}&=\Big[\frac{b}{\sqrt{\xbar g}}\Pi\Big]_{\ell=0}+\Big[\mathcal{L}_Y \mathcal{T}\Big]_{\ell=0}\,,\\
\tilde{\delta}_{b,Y}U&=-b (V)_{l=0}+Y^A\partial_A (U)_{\ell=1}\,,\\
\tilde{\delta}_{b,Y}V&=-b (U)_{\ell=1}-\partial_A b \xbar D^A (U)_{\ell=1}\,,
\end{align} 
where, for convenience, we have defined
\begin{equation}
\Pi\equiv\overline{\pi}^{rr}-\overline{\pi}+\pi^{rr}_{\log}\,,\qquad
\mathcal{T}\equiv \overline{h}_{rr}+\overline{D}_A \overline{\lambda}^A-\frac{1}{2}\theta\,.
\end{equation}
We will compute the Poisson brackets of the new Lorentz generators case by case.

\begin{itemize}

\item We start with the computation of the Poisson bracket between two Lorentz boosts:
\begin{equation}\label{eq:2boosts}
\{\tilde{Q}_{b_1},\tilde{Q}_{b_2}\}=\tilde{\delta}_{b_2} \tilde{Q}_{b_1}=\tilde{\delta}_{b_2} Q_{b_1}+\tilde{\delta}_{b_2} Q^{\text{extra}}_{b_1}\,.
\end{equation}
The last term in the above expression can be directly obtained using the new transformation laws of the fields:
\begin{align}
\tilde{\delta}_{b_2} Q^{\text{extra}}_{b_1}&=\oint d^2x \Big\{2 \sqrt{\xbar g}\,b_1 (V)_{\ell \geq 2} \Big[3b_2\mathcal{T} +\partial_A b_2 \xbar D^A\mathcal{T}+b_2\xbar \triangle \mathcal{T}\Big]_{\ell=1} \\
&\qquad \qquad \quad +2\left(3b_1+\partial_A b_1 \xbar D^A+b_1\xbar \triangle\right)(U)_{\ell\geq 3}\Big[b_2\Pi\Big]_{\ell=0}\Big\}\,.
\end{align}
The first term on the right hand side of \eqref{eq:2boosts} must be carefully computed taking into account the field-dependent gauge parameters given at the beginning of this section:
\begin{equation}
\tilde{\delta}_{b_2} Q_{b_1}=\delta_{b_2}Q_{b_1}+\delta_{W^{(b_2)}}Q_{b_1}+\delta_{T^{(b_2)}}Q_{b_1}+\delta_{W_{\text{log}}^{(b_2)}}Q_{b_1}+\delta_{T_{\text{log}}^{(b_2)}}Q_{b_1}\,.
\end{equation}
The second and fifth terms give
\begin{align}
\delta_{W^{(b_2)}}Q_{b_1}+\delta_{T_{\text{log}}^{(b_2)}}Q_{b_1}=&2\oint d^2x\sqrt{\xbar g}(b_2 \partial_A b_1-b_1 \partial_A b_2)\xbar D^A (V)_{\ell \geq 2}\mathcal{T}\\
&-2\oint d^2x \sqrt{\xbar g}\,b_1 (V)_{\ell\geq 2} \Big[3b_2\mathcal{T} +\partial_A b_2 \xbar D^A\mathcal{T}+b_2\xbar \triangle \mathcal{T}\Big]_{\ell=1}\,,
\end{align}
while the third and fourth terms reduce to
\begin{align}
\delta_{T^{(b_2)}}Q_{b_1}+\delta_{W_{\text{log}}^{(b_2)}}Q_{b_1}=&2\oint d^2x(b_2 \partial_A b_1-b_1 \partial_A b_2)\xbar D^A (U)_{\ell \geq 3}\Pi\\
&-2\oint d^2x \left(3b_1+\partial_A b_1 \xbar D^A+b_1\xbar \triangle\right)(U)_{\ell\geq 3}\Big[b_2\Pi\Big]_{\ell=0}\,.
\end{align}
Thus, putting all the terms together we obtain the expected result
\begin{equation}
\{\tilde{Q}_{b_1},\tilde{Q}_{b_2}\}=Q_{\hat{Y}}+Q^{\text{extra}}_{\hat{Y}}=\tilde{Q}_{\hat{Y}}\,,
\end{equation}
where we used the bracket of the original generators, $\delta_{b_2}Q_{b_1}=Q_{\hat{Y}}$, and
\begin{equation}
\hat{Y}^A=b_1\xbar D^Ab_2-b_2\xbar D^Ab_1\,.
\end{equation}

\item We continue with the computation of the Poisson bracket between two spatial rotations:
\begin{equation}\label{eq:2rotations}
\{\tilde{Q}_{Y_1},\tilde{Q}_{Y_2}\}=\tilde{\delta}_{Y_2} \tilde{Q}_{Y_1}=\tilde{\delta}_{Y_2} Q_{Y_1}+\tilde{\delta}_{Y_2} Q^{\text{extra}}_{Y_1}\,.
\end{equation}
The last term can also be directly computed, which gives
\begin{align}
\tilde{\delta}_{Y_2}Q^{\text{extra}}_{Y_1}=2\oint d^2x \mathcal{L}_{Y_1}(\mathcal{L}_{Y_2}\Pi)_{\ell=1}U_{\ell \geq 3}\,.
\end{align}
Decomposing $\tilde \delta_{Y_2} Q_{Y_1}$ as before
\begin{equation}
\tilde{\delta}_{Y_2} Q_{Y_1}=\delta_{Y_2}Q_{Y_1}+\delta_{W^{(Y_2)}}Q_{Y_1}+\delta_{T^{(Y_2)}}Q_{Y_1}+\delta_{W_{\text{log}}^{(Y_2)}}Q_{Y_1}+\delta_{T_{\text{log}}^{(Y_2)}}Q_{Y_1}\,,
\end{equation}
we can can combine the third and fourth terms to get
\begin{align}
\delta_{T^{(Y_2)}}Q_{Y_1}+\delta_{W_{\text{log}}^{(Y_2)}}Q_{Y_1}&=-2\oint d^2x\sqrt{\xbar g} \,\mathcal {L}_{Y_2}\mathcal {L}_{Y_1}\mathcal{T} V_{\ell \geq 2}+2\oint d^2x \sqrt{\xbar g}\,\mathcal{L}_{Y_1}(\mathcal{L}_{Y_2}\mathcal{T})_{l \geq 2}V_{\ell\geq 2} \nonumber\\
&=2\oint d^2x\sqrt{\xbar g}\,[\mathcal{L}_{Y_1},\mathcal{L}_{Y_2}]\mathcal{T}V_{\ell\geq2}\,.
\end{align}
while the second and fifth terms reduce to
\begin{align}
\delta_{W^{(Y_2)}}Q_{Y_1}+\delta_{T_{\text{log}}^{(Y_2)}}Q_{Y_1}&=-2\oint d^2x \sqrt{\xbar g}\,\mathcal{L}_{Y_2}\mathcal{L}_{Y_1}\Pi U_{\ell \geq 3}+2\oint d^2x\mathcal{L}_{Y_1}(\mathcal{L}_{Y_2}\Pi)_{\ell\geq 3} U_{\ell \geq 3}\nonumber \\
&=2\oint d^2x [\mathcal{L}_{Y_1},\mathcal{L}_{Y_2}]\Pi U_{\ell\geq 3}-2\oint \mathcal{L}_{Y_1}(\mathcal{L}_{Y_2}\Pi)_{\ell=1} U_{\ell \geq 3}\,.
\end{align}
Thus, adding all the contributions we have shown that
\begin{align}
\{\tilde{Q}_{Y_1},\tilde{Q}_{Y_2}\}&=Q_{\hat{Y}}+Q^{\text{extra}}_{\hat{Y}}=\tilde Q_{\hat{Y}}\,,
\end{align}
where we used the original bracket $\delta_{Y_2}Q_{Y_1}=Q_{\hat{Y}}$, with
\begin{equation}
\hat{Y}^B=Y_1^A\partial_A Y^B_2-Y_2^A\partial_A Y^B_1\,.
\end{equation}

\item A similar strategy allows us to compute the Poisson bracket between a Lorentz boost and a spatial rotation, obtaining
\begin{align}
\{\tilde Q_Y, \tilde Q_b\}&=\tilde Q_{\hat{b}}\,,\qquad \text{with} \qquad \hat{b}=Y^A\partial_A b\,,\\
\{\tilde Q_b, \tilde Q_Y\}&=\tilde Q_{\hat{b}}\,,\qquad \text{with} \qquad \hat{b}=-Y^A\partial_A b\,.
\end{align}
\end{itemize}

We have thus shown that the new contributions to the Lorentz generators do not modify the homogeneous Lorentz subalgebra. The Poincar\'e algebra remains therefore untouched.

\subsubsection{Alternative redefinition of the Lorentz charges}
As mentioned above, there exists an alternative redefinition of the Lorentz generators \eqref{eq:M-mod2}, which decouples the pure BMS supertranslations from the translations, while keeping them in a non-trivial infinite-dimensional representation of the Lorentz algebra. The canonical realization for these charges can be obtained by adding a field-dependent improper gauge transformation with the following parameters
\begin{equation}
W_{\log}^{(b)}=-\Big[\frac{b}{\sqrt{\xbar g}}(\xbar \pi^{rr}-\xbar \pi+\pi^{rr}_{\log})_{\ell=1}\Big]_{\ell\geq 2}  \qquad 
W^{(b)}=\Big[b (V)_{\ell \geq 2}\Big]_{\ell=1}\,.
\end{equation}
This transformation leads to an extra term in the Lorentz boost charge of the form
\begin{equation}
Q^{\text{extra}}_b=\oint d^2x \Big[2b(V)_{\ell \geq 2}(\xbar \pi^{rr}-\xbar \pi+\pi^{rr}_{\log})_{\ell=1}\Big]\,,
\end{equation}
while keeping the spatial rotation charge invariant. One can check that the Poisson brackets of these new Lorentz charges with pure supertranslations involve only pure supertranslations and keep the brackets with all other charges unchanged.

\section{Conclusions}
\label{sec:Conclusions}

In this paper, we have shown that a special class of logarithmic supertranslations, characterized by definite parity properties under the antipodal map, can be consistently included in the Hamiltonian formulation of asymptotically flat spaces. One can enlarge the boundary conditions while keeping the action finite in such a way that these logarithmic supertranslations are symmetry transformations  with well-defined  (integrable, finite) canonical generators.  We have also computed the algebra of the logarithmic supertranslation generators with the other generators of the BMS algebra (``logarithmic BMS algebra'') and found non-trivial  central terms.  Previous instances of central terms appearing in the asymptotic symmetry algebra occur in AdS$_3$ gravity \cite{Brown:1986nw,Brown:1986ed}, or  in  five-dimensional gravity, which shares a very similar structure \cite{Fuentealba:2021yvo,Fuentealba:2022yqt}.

We insisted throughout our approach both on finiteness of the symplectic form and on exact invariance of the symplectic form under the transformations of the logarithmic BMS algebra -- and not just invariance up to (possibly divergent) surface terms at spatial infinity.  These strong requirements lead to rather lengthy developments, but are worth being pursued for at least two reasons. (i) First, the very fact that they can be successfully implemented ``on the nose'', making regularizations unnecessary, reveals non-trivial properties of the Einstein theory on asymptotically flat spaces.  The detailed non-trivial cancellations underlying this success teaches us a lesson about the theory which might be less apparent in approaches where these requirements are not put in the forefront. (ii) Second, our approach guarantees that standard Hamiltionian methods can be used and, in particular, that the symmetry generators form a true (possibly non-linear) algebra under the Poisson brackets, which fulfills the Jacobi identity exactly.

Even though the logarithmic translations (\ref{eq:LogTran}) are not contained in it,  the class of logarithmic supertranslations that are consistently included in our approach is rather huge.  Indeed, it involves one function of the angles, as do the supertranslations.  The odd part of this function of the angles is related to the logarithmic supertranslations in time, while the even part is related to new supertranslations in space, which become improper with the new boundary conditions and which we also call ``logarithmic supertranslations'' since their presence is related to the logarithmic enlargement of the boundary conditions. As we have also shown, logarithmic supertranslations in space parametrized by the function $\tilde{U}$ are proper gauge transformations with zero charge.

Furthermore, the logarithmic supertranslation charges are canonically conjugate to the pure supertranslation charges. This implies that one can generate shifts of the pure supertranslation charges by performing logarithmic supertranslations.  This remarkable feature enables one to completely decouple in the asymptotic symmetry algebra pure supertranslations and logarithmic supertranslations from the Poincar\'e generators, without having to fix the gauge or truncating the theory.  

The new Poincar\'e generators are invariant under supertranslations.   As we have pointed out, this decoupling is conceptually very similar to the decoupling achieved in \cite{Mirbabayi:2016axw,Bousso:2017dny,Javadinezhad:2018urv,Javadinezhad:2022hhl}, but our approach is entirely classical (non-quantum). The decoupling amounts to add to the Lorentz transformations appropriately chosen supertranslations and logarithmic supertranslations with coefficients that depend on the charges.  These nonlinear redefinitions evade the obstructions found earlier on the impossibility to extract a natural Poincar\'e subalgebra through linear methods. An intermediate decoupling can be achieved, such that the brackets of the pure supertranslations with the Lorentz generators involve only the pure supertranslations and not the ordinary translations.

The logarithmic supertranslations are allowed in the formalism by including in the asymptotic form of the fields terms that take the precise form of logarithmic supertranslation variations. This is reminiscent of the method of orbits, in which one parametrizes the fields in terms of an orbit representative (which would in our case be the configuration with no $\log$-terms) and the symmetry element that brings that reference representative to the given configuration (here, the improper logarithmic diffeomorphisms).  Such a presentation can in principle always be given for any symmetry.

We conclude this section with six open questions:
\begin{itemize}
\item What makes the decoupling possible between the Poincar\'e algebra and the supertranslations is the invertible central charge appearing in the brackets of the logarithmic supertranslations with the pure supertranslations.  Such a mechanism would thus be in principle available in any similar situation where such a central charge would be present.  It would be interesting to investigate this mechanism in the nonlinear context of the BMS(5) algebra emerging in five spacetime dimensions \cite{Fuentealba:2021yvo,Fuentealba:2022yqt}, or in the case of supersymmetry where a similar central charge also appears \cite{Fuentealba:2021xhn}.  
\item Also of interest would be to study the analogs of the logarithmic supertranslations in the Maxwell theory, which would be angle-dependent logarithmic $u(1)$ gauge transformations \cite{Henneaux:2018gfi}. 
\item One might wonder if the procedure followed to include logarithmic supertranslations (i.e., introducing slowlier decaying terms in the metric that take the form of an (improper) diffeomorphism) can be pushed further to include superrotations  \cite{Banks:2003vp,Barnich:2009se,Barnich:2010eb} or diffeomorphisms of the sphere
\cite{Campiglia:2014yka,Campiglia:2015yka}, which would need $\mathcal O(1)$ deviations from the flat metric.  The answer to this question is not immediate, precisely because the allowed new terms would not be proper gauge transformations but would be (if successfully included) improper gauge ones.   In that context, adding new surface degrees of freedom might perhaps help for finiteness or integrability of the charges \cite{Henneaux:2018gfi}. 
\item Since our new boundary conditions allow logarithmic supertranslations, a natural question is: what are the Ward identities associated with these new symmetries \cite{Strominger:2013jfa,He:2014laa,Strominger:2017zoo}? Furthermore, in view of the decoupling mechanism, one might wonder whether these Ward identities would provide new insightful information on ``hard processes'', or merely constraint ``soft processes'', see \cite{Mirbabayi:2016axw,Bousso:2017dny,Javadinezhad:2018urv,Javadinezhad:2022hhl}.  
\item  In the same vein, one would like to repeat the complete analysis at null infinity and, in particular, write down the action of the logarithmic supertranslations there and study their matching with logarithmic supertranslations at spatial infinity along the lines of  \cite{Troessaert:2017jcm,Henneaux:2018hdj}.
\item Finally we point out that our boundary conditions yield a Weyl tensor which contains no log-type singularity as one integrates the equations to null infinity \cite{Henneaux:2018hdj,Fried1,Friedrich:1999wk,Friedrich:1999ax,Mohamed:2021rfg}, since the boundary conditions differ from those of \cite{Henneaux:2018hdj} or \cite{Regge:1974zd} only by diffeomorphism terms to which the Weyl tensor is blind\footnote{In fact, the regularity and parity conditions on the Weyl tensor imply the boundary conditions given in this paper if one allows $\log$-terms in the asymptotic expansion of the metric.  This is an extension of the results of \cite{Henneaux:2018hdj}.}.  In that respect, it would be of interest to extend the analysis to the alternative BMS-invariant boundary conditions of \cite{Henneaux:2018cst}, which do generically lead to log-type singularities at null infinity.  A motivation for achieving this task is given by \cite{Winicour1}.
\end{itemize}

Work along these lines is currently in progress.

\section*{Notes added}

After our paper was completed, the reference \cite{Javadinezhad:2022ldc} was posted on the archive.  That work studies how to redefine angular momentum flux at null infinity in order to make it free from supertranslation ambiguities.  It would be interesting to study the connection with our construction.

We should also mention the existence of similarities of our symmetry structure with the direct product structure of the symmetry algebra derived in  \cite{Adami:2021nnf} on null boundaries.  We thank Shahin Sheikh-Jabbari for a discussion on this point.

\section*{Acknowledgments}

It is a pleasure to thank Sucheta Majumdar and Javier Matulich for discussions at an early stage of this work. O. F. is grateful to the Coll\`ege de France for kind hospitality while this article was completed.  This work was partially supported by  FNRS-Belgium (conventions FRFC PDRT.1025.14 and IISN 4.4503.15), as well as by funds from the Solvay Family. 
%%%*************************************

\appendix
%dummy comment inserted by tex2lyx to ensure that this paragraph is not empty%dummy comment inserted by tex2lyx to ensure that this paragraph is not empty%dummy comment inserted by tex2lyx to ensure that this paragraph is not empty

\section*{Appendices}

\section{$2+1$ Decomposition of the spatial metric and spatial curvature
\label{app-decomp}}

This appendix provides useful formulas related to the $2+1$ slicing of the spatial equal time hypersurfaces by  spheres of constant radius $r$.
The ``lapse'' is denoted $\lambda$
while the ``shift'' is $\lambda^{A}$, 
\begin{equation}
\gamma_{AB}\equiv g_{AB},\quad\lambda_{A}\equiv g_{rA},\quad\lambda\equiv\frac{1}{\sqrt{g^{rr}}}.
\end{equation}
The three-dimensional spatial metric and its inverse
take the form
\begin{equation}
g_{ij}=\left(\begin{array}{cc}
\lambda^{2}+\lambda_{C}\lambda^{C} & \lambda_{B}\\
\lambda_{A} & \gamma_{AB}
\end{array}\right),\quad g^{ij}=\left(\begin{array}{cc}
\frac{1}{\lambda^{2}} & -\frac{\lambda^{B}}{\lambda^{2}}\\
-\frac{\lambda^{A}}{\lambda^{2}} & \gamma^{AB}+\frac{\lambda^{A}\lambda^{B}}{\lambda^{2}}
\end{array}\right)\,.
\end{equation}
Here, the $2$-dimensional metric $\gamma_{AB}$ and its inverse $\gamma^{AB}$ have been used to
raise and lower the angular indices $A,B,...$. The Christoffel symbols
can be written in terms of the extrinsic curvature of the spheres
$K_{AB}$:
\begin{eqnarray}
K_{AB} & = & \frac{1}{2\lambda}\left(-\partial_{r}g_{AB}+D_{A}\lambda_{B}+D_{B}\lambda_{A}\right)\,,\\
\Gamma_{AB}^{r} & = & \frac{1}{\lambda}K_{AB}\,,\\
\Gamma_{BC}^{A} & = & ^{\gamma}\Gamma_{BC}^{A}-\frac{\lambda^{A}}{\lambda}K_{BC}\,,\\
\Gamma_{rA}^{r} & = & \frac{1}{\lambda}\left(\partial_{A}\lambda+K_{AB}\lambda^{B}\right)\,,\\
\Gamma_{rr}^{r} & = & \frac{1}{\lambda}\partial_{r}\lambda+\frac{\lambda^{A}}{\lambda}\left(\partial_{A}\lambda+K_{AB}\lambda^{B}\right)\,,\\
\Gamma_{rB}^{A} & = & -\frac{\lambda^{A}}{\lambda}\left(\partial_{B}\lambda+K_{BC}\lambda^{C}\right)+D_{B}\lambda^{A}-\lambda K_{B}^{A}\,,\\
\Gamma_{rr}^{A} & = & -\lambda\left(\gamma^{AB}+\frac{\lambda^{A}\lambda^{B}}{\lambda^{2}}\right)\left(\partial_{B}\lambda+K_{BC}\lambda^{C}\right)-\lambda^{C}\left(D^{A}\lambda_{C}-\lambda K_{C}^{A}\right)\nonumber \\
 &  & \qquad-\frac{\lambda^{A}}{\lambda}\partial_{r}\lambda+\gamma^{AB}\partial_{r}\lambda_{B}\,.
\end{eqnarray}
$D_{A}$ denotes the covariant derivative associated to $\gamma_{AB}$. The
components of the Ricci tensor can be obtained from
\begin{eqnarray}
^{(3)}R_{AB} & = & \frac{1}{\lambda}\partial_{r}K_{AB}+2K_{AC}K_{B}^{C}-KK_{AB}-\frac{1}{\lambda}D_{A}D_{B}\lambda\nonumber \\
 &  & +^{\gamma}R_{AB}-\frac{1}{\lambda}\mathcal{L}_{\lambda}K_{AB}\,,\\
^{(3)}R_{rA} & = & \lambda\left(\partial_{A}K-D_{B}K_{A}^{B}\right)+{}^{(3)}R_{AB}\lambda^{B}\,,\\
^{(3)}R_{rr} & = & \lambda(\partial_{r}K-\lambda^{A}\partial_{A}K)-\lambda^{2}K_{B}^{A}K_{A}^{B}-\lambda D_{A}D^{A}\lambda\nonumber \\
 &  & -^{(3)}R_{AB}\lambda^{A}\lambda^{B}+2\,{}^{(3)}R_{rB}\lambda^{B}\,,
\end{eqnarray}
which implies the expression
\begin{equation}
^{(3)}R=\frac{2}{\lambda}(\partial_{r}K-\lambda^{A}\partial_{A}K)+{}^{\gamma}R-K_{B}^{A}K_{A}^{B}-K^{2}-\frac{2}{\lambda}D_{A}D^{A}\lambda\,
\end{equation}
for the Ricci scalar.

\section{ Vanishing of the logarithmic divergences in the boost
generator \label{app-van}}

In this appendix we  show that the remaining divergences (proportional to $\ln^{2}r$
and $\ln r$) in the boost generator vanish. These are, taking into account that many terms obviously vanish due to parity properties,
\begin{equation}
\delta Q_{\log}^{(2)}=\ln^{2}r\oint d^{2}x\sqrt{\xbar g}\,b\delta\left(2\theta^{(2)}+2k_{\log(2)}^{(2)}+\frac{1}{4}\theta^{2}-\frac{3}{4}\theta_{A}^{B}\theta_{B}^{A}\right)\,,
\end{equation}
and
\begin{align}
\delta Q_{\log}^{(1)} & =\ln r\oint d^{2}x\sqrt{\xbar g}\Big[b\Big(2\delta\sigma+2\delta k_{\log(1)}^{(2)}+\frac{1}{2}\delta(-3\xbar h_{A}^{B}\theta_{B}^{A}+h\theta)+\frac{1}{4}\delta(3\theta_{A}^{B}\theta_{B}^{A}-\theta^{2})\nonumber\\
 & \quad-\xbar h_{rr}\delta\theta+\theta\xbar D_{A}\delta\xbar\lambda^{A}-2\theta_{AB}\xbar D^{A}\delta\xbar\lambda^{B}-\delta\theta_{AB}\xbar D^{A}\xbar\lambda^{B}\Big)-\partial_{A}b\xbar\lambda^{A}\delta\theta+4b\delta\Big(\xbar D_{A}\tilde{V}\xbar D^{A}V\Big)\Big]\,,
\end{align}
(the term proportional to $\tilde T$ in $Q_{\log}$ in \eqref{eq:Qlog} is manifestly zero as follows from similar parity considerations).

The critical tool in establishing the absence of divergences is the fast decay of the constraints imposed in Section \ref{constraints}.

\begin{itemize}
\item Let us first focus on the $\ln^{2}r$-divergent term, which reads
\begin{equation}
\oint d^{2}x\sqrt{\xbar g}\,b\delta\left(2\theta^{(2)}+2k_{\log(2)}^{(2)}+\frac{1}{4}\theta^{2}-\frac{3}{4}\theta_{A}^{B}\theta_{B}^{A}\right)\,.
\end{equation}
We can see that the quadratic terms in $\theta_{AB}$ vanish under
the integral because of parity (recall that $\theta_{AB}$ is even).
Then, we must only deal with the integral
\begin{equation}
\oint d^{2}x\sqrt{\xbar g}\,b\left(\theta^{(2)}+k_{\log(2)}^{(2)}\right)\,.\label{eq:Log2-1}
\end{equation}
We now make use of the condition $\mathcal{H}_{\log(1)}=0$ in \eqref{eq:Hlog(1)},
which allows to relate $\theta^{(2)}$ with other coefficients of
the fall-off of the fields. Thus, we find that
\begin{align}
\theta^{(2)} & =-\frac{1}{4}\Big[4k_{\log(2)}^{(2)}-2h_{rr}^{\log(1)}-\xbar\triangle\,h_{rr}^{\log(1)}-\sigma+\xbar D^{A}\xbar D^{B}\sigma_{AB}-\xbar\triangle\,\sigma+\frac{1}{2}\Big(-3\theta_{B}^{A}\theta_{A}^{B}+\theta^{2}\Big)\nonumber\\
 & \quad+\frac{3}{2}\xbar h_{B}^{A}\theta_{A}^{B}-\frac{1}{2}\xbar h_{rr}\theta-\frac{1}{2}\xbar h\,\theta-\xbar\lambda_{A}\,\xbar D^{A}\theta-\theta\xbar D^{A}\xbar\lambda_{A}+\frac{1}{2}\xbar D_{A}\theta\xbar D^{A}h_{rr}-\xbar D_{A}\xbar h_{B}^{A}\xbar D^{B}\theta\nonumber\\
 & \quad+\theta_{B}^{A}\xbar D_{A}\xbar D^{B}\xbar h-\theta_{B}^{A}\xbar D^{B}\xbar D_{C}\xbar h_{A}^{C}+\frac{1}{2}\xbar D_{A}\theta\xbar D^{A}\xbar h-\theta_{B}^{A}\xbar D^{B}\xbar\lambda_{A}+\theta_{B}^{A}\xbar D_{A}\xbar D^{B}\xbar h_{rr}\nonumber\\
 & \quad-\theta_{B}^{A}\xbar D_{C}\xbar D^{B}\xbar h_{A}^{C}+\theta_{B}^{A}\xbar\triangle\,\xbar h_{A}^{B}+\frac{1}{2}\xbar D_{A}\theta_{C}^{B}\xbar D^{A}\xbar h_{B}^{C}\Big]\nonumber\\
 &\quad+\frac{1}{4\xbar g}\Big(\xbar\pi^{rr}\pi_{\log}^{rr}-\xbar\pi\,\pi_{\log}^{rr}+4\xbar\pi^{rA}\pi_{\log A}^{r}+2\xbar\pi_{B}^{A}\pi_{\log A}^{B}-\xbar\pi^{rr}\pi_{\log}-\xbar\pi\,\pi_{\log}\Big)\,.
\end{align}
Then, the integral in \eqref{eq:Log2-1} becomes
\begin{align}
 & \frac{1}{4}\oint d^{2}x\sqrt{\xbar g}\,b\left[2h_{rr}^{\log(1)}+\xbar\triangle\,h_{rr}^{\log(1)}+\sigma-\xbar D^{A}\xbar D^{B}\sigma_{AB}+\xbar\triangle\,\sigma+\frac{1}{2}\Big(3\theta_{B}^{A}\theta_{A}^{B}-\theta^{2}\Big)\right]\label{eq:Log2-2}\\
 & +\frac{1}{4}\oint d^{2}x\frac{b}{\sqrt{\xbar g}}\Big(\xbar\pi^{rr}\pi_{\log}^{rr}-\xbar\pi\,\pi_{\log}^{rr}+4\xbar\pi^{rA}\pi_{\log A}^{r}+2\xbar\pi_{B}^{A}\pi_{\log A}^{B}-\xbar\pi^{rr}\pi_{\log}-\xbar\pi\,\pi_{\log}\Big)\label{eq:Log2-3}\\
 & -\frac{1}{4}\oint d^{2}x\sqrt{\xbar g}\,b\Big(\frac{3}{2}\xbar h_{B}^{A}\theta_{A}^{B}-\frac{1}{2}\xbar h_{rr}\theta-\frac{1}{2}\xbar h\,\theta-\xbar\lambda_{A}\,\xbar D^{A}\theta-\theta\xbar D^{A}\xbar\lambda_{A}+\frac{1}{2}\xbar D_{A}\theta\xbar D^{A}h_{rr}\nonumber \\
 & \qquad-\xbar D_{A}\xbar h_{B}^{A}\xbar D^{B}\theta+\theta_{B}^{A}\xbar D_{A}\xbar D^{B}\xbar h-\theta_{B}^{A}\xbar D^{B}\xbar D_{C}\xbar h_{A}^{C}+\frac{1}{2}\xbar D_{A}\theta\xbar D^{A}\xbar h-\theta_{B}^{A}\xbar D^{B}\xbar\lambda_{A}\nonumber \\
 & \qquad+\theta_{B}^{A}\xbar D_{A}\xbar D^{B}\xbar h_{rr}-\theta_{B}^{A}\xbar D_{C}\xbar D^{B}\xbar h_{A}^{C}+\theta_{B}^{A}\xbar\triangle\,\xbar h_{A}^{B}+\frac{1}{2}\xbar D_{A}\theta_{C}^{B}\xbar D^{A}\xbar h_{B}^{C}\Big)\,.\label{eq:Log2-4}
\end{align}
It is easy to show that the integral in \eqref{eq:Log2-2} is zero
by virtue of the equation for the boost parameter $\xbar D_{A}\xbar D_{B}b+\xbar g_{AB}b=0$
(after integrating by parts) and the fact that $\theta_{AB}$ is even
and $b$ is odd.

For the integral in \eqref{eq:Log2-3}, we can see that only the ``pure
gauge part'' of the subleading terms in the asymptotic expansion  of the conjugate momentum,  i.e.,
\begin{align}
\xbar\pi_{\text{even}}^{rr} & =-\sqrt{\xbar g}\,\xbar\triangle\,V_{\text{even}\,,}\\
\xbar\pi_{\text{odd}}^{rA} & =-\sqrt{\xbar g}\,\xbar D^{A}V_{\text{even}}\,,\\
\xbar\pi_{\text{even}}^{AB} & =\sqrt{\xbar g}\,(\xbar D^{A}\xbar D^{B}V_{\text{even}}-\xbar g^{AB}\xbar\triangle V_{\text{even}})\,,
\end{align}
contributes
to the surface integral.
Thus, \eqref{eq:Log2-3} reduces to
\begin{equation}
\oint d^{2}xb\left(-4\xbar D_{A}V\pi_{\log}^{rA}+2\xbar D_{A}\xbar D_{B}V\pi_{\log}^{AB}\right)\,.
\end{equation}
If we integrate
by parts, we get
\begin{equation}
\oint d^2x \Big[4\xbar D_A b(\pi^{rA}_{\text{log}}+\xbar D_B \pi^{AB}_{\text{log}})+2b(\xbar D_A\xbar D_B\pi^{AB}_{\text{log}}-\pi_{\text{log}}+2\xbar D_A \pi^{rA}_{\text{log}})\Big]\,,
\end{equation}
where the equation for the boost parameter was used. If we now consider the asymptotic constraint equations $\xbar D_{A}\pi_{\log}^{AB}+\pi_{\log}^{rA}=0$
and $\xbar D_{A}\xbar D_{B}\pi_{\log}^{AB}+\pi_{\log}=0$, the above
integral is straightforwardly seen to vanish.

Accordingly, it only remains to show that \eqref{eq:Log2-4} is zero.
Again parity considerations imply that \eqref{eq:Log2-4} reduces to
\begin{align}
 & -\frac{1}{4}\oint d^{2}x\sqrt{\xbar g}\,b\Big(\frac{3}{2}\xbar h_{B}^{A}\theta_{A}^{B}-\frac{1}{2}\xbar h\,\theta-\xbar D_{A}\xbar h_{B}^{A}\xbar D^{B}\theta+\theta_{B}^{A}\xbar D_{A}\xbar D^{B}\xbar h-\theta_{B}^{A}\xbar D^{B}\xbar D_{C}\xbar h_{A}^{C}\\
 & \qquad\qquad\qquad\qquad+\frac{1}{2}\xbar D_{A}\theta\xbar D^{A}\xbar h-\theta_{B}^{A}\xbar D_{C}\xbar D^{B}\xbar h_{A}^{C}+\theta_{B}^{A}\xbar\triangle\,\xbar h_{A}^{B}+\frac{1}{2}\xbar D_{A}\theta_{C}^{B}\xbar D^{A}\xbar h_{B}^{C}\Big)\,, \label{eq:Log2-5}
\end{align}
where only the pure gauge part of $h_{AB}$ (odd) contributes,
\begin{equation}
\xbar h_{AB}^{\text{odd}}=2\left(\xbar D_{A}\xbar D_{B}U_{\text{odd}}+\xbar g_{AB}U_{\text{odd}}\right)\,.
\end{equation}
Integrating by parts we get that \eqref{eq:Log2-5} becomes
\begin{align}
-\frac{1}{4}\oint d^{2}x\sqrt{\xbar g}\,U_{\text{odd}}\Big[&b\Big(2\xbar D_{A}\xbar D^{B}\theta_{B}^{A}-\xbar D_{A}\xbar D^{B}\xbar\triangle\,\theta_{B}^{A}+2\xbar D_{A}\xbar\triangle\,\xbar D^{B}\theta_{B}^{A}-2\xbar D_{C}\xbar D_{A}\xbar D^{C}\xbar D^{B}\theta_{B}^{A}\nonumber \\
&+\xbar\triangle\,\xbar D_{A}\xbar D^{B}\theta_{B}^{A}\Big)+\partial_{A}b\Big(\xbar D^{A}\theta-2\xbar D^{A}\xbar D_{B}\xbar D^{C}\theta_{C}^{B}+2\xbar D_{B}\xbar D^{A}\xbar D^{C}\theta_{C}^{B}\nonumber\\
&-3\xbar D_{B}\xbar D^{C}\xbar D^{A}\theta_{C}^{B}+3\xbar\triangle\,\xbar D^{A}\theta\Big)\Big]\,.\label{eq:log2-6}
\end{align}
Using that $\xbar D_{A}\theta_{BC}=\xbar D_{B}\theta_{AC}$ and the
commutators
\begin{align}
[\xbar\triangle,\xbar D_{B}]\theta^{AB} & =2\xbar D^{A}\theta-3\xbar D_{B}\theta^{AB}\,,\\{}
[\xbar D^{A},\xbar\triangle]\xbar D^{B}\theta_{AB} & =\xbar D_{A}\xbar D_{B}\theta^{AB}\,,
\end{align}
we conclude that \eqref{eq:log2-6} vanishes (and thus also \eqref{eq:Log2-4}).
\newline

\item  We will now show that the $\ln r$-divergent term
\begin{align}
 & \oint d^{2}x\sqrt{\xbar g}\Big[b\Big(2\delta\sigma+2\delta k_{\log(1)}^{(2)}+\frac{1}{2}\delta(-3\xbar h_{A}^{B}\theta_{B}^{A}+h\theta)-\xbar h_{rr}\delta\theta+\frac{1}{4}\delta(3\theta_{A}^{B}\theta_{B}^{A}-\theta^{2})\nonumber \\
 & \qquad\qquad\quad+\theta\xbar D_{A}\delta\xbar\lambda^{A}-2\theta_{AB}\xbar D^{A}\delta\xbar\lambda^{B}-\delta\theta_{AB}\xbar D^{A}\xbar\lambda^{B}\Big)-\partial_{A}b\xbar\lambda^{A}\delta\theta-4b\delta\Big(\xbar D_{A}\tilde{V}\xbar D^{A}V\Big)\Big]
\end{align}
is also zero. Parity conditions of the fields imply that the above integral
reduces to

\begin{equation}
\oint d^{2}x\sqrt{\xbar g}\,b\delta\Big(2\sigma+2k_{\log(1)}^{(2)}-\frac{3}{2}\xbar h_{A}^{B}\theta_{B}^{A}+\frac{1}{2}h\theta\Big)\,.\label{eq:Log2-7}
\end{equation}
The condition $\mathcal{H}^{(2)}=0$ in \eqref{eq:H(2)} allows one to obtain an expression
for $\sigma$ in terms of the other asymptotic coefficients in the decay of the fields
\begin{align}
\sigma & =-\frac{1}{2}\Big(2k_{\log(1)}^{(2)}-2h_{rr}^{(2)}-\xbar\triangle\,h_{rr}^{(2)}-h^{(2)}+\xbar D^{A}\xbar D_{B}h_{A}^{(2)B}-\xbar\triangle\,h^{(2)}+2\xbar h_{rr}^{2}+\xbar h_{rr}\xbar\triangle\,\xbar h_{rr}\nonumber\\
 & \quad+\frac{1}{2}\xbar D_{A}\xbar h_{rr}\xbar D^{A}\xbar h_{rr}+\frac{3}{4}\xbar h_{B}^{A}\xbar h_{A}^{B}-\frac{1}{4}\xbar h^{2}+\xbar h_{B}^{A}\xbar D^{B}\xbar D_{A}\xbar h-\xbar h_{B}^{A}\xbar D_{A}\xbar D^{C}\xbar h_{C}^{B}-\frac{1}{4}\xbar D_{A}\xbar h\,\xbar D^{A}\xbar h\nonumber\\
 & \quad-\xbar D_{A}\xbar h_{B}^{A}\xbar D^{C}\xbar h_{C}^{B}+\xbar D^{A}\xbar h\,\xbar D_{B}\xbar h_{A}^{B}-\xbar h_{B}^{A}\xbar D_{C}\xbar D^{B}\xbar h_{A}^{C}+\xbar h_{B}^{A}\xbar\triangle\,\xbar h_{A}^{B}-\frac{1}{2}\xbar D_{A}\xbar h_{C}^{B}\xbar D^{C}\xbar h_{B}^{A}\nonumber\\
 & \quad+\frac{3}{4}\xbar D_{C}\xbar h_{A}^{B}\xbar D^{C}\xbar h_{B}^{A}-\frac{1}{2}\xbar h_{rr}\xbar h-\frac{1}{2}\xbar D_{A}\xbar h\xbar D^{A}\xbar h_{rr}+\xbar D_{A}\xbar h_{rr}\xbar D^{B}\xbar h_{B}^{A}+\xbar h_{A}^{B}\xbar D^{A}\xbar D_{B}\xbar h_{rr}\nonumber\\
 & \quad-\xbar h_{rr}\xbar D_{A}\xbar\lambda^{A}-2\xbar\lambda_{A}\xbar D^{A}\xbar h_{rr}-\xbar\lambda_{A}\xbar D^{A}\xbar h-\xbar h\xbar D_{A}\xbar\lambda^{A}-\xbar h_{A}^{B}\xbar D^{A}\xbar\lambda_{B}+2\xbar\lambda_{A}\xbar\lambda^{A}\nonumber\\
 & \quad-2\xbar\lambda^{A}\xbar D_{A}\xbar D_{B}\xbar\lambda^{A}-\xbar D_{A}\xbar\lambda^{A}\xbar D_{B}\xbar\lambda^{B}+2\xbar\lambda_{A}\xbar\triangle\,\xbar\lambda^{A}-\frac{1}{2}\xbar D_{A}\xbar\lambda^{B}\xbar D_{B}\xbar\lambda^{A}+\frac{3}{2}\xbar D_{A}\xbar\lambda_{B}\xbar D^{A}\xbar\lambda^{B}\nonumber\\
 & \quad+\frac{3}{4}\theta_{A}^{B}\theta_{B}^{A}-\frac{1}{4}\theta^{2}-\frac{3}{2}\xbar h_{A}^{B}\theta_{B}^{A}+\frac{1}{2}\xbar h\,\theta-\xbar h_{rr}\theta+\xbar\lambda_{A}\xbar D^{A}\theta+\theta\xbar D_{A}\xbar\lambda^{A}-\theta_{A}^{B}\xbar D_{B}\xbar\lambda^{A}\Big)\nonumber\\
 & \quad+\frac{1}{2\xbar g}\Big[\frac{1}{2}(\xbar\pi^{rr})^{2}+2\xbar\pi^{rA}\xbar\pi_{A}^{r}+\xbar\pi^{AB}\xbar\pi_{AB}-\xbar\pi^{rr}\xbar\pi-\xbar\pi^{2}\Big]\,.
\end{align}
Using parity conditions and the above relation, we find that \eqref{eq:Log2-7}
can be written as

\begin{align}
 & -\oint d^{2}x\sqrt{\xbar g}\,b\Big(-2h_{rr}^{(2)}-\xbar\triangle\,h_{rr}^{(2)}-h^{(2)}+\xbar D^{A}\xbar D_{B}h_{A}^{(2)B}-\xbar\triangle\,h^{(2)}\Big)\label{eq:Log2-8}\\
 & \quad-\oint d^{2}x\sqrt{\xbar g}\,b\Big(\frac{3}{4}\xbar h_{B}^{A}\xbar h_{A}^{B}-\frac{1}{4}\xbar h^{2}+\xbar h_{B}^{A}\xbar D^{B}\xbar D_{A}\xbar h-\xbar h_{B}^{A}\xbar D_{A}\xbar D^{C}\xbar h_{C}^{B}-\frac{1}{4}\xbar D_{A}\xbar h\,\xbar D^{A}\xbar h\nonumber\\
& \quad -\xbar D_{A}\xbar h_{B}^{A}\xbar D^{C}\xbar h_{C}^{B}+\xbar D^{A}\xbar h\,\xbar D_{B}\xbar h_{A}^{B}-\xbar h_{B}^{A}\xbar D_{C}\xbar D^{B}\xbar h_{A}^{C}+\xbar h_{B}^{A}\xbar\triangle\,\xbar h_{A}^{B}-\frac{1}{2}\xbar D_{A}\xbar h_{C}^{B}\xbar D^{C}\xbar h_{B}^{A}\nonumber \\
 & \quad+\frac{3}{4}\xbar D_{C}\xbar h_{A}^{B}\xbar D^{C}\xbar h_{B}^{A}-\frac{1}{2}\xbar h_{rr}\xbar h-\frac{1}{2}\xbar D_{A}\xbar h\xbar D^{A}\xbar h_{rr}+\xbar D_{A}\xbar h_{rr}\xbar D^{B}\xbar h_{B}^{A}+\xbar h_{A}^{B}\xbar D^{A}\xbar D_{B}\xbar h_{rr}\nonumber \\
 & \quad-\xbar\lambda_{A}\xbar D^{A}\xbar h-\xbar h\xbar D_{A}\xbar\lambda^{A}-\xbar h_{A}^{B}\xbar D^{A}\xbar\lambda_{B}\Big)\label{eq:Log2-9}\\
 & \quad+\oint d^{2}xb\Big[\frac{1}{\sqrt{\xbar g}}\Big(\frac{1}{2}(\xbar\pi^{rr})^{2}+2\xbar\pi^{rA}\xbar\pi_{A}^{r}+\xbar\pi^{AB}\xbar\pi_{AB}-\xbar\pi^{rr}\xbar\pi-\xbar\pi^{2}\Big)-4\sqrt{\xbar g}\,\xbar D_{A}\tilde{V}\xbar D^{A}V\Big]\,.\label{eq:Log2-10}
\end{align}

The integral in line \eqref{eq:Log2-8}
is zero as it follows from integration by parts and using the relation $\xbar D_{A}\xbar D_{B}b+\xbar g_{AB}b=0$ fulfilled by the Lorentz boost parameters. 

The computation of the integral \eqref{eq:Log2-9} is bit a more involved,
but the idea is again simple. We make use of the definition of $\xbar h_{AB}$
:
\begin{align}
\xbar h_{AB} & =(\xbar h_{AB})^{\text{even}}+2(\xbar D_{A}D_{B}U_{\text{odd}}+\xbar g_{AB}U_{\text{odd}})\,.\label{eq:hbarAB-1}
\end{align}
Introducing this expression into \eqref{eq:Log2-9} and integrating by parts,
we obtain that \eqref{eq:Log2-9} reduces to
\begin{align}
\oint d^{2}x\sqrt{\xbar g}\Big\{&\xbar D_{A}\Big[b\xbar D^{A}\Big(\xbar\triangle\,\xbar h-\xbar D_{A}\xbar D^{B}\xbar h_{B}^{A}+\xbar\triangle\,\xbar h_{rr}-2\xbar D_{B}\xbar\lambda^{B}\Big)\Big]\nonumber\\
&+3b\Big(\xbar\triangle\,\xbar h-\xbar D_{A}\xbar D^{B}\xbar h_{B}^{A}+\xbar\triangle\,\xbar h_{rr}-2\xbar D_{B}\xbar\lambda^{B}\Big)\Big\} U_{\text{odd}}\,,
\end{align}
which clearly vanishes by virtue of the subleading order of the Hamiltonian
constraint.

Finally, in the integral \eqref{eq:Log2-10} we make use of the definition
of the subleading part of the conjugate momentum
\begin{align}
\xbar\pi^{rr} & =(\xbar\pi^{rr})^{\text{odd}}-\sqrt{\xbar g}\,\xbar\triangle V_{\text{even}}\,,\label{eq:pibrr-1}\\
\xbar\pi^{rA} & =(\xbar\pi^{rA})^{\text{even}}-\sqrt{\xbar g}\,\xbar D^{A}V_{\text{even}}\,,\label{eq:pibrA-1}\\
\xbar\pi^{AB} & =(\xbar\pi^{AB})^{\text{odd}}+\sqrt{\xbar g}\,(\xbar D^{A}\xbar D^{B}V_{\text{even}}-\xbar g^{AB}\xbar\triangle\,V_{\text{even}})\,.\label{eq:pibarAB-1}
\end{align}
Replacing these definitions in \eqref{eq:Log2-10} and using the relations
coming from the subleading order of the momentum constraint
\begin{align}
\xbar D_{A}\xbar\pi^{rA}-\xbar\pi_{A}^{A}+\pi_{\text{\ensuremath{\log}}}^{rr} & =0\,,\\
\xbar D_{B}\xbar\pi_{\,\,\,A}^{B}+\xbar\pi_{\,\,\,A}^{r}+\pi_{\text{log}\,A}^{r} & =0\,.
\end{align}
we find that
\begin{equation}
\oint d^{2}x\frac{b}{\sqrt{\xbar g}}\Big(\frac{1}{2}(\xbar\pi^{rr})^{2}+2\xbar\pi^{rA}\xbar\pi_{A}^{r}+\xbar\pi^{AB}\xbar\pi_{AB}-\xbar\pi^{rr}\xbar\pi-\xbar\pi^{2}\Big)=\oint d^{2}x\sqrt{\xbar g}\left(4b\xbar D_{A}\tilde{V}\xbar D^{A}V_{\text{even}}\right)\,.
\end{equation}
In consequence, \eqref{eq:Log2-10} vanishes. Thus, we have shown
that the $\ln r$-divergence in the boost charge is  zero.
\end{itemize}

This completes the proof of convergence of $\delta Q_{\xi}$.

\section{Transformation laws of the subleading terms \label{app-transf}}

In this appendix we list without comment the transformation laws of the subleading
terms in the fall-off of the fields that are useful to compute the
algebra.
\begin{itemize}
\item Subleading terms of the metric
\begin{align}
\delta_{\xi,\xi^{i}}h_{rr}^{(2)} & =Y^{A}\partial_{A}h_{rr}^{(2)}+I^{A}\partial_{A}\xbar h_{rr}-2I^{A}\xbar\lambda_{A}+2\tilde{I}^{A}\xbar\lambda_{A}+2\tilde{W}\xbar h_{rr}-W\xbar h_{rr}-2W^{(1)}\nonumber\\
&\quad+\frac{T}{\sqrt{\xbar g}}\left(\xbar\pi^{rr}-\xbar\pi\right)
+\frac{b}{2\sqrt{\xbar g}}\Big(2\pi_{(2)}^{rr}-2\pi_{(2)}+3\xbar h_{rr}\xbar\pi^{rr}-\xbar h_{rr}\xbar\pi+4\xbar\lambda_{A}\xbar\pi^{rA}\nonumber\\
&\quad-2\xbar h_{AB}\xbar\pi^{AB}-\xbar h\xbar\pi^{rr}+\xbar h\xbar\pi\Big)\,,
\end{align}
\begin{align}
\delta_{\xi,\xi^{i}}h_{rA}^{(2)} & =\mathcal{L}_{Y}h_{rA}^{(2)}+\mathcal{L}_{I}\xbar\lambda_{A}-I^{B}\xbar h_{AB}+\tilde{I}^{B}\xbar h_{AB}+\tilde{W}\xbar\lambda_{A}+\partial_{A}W\xbar h_{rr}+\partial_{A}W^{(1)}-2I_{A}^{(1)}\nonumber\\
 & \quad+\frac{2T}{\sqrt{\xbar g}}\xbar\pi_{A}^{r}+\frac{b}{\sqrt{\xbar g}}\Big(2\pi_{(2)A}^{r}+\xbar h_{rr}\xbar\pi_{A}^{r}-\xbar h\xbar\pi_{A}^{r}+2\xbar h_{AB}\xbar\pi^{rB}+2\xbar\lambda_{B}\xbar\pi_{A}^{B}+\xbar\lambda_{A}\xbar\pi^{rr}-\xbar\lambda_{A}\xbar\pi\Big)\,,
\end{align}
\begin{align}
\delta_{\xi,\xi^{i}}h_{AB}^{(2)} & =\mathcal{L}_{Y}h_{AB}^{(2)}+\mathcal{L}_{I}\xbar h_{AB}+2\partial_{(A}W\xbar\lambda_{B)}+W\theta_{AB}+W\xbar h_{AB}+2\left(\xbar D_{(A}I_{B)}^{(1)}+\xbar g_{AB}W^{(1)}\right)\nonumber\\
 & \quad+\frac{T}{\sqrt{\xbar g}}\left[2\xbar\pi_{AB}-\xbar g_{AB}\left(\xbar\pi^{rr}+\xbar\pi\right)\right]+\frac{b}{\sqrt{\xbar g}}\Big[2\pi_{AB}^{(2)}-\xbar h_{AB}\xbar\pi_{rr}+4\xbar\lambda_{(A}\xbar\pi_{B)}^{r}-\xbar h_{rr}\xbar\pi_{AB}\nonumber\\
 & \quad-\xbar h\xbar\pi_{AB}+4\xbar h_{(A}^{C}\xbar\pi_{B)C}-\xbar h_{AB}\xbar\pi+\xbar g_{AB}\Big(-\pi_{rr}^{(2)}-\pi^{(2)}-\frac{1}{2}\xbar h_{rr}\xbar\pi^{rr}+\frac{1}{2}\xbar h\,\xbar\pi^{rr}\nonumber\\
 & \quad-2\xbar\lambda_{A}\xbar\pi^{rA}-\xbar h_{AB}\xbar\pi^{AB}+\frac{1}{2}\xbar h_{rr}\xbar\pi+\frac{1}{2}\xbar h\,\xbar\pi\Big)\Big]\,,
\end{align}
\begin{align}
\delta_{\xi,\xi^{i}}\sigma_{AB} & =\mathcal{L}_{Y}\sigma_{AB}+\mathcal{L}_{I}\theta_{AB}+\mathcal{L}_{\tilde{I}}\xbar h_{AB}+2\partial_{(A}\tilde{W}\xbar\lambda_{B)}+\tilde{W}\theta_{AB}+\tilde{W}\xbar h_{AB}+W\theta_{AB}\nonumber\\
 & \quad+\frac{2\tilde{T}}{\sqrt{\xbar g}}\xbar\pi_{A}^{r}+\frac{2T}{\sqrt{\xbar g}}\pi_{\log A}^{r}+\frac{b}{\sqrt{\xbar g}}\Big(2\pi_{\log(1)A}^{r}+\xbar\lambda_{A}\pi_{\log}^{rr}-\theta\xbar\pi_{A}^{r}+2\theta_{AB}\xbar\pi^{rB}\nonumber\\
 & \quad+\xbar h_{rr}\pi_{\log A}^{r}-\xbar h\,\pi_{\log A}^{r}+2\xbar h_{AB}\pi_{\log}^{rB}+2\xbar\lambda_{B}\pi_{\log A}^{B}-\xbar\lambda_{A}\xbar\pi\Big)\,.
\end{align}
 
\item Subleading term of the momentum
\begin{align}
\delta_{\xi,\xi^{i}}\pi_{(2)}^{rA} & =\mathcal{L}_{Y}\pi_{(2)}^{rA}+\mathcal{L}_{I}\xbar\pi^{rA}+I^{A}\xbar\pi^{rr}-\tilde{I}^{A}\xbar\pi^{rr}-\partial_{B}W\xbar\pi^{AB}-W\xbar\pi^{rA}+W\pi_{\log}^{rA}\nonumber\\
 & +\frac{\sqrt{\xbar g}\,T}{2}\Big(\xbar D_{B}\xbar h^{AB}+\xbar\triangle\,\xbar\lambda^{A}+\xbar D_{B}\xbar D^{A}\xbar\lambda^{B}-2\xbar D^{A}\xbar D_{B}\xbar\lambda^{B}-\xbar D^{A}\xbar h_{rr}-\xbar D^{A}\xbar h\Big)\nonumber\\
 & +\frac{\sqrt{\xbar g}}{2}\xbar D^{A}T\left(\xbar h_{rr}-\xbar h\right)+\frac{\sqrt{\xbar g}}{2}\partial_{B}T\left(3\xbar h^{AB}-\theta^{AB}+\xbar D^{B}\xbar\lambda^{A}-\xbar D^{A}\xbar\lambda^{B}\right)\nonumber\\
 & +\sqrt{\xbar g}\left(\xbar\lambda^{A}\xbar\triangle\,T-\xbar\lambda_{B}\xbar D^{B}\xbar D^{A}T\right)+\frac{\sqrt{\xbar g}\,\tilde{T}}{2}\left(4\xbar\lambda^{A}+\xbar D^{A}\xbar h_{rr}\right)+\frac{\sqrt{\xbar g}}{2}\xbar D^{A}\tilde{T}\left(\xbar h-\xbar h_{rr}\right)\nonumber\\
 & -\sqrt{\xbar g}\,\partial_{B}\tilde{T}\xbar h^{AB}-2\sqrt{\xbar g}\,\xbar D^{A}T^{(1)}+\frac{b}{\sqrt{\xbar g}}\left(-\xbar\pi^{rr}\xbar\pi^{rA}+\xbar\pi\,\xbar\pi^{rA}-2\xbar\pi^{rB}\xbar\pi_{B}^{A}\right)\nonumber\\
 &+\sqrt{\xbar g}\,\partial_A b\Big(h_{r}^{(2)A}+\xbar D_{B}k_{(2)}^{AB}-\xbar D^{A}k_{(2)}-2\xbar h_{B}^{A}\xbar\lambda^{B}-\frac{7}{2}\xbar h_{rr}\xbar\lambda^{A}+\xbar h\,\xbar\lambda^{A}-\frac{1}{2}R_{(1)}\xbar\lambda^{A}\nonumber\\
 & +\theta\xbar\lambda^{A}+\frac{3}{4}\xbar h_{B}^{A}\xbar D^{B}\xbar h-\frac{1}{4}\theta_{B}^{A}\xbar D^{B}\xbar h-\frac{1}{2}\xbar h_{rr}\xbar D^{B}\xbar h_{B}^{A}+\xbar D^{B}h_{B}^{(2)A}-\frac{1}{2}\xbar h_{B}^{A}\xbar D^{B}\theta\nonumber\\
 & -2\xbar\lambda^{A}\xbar D_{B}\xbar\lambda^{B}+\frac{1}{4}\xbar h\,\xbar\triangle\,\xbar\lambda^{A}-\frac{1}{2}\xbar h_{B}^{A}\xbar D_{C}\xbar D^{B}\xbar\lambda^{C}+\frac{1}{4}\xbar h\,\xbar D_{B}\xbar D^{A}\xbar\lambda^{B}-\frac{1}{2}\xbar h_{B}^{A}\xbar D^{B}\xbar h_{rr}\nonumber\\
 & +\frac{1}{4}\partial_{B}\xbar h\,\xbar D^{B}\xbar\lambda^{A}-\frac{1}{2}\xbar\lambda_{B}\xbar D^{B}\xbar D^{A}\xbar h_{rr}-\xbar h_{B}^{A}\xbar D^{C}\xbar h_{C}^{B}+\frac{1}{2}\theta_{B}^{A}\xbar D^{C}\xbar h_{C}^{B}-\frac{1}{2}\xbar D_{B}\xbar\lambda^{A}\xbar D^{C}\xbar h_{C}^{B}\nonumber\\
 & -\frac{1}{2}\xbar h_{C}^{B}\xbar D^{C}\xbar h_{B}^{A}+\frac{1}{4}\xbar h\,\xbar D^{B}\xbar h_{B}^{A}+\frac{1}{2}\xbar h_{B}^{A}\xbar D^{C}\xbar h_{C}^{B}+\frac{1}{2}\xbar h_{C}^{B}\xbar D^{C}\theta_{B}^{A}-\frac{1}{4}\xbar h\,\xbar D^{B}\theta_{B}^{A}-\frac{1}{2}\xbar h_{B}^{A}\xbar\triangle\,\xbar\lambda^{B}\nonumber\\
 & +\xbar h_{B}^{A}\xbar D^{B}\xbar D_{C}\xbar\lambda^{C}-\frac{1}{2}\xbar h_{C}^{B}\xbar D^{C}\xbar D_{B}\xbar\lambda^{A}-\frac{1}{2}\xbar h_{C}^{B}\xbar D^{C}\xbar D^{A}\xbar\lambda_{B}+\frac{1}{2}\xbar D^{A}h_{rr}^{(2)}-\frac{5}{4}\xbar h_{rr}\xbar D^{A}\xbar h_{rr}\nonumber\\
 &+\frac{1}{2}\xbar h\,\xbar D^{A}\xbar h_{rr}+\frac{3}{4}\xbar h_{C}^{B}\xbar D^{A}\xbar h_{B}^{C}-\frac{1}{4}\theta_{C}^{B}\xbar D^{A}\xbar h_{B}^{C}+\frac{1}{2}\xbar D_{B}\xbar\lambda^{C}\xbar D^{A}\xbar h_{C}^{B}+\frac{1}{2}\xbar h_{rr}\xbar D^{A}\xbar h-\frac{1}{4}\xbar h\,\xbar D^{A}\xbar h\nonumber\\
 & -\xbar D^{A}h^{(2)}-\frac{1}{2}\xbar h_{C}^{B}\xbar D^{A}\theta_{B}^{C}+\frac{1}{4}\xbar h\,\xbar D^{A}\theta-\xbar\lambda_{B}\xbar D^{A}\xbar\lambda^{B}+\frac{1}{4}\xbar D_{B}\xbar h\,\xbar D^{A}\xbar\lambda^{B}-\frac{1}{2}\xbar D^{C}\xbar h_{C}^{B}\xbar D^{A}\xbar\lambda_{B}\nonumber\\
 & -\frac{1}{2}\xbar h\xbar D^{A}\xbar D_{B}\xbar\lambda^{B}+\frac{1}{2}\xbar\lambda_{B}\xbar D^{A}\xbar D^{B}\xbar h_{rr}+\xbar h_{C}^{B}\xbar D^{A}\xbar D_{B}\xbar\lambda^{C}\Big)\,.
\end{align}
\end{itemize}
Note that 
\begin{align}
\tilde{T} & \rightarrow\tilde{T}+\tilde{T}_{(b)}\,,\\
T & \rightarrow T+T_{(b)}\,,\\
\tilde{I}^{A} & \rightarrow\xbar D^{A}\tilde{W}+\tilde{I}_{(b)}^{A}\,,\\
I^{A} & \rightarrow I^{A}+I_{(b)}^{A}\,.
\end{align}
where
\begin{align}
\tilde{T}_{(b)} & =\partial_{A}b\xbar\lambda^{A}-\left(\xbar\triangle+2\right)^{-1}\left(\xbar D_{A}\xbar D_{B}+\xbar g_{AB}\right)\left[b\,\left(\xbar D^{A}\xbar\lambda^{B}-\frac{1}{2}\theta^{AB}\right)\right]\,,\\
T_{(b)} & =-\frac{1}{2}b\xbar h\,,\\
\tilde{I}^{A}_{(b)} & =\frac{2b}{\sqrt{\xbar g}}\pi_{\log}^{rA}\,,\\
I^{A}_{(b)} & =\frac{2b}{\sqrt{\xbar g}}\xbar\pi^{rA}\,.
\end{align}

%%***************************************************


\begin{thebibliography}{99}

%\cite{Bergmann:1961zz}
\bibitem{Bergmann:1961zz}
P.~G.~Bergmann,
``'Gauge-Invariant' Variables in General Relativity,''
Phys. Rev. \textbf{124} (1961), 274-278
doi:10.1103/PhysRev.124.274

%\cite{AshtekarLog85}
\bibitem{AshtekarLog85}
A. Ashtekar, ``Logarithmic Ambiguities in the Description of Spatial Infinity,'' Found. Phys. \textbf{15} (1985), 419

%\cite{AARP95}
 \bibitem{AARP95} A. Ashtekar and R. Penrose, ``Mass positivity from focussing and the structure of spacelike infinity'', in
{\em Further advances in Twistor Theory, Vol II}, edited by L. Mason, L.P. Hughston and P.Z. Kobak, pp169-173, Longman (Harlow: 1995)

 %\cite{BeigSchmidt82}
  \bibitem{BeigSchmidt82}
  R.~Beig and B.~Schmidt, ``Einstein's equations near spatial infinity,'' Commun. Math. Phys. {\bf 87} (1982) 65.
  
      %\cite{Beig:1983sw}
\bibitem{Beig:1983sw}
  R.~Beig,
  ``Integration Of Einstein's Equations Near Spatial Infinity,''
  Proc.  Royal Soc. A
{\bf 1801} (1984) 295--304.
  
   %\cite{Chrusciel89}
  \bibitem{Chrusciel89}
  P. T. Chru\'sciel, ``On the structure of spatial infinity. II. Geodesically regular Ashtekar-Hansen
structures,'' J. Math. Phys. \textbf{30} (1989) 2094.

%\cite{Ashtekar:1990gc}
\bibitem{Ashtekar:1990gc}
A.~Ashtekar, L.~Bombelli and O.~Reula,
``The covariant phase space of asymptotically flat gravitational fields," PRINT-90-0318 (Syracuse), published in 
``Mechanics, Analysis and Geometry: 200 Years After Lagrange'', ed. Mauro Francaviglia, North-Holland Delta Series, Elsevier (Amsterdam: 1991)

%\cite{Compere:2011ve}
\bibitem{Compere:2011ve}
G.~Comp\`ere and F.~Dehouck,
``Relaxing the Parity Conditions of Asymptotically Flat Gravity,''
Class. Quant. Grav. \textbf{28} (2011), 245016
[erratum: Class. Quant. Grav. \textbf{30} (2013), 039501]
doi:10.1088/0264-9381/28/24/245016
[arXiv:1106.4045 [hep-th]].

 %\cite{Troessaert:2017jcm}
\bibitem{Troessaert:2017jcm}
C.~Troessaert,
``The BMS4 algebra at spatial infinity,''
Class. Quant. Grav. \textbf{35} (2018) no.7, 074003
doi:10.1088/1361-6382/aaae22
[arXiv:1704.06223 [hep-th]].

%\cite{Fuentealba:2021yvo}
\bibitem{Fuentealba:2021yvo}
O.~Fuentealba, M.~Henneaux, J.~Matulich and C.~Troessaert,
``Bondi-Metzner-Sachs Group in Five Spacetime Dimensions,''
Phys. Rev. Lett. \textbf{128} (2022) no.5, 051103
doi:10.1103/PhysRevLett.128.051103
[arXiv:2111.09664 [hep-th]].

%\cite{Fuentealba:2022yqt}
\bibitem{Fuentealba:2022yqt}
O.~Fuentealba, M.~Henneaux, J.~Matulich and C.~Troessaert,
``Asymptotic structure of the gravitational field in five spacetime dimensions: Hamiltonian analysis,''
JHEP \textbf{07} (2022), 149
doi:10.1007/JHEP07(2022)149
[arXiv:2206.04972 [hep-th]].

    %\cite{Benguria:1976in}
\bibitem{Benguria:1976in}
  R.~Benguria, P.~Cordero and C.~Teitelboim,
  ``Aspects of the Hamiltonian Dynamics of Interacting Gravitational Gauge and Higgs Fields with Applications to Spherical Symmetry,''
  Nucl.\ Phys.\ B {\bf 122} (1977) 61.
  %%CITATION = doi:10.1016/0550-3213(77)90426-6;%% 
  
  %\cite{Bondi:1962px}
\bibitem{Bondi:1962px}
H.~Bondi, M.~G.~J.~van der Burg and A.~W.~K.~Metzner,
``Gravitational waves in general relativity. 7. Waves from axisymmetric isolated systems,''
Proc. Roy. Soc. Lond. A \textbf{269} (1962), 21-52
doi:10.1098/rspa.1962.0161

%\cite{Sachs:1962wk}
\bibitem{Sachs:1962wk}
R.~K.~Sachs,
``Gravitational waves in general relativity. 8. Waves in asymptotically flat space-times,''
Proc. Roy. Soc. Lond. A \textbf{270} (1962), 103-126
doi:10.1098/rspa.1962.0206

 %\cite{Sachs:1962zza}
\bibitem{Sachs:1962zza}
  R.~Sachs,
  ``Asymptotic symmetries in gravitational theory,''
  Phys.\ Rev.\  {\bf 128} (1962) 2851.
  %%CITATION = doi:10.1103/PhysRev.128.2851;%%

  %\cite{Penrose:1962ij}
\bibitem{Penrose:1962ij}
  R.~Penrose,
  ``Asymptotic properties of fields and space-times,''
  Phys.\ Rev.\ Lett.\  {\bf 10} (1963) 66.
  %%CITATION = doi:10.1103/PhysRevLett.10.66;%%
  
  %\cite{Madler:2016xju}
\bibitem{Madler:2016xju}
T.~M\"adler and J.~Winicour,
``Bondi-Sachs Formalism,''
Scholarpedia \textbf{11} (2016), 33528
doi:10.4249/scholarpedia.33528
[arXiv:1609.01731 [gr-qc]].
  
  %\cite{Alessio:2017lps}
\bibitem{Alessio:2017lps}
F.~Alessio and G.~Esposito,
``On the structure and applications of the Bondi\textendash{}Metzner\textendash{}Sachs group,''
Int. J. Geom. Meth. Mod. Phys. \textbf{15} (2018) no.02, 1830002
doi:10.1142/S0219887818300027
[arXiv:1709.05134 [gr-qc]].
  
  %\cite{Ashtekar:2018lor}
\bibitem{Ashtekar:2018lor}
A.~Ashtekar, M.~Campiglia and A.~Laddha,
``Null infinity, the BMS group and infrared issues,''
Gen. Rel. Grav. \textbf{50} (2018) no.11, 140-163
doi:10.1007/s10714-018-2464-3
[arXiv:1808.07093 [gr-qc]].

   %\cite{Henneaux:2018hdj}
\bibitem{Henneaux:2018hdj}
  M.~Henneaux and C.~Troessaert,
  ``Hamiltonian structure and asymptotic symmetries of the Einstein-Maxwell system at spatial infinity,''
  JHEP {\bf 1807} (2018) 171
  [arXiv:1805.11288 [gr-qc]].
  %%CITATION = doi:10.1007/JHEP07(2018)171;%%     
  
  %\cite{Henneaux:2019yax}
\bibitem{Henneaux:2019yax}
M.~Henneaux and C.~Troessaert, ``The asymptotic structure of gravity at spatial infinity in four spacetime dimensions,'' Proc. Steklov Inst.  Math.  {\bf 309} (2020) 127-149 [arXiv:1904.04495 [hep-th]].    

%\cite{Mirbabayi:2016axw}
\bibitem{Mirbabayi:2016axw}
M.~Mirbabayi and M.~Porrati,
``Dressed Hard States and Black Hole Soft Hair,''
Phys. Rev. Lett. \textbf{117}, no.21, 211301 (2016)
doi:10.1103/PhysRevLett.117.211301
[arXiv:1607.03120 [hep-th]].

  %\cite{Bousso:2017dny}
\bibitem{Bousso:2017dny}
R.~Bousso and M.~Porrati,
``Soft Hair as a Soft Wig,''
Class. Quant. Grav. \textbf{34} (2017) no.20, 204001
doi:10.1088/1361-6382/aa8be2
[arXiv:1706.00436 [hep-th]].  

%\cite{Javadinezhad:2018urv}
\bibitem{Javadinezhad:2018urv}
R.~Javadinezhad, U.~Kol and M.~Porrati,
``Comments on Lorentz Transformations, Dressed Asymptotic States and Hawking Radiation,''
JHEP \textbf{01} (2019), 089
doi:10.1007/JHEP01(2019)089
[arXiv:1808.02987 [hep-th]].

%\cite{Javadinezhad:2022hhl}
\bibitem{Javadinezhad:2022hhl}
R.~Javadinezhad, U.~Kol and M.~Porrati,
``Supertranslation-invariant dressed Lorentz charges,''
JHEP \textbf{04} (2022), 069
doi:10.1007/JHEP04(2022)069
[arXiv:2202.03442 [hep-th]].
  

  %\cite{Ashtekar:1978zz}
\bibitem{Ashtekar:1978zz}
A.~Ashtekar and R.~O.~Hansen,
``A unified treatment of null and spatial infinity in general relativity. I - Universal structure, asymptotic symmetries, and conserved quantities at spatial infinity,''
J. Math. Phys. \textbf{19} (1978), 1542-1566
doi:10.1063/1.523863

%\cite{Dirac:1958sc}
\bibitem{Dirac:1958sc}
  P.~A.~M.~Dirac,
  ``The Theory of gravitation in Hamiltonian form,''
  Proc.\ Roy.\ Soc.\ Lond.\ A {\bf 246} (1958) 333.
  %%CITATION = doi:10.1098/rspa.1958.0142;%%
  
  %\cite{Dirac:1958jc}
\bibitem{Dirac:1958jc}
  P.~A.~M.~Dirac,
  ``Fixation of coordinates in the Hamiltonian theory of gravitation,''
  Phys.\ Rev.\  {\bf 114} (1959) 924.
  %%CITATION = doi:10.1103/PhysRev.114.924;%%

%\cite{Arnowitt:1962hi}
\bibitem{Arnowitt:1962hi}
R.~L.~Arnowitt, S.~Deser and C.~W.~Misner,
``The Dynamics of general relativity,''
Gen. Rel. Grav. \textbf{40} (2008), 1997-2027
doi:10.1007/s10714-008-0661-1
[arXiv:gr-qc/0405109 [gr-qc]].

%\cite{Regge:1974zd}
\bibitem{Regge:1974zd}
  T.~Regge and C.~Teitelboim,
  ``Role of Surface Integrals in the Hamiltonian Formulation of General Relativity,''
  Annals Phys.\  {\bf 88} (1974) 286.
  %%CITATION = doi:10.1016/0003-4916(74)90404-7;%%
  
  %\cite{Fuentealba:2020ghw}
\bibitem{Fuentealba:2020ghw}
O.~Fuentealba, M.~Henneaux, S.~Majumdar, J.~Matulich and C.~Troessaert,
``Asymptotic structure of the Pauli-Fierz theory in four spacetime dimensions,''
Class. Quant. Grav. \textbf{37} (2020) no.23, 235011
doi:10.1088/1361-6382/abbe6e
[arXiv:2007.12721 [hep-th]].

%\cite{Henneaux:2018cst}
\bibitem{Henneaux:2018cst} 
M.~Henneaux and C.~Troessaert, ``BMS Group at Spatial Infinity: the Hamiltonian (ADM) approach,'' JHEP \textbf{03}, 147 (2018) doi:10.1007/JHEP03(2018)147 [arXiv:1801.03718 [gr-qc]].

%\cite{Henneaux:2018gfi}
 \bibitem{Henneaux:2018gfi}
  M.~Henneaux and C.~Troessaert, ``Asymptotic symmetries of electromagnetism at spatial infinity,'' JHEP \textbf{05} (2018), 137 doi:10.1007/JHEP05(2018)137 [arXiv:1803.10194 [hep-th]].

\bibitem{Naimark62} M.A. Naimark, ``Les repr\'esentations lin\'eaires du groupe de Lorentz,'' Dunod (Paris: 1962) (French translation); ``Linear Representations of the Lorentz Group,'' Pergamon (Oxford: 1964) (English translation).
  
\bibitem{Gel'fand63} I.M. Gel'fand, R.A. Minlos and Z. Ya. Shapiro, ``Representations of the rotation and Lorentz groups and their applications,'' Pergamon Press (1963)

\bibitem{HarishChandra47} Harish-Chandra, ``Infinite irreducible representations of the Lorentz group,'' Proc. Roy. Soc. A 189 (1947) 372.

%\cite{Brown:1986nw}
\bibitem{Brown:1986nw}
J.~Brown and M.~Henneaux,
``Central Charges in the Canonical Realization of Asymptotic Symmetries: An Example from Three-Dimensional Gravity,''
Commun. Math. Phys. \textbf{104} (1986) 207-226
doi:10.1007/BF01211590

  %\cite{Brown:1986ed}
\bibitem{Brown:1986ed}
  J.~D.~Brown and M.~Henneaux,
  ``On the Poisson Brackets of Differentiable Generators in Classical Field Theory,''
  J.\ Math.\ Phys.\  {\bf 27} (1986) 489.
  %%CITATION = doi:10.1063/1.527249;%%
  
  %\cite{Fuentealba:2021xhn}
\bibitem{Fuentealba:2021xhn}
O.~Fuentealba, M.~Henneaux, S.~Majumdar, J.~Matulich and T.~Neogi,
``Local supersymmetry and the square roots of Bondi-Metzner-Sachs supertranslations,''
Phys. Rev. D \textbf{104} (2021) no.12, L121702
doi:10.1103/PhysRevD.104.L121702
[arXiv:2108.07825 [hep-th]].

   %\cite{Banks:2003vp}
\bibitem{Banks:2003vp}
  T.~Banks,
  ``A Critique of pure string theory: Heterodox opinions of diverse dimensions,''
  hep-th/0306074.
  %%CITATION = HEP-TH/0306074;%%

  %\cite{Barnich:2009se}
\bibitem{Barnich:2009se}
  G.~Barnich and C.~Troessaert,
  ``Symmetries of asymptotically flat 4 dimensional spacetimes at null infinity revisited,''
  Phys.\ Rev.\ Lett.\  {\bf 105} (2010) 111103
    [arXiv:0909.2617 [gr-qc]].
  %%CITATION = doi:10.1103/PhysRevLett.105.111103;%%

%\cite{Barnich:2010eb}
\bibitem{Barnich:2010eb}
G.~Barnich and C.~Troessaert,
``Aspects of the BMS/CFT correspondence,''
JHEP \textbf{05} (2010), 062
doi:10.1007/JHEP05(2010)062
[arXiv:1001.1541 [hep-th]].

%\cite{Campiglia:2014yka}
\bibitem{Campiglia:2014yka}
M.~Campiglia and A.~Laddha,
``Asymptotic symmetries and subleading soft graviton theorem,''
Phys. Rev. D \textbf{90} (2014) no.12, 124028
doi:10.1103/PhysRevD.90.124028
[arXiv:1408.2228 [hep-th]].

%\cite{Campiglia:2015yka}
\bibitem{Campiglia:2015yka}
M.~Campiglia and A.~Laddha,
``New symmetries for the Gravitational S-matrix,''
JHEP \textbf{04} (2015), 076
doi:10.1007/JHEP04(2015)076
[arXiv:1502.02318 [hep-th]].

 %\cite{Strominger:2013jfa}
\bibitem{Strominger:2013jfa}
A.~Strominger,
``On BMS Invariance of Gravitational Scattering,''
JHEP \textbf{07} (2014), 152
doi:10.1007/JHEP07(2014)152
[arXiv:1312.2229 [hep-th]].

%\cite{He:2014laa}
\bibitem{He:2014laa}
T.~He, V.~Lysov, P.~Mitra and A.~Strominger,
``BMS supertranslations and Weinberg\textquoteright{}s soft graviton theorem,''
JHEP \textbf{05} (2015), 151
doi:10.1007/JHEP05(2015)151
[arXiv:1401.7026 [hep-th]].

%\cite{Strominger:2017zoo}
\bibitem{Strominger:2017zoo}
A.~Strominger,
``Lectures on the Infrared Structure of Gravity and Gauge Theory,''
[arXiv:1703.05448 [hep-th]].


%\cite{Fried1}
 \bibitem{Fried1}
  H. Friedrich, ``Gravitational fields near space-like and null infinity,'' J. Geom. Phys. {\bf 24} (1998) 83-163.
  
 %\cite{Friedrich:1999wk}
\bibitem{Friedrich:1999wk}
H.~Friedrich and J.~Kannar,
``Bondi type systems near space - like infinity and the calculation of the NP constants,''
J. Math. Phys. \textbf{41} (2000), 2195-2232
doi:10.1063/1.533235
[arXiv:gr-qc/9910077 [gr-qc]].

%\cite{Friedrich:1999ax}
\bibitem{Friedrich:1999ax}
H.~Friedrich and J.~Kannar,
``Calculating asymptotic quantities near space - like and null infinity from Cauchy data,''
Annalen Phys. \textbf{9} (2000), 321-330
doi:10.1002/(SICI)1521-3889(200005)9:3/5\ensuremath{<}321::AID-ANDP321\ensuremath{>}3.0.CO
[arXiv:gr-qc/9911103 [gr-qc]].

%\cite{Mohamed:2021rfg}
\bibitem{Mohamed:2021rfg}
M.~M.~A.~Mohamed and J.~A.~Valiente~Kroon,
``Asymptotic charges for spin-1 and spin-2 fields at the critical sets of null infinity,''
[arXiv:2112.03890 [gr-qc]].

%\cite{Winicour1}
\bibitem{Winicour1}
J.~Winicour,
``Logarithmic Asymptotic Flatness,''
Found. Phys. \textbf{15} (1985), 605

%\cite{Javadinezhad:2022ldc}
\bibitem{Javadinezhad:2022ldc}
R.~Javadinezhad and M.~Porrati,
``A Supertranslation-Invariant Formula for the Angular Momentum Flux in Gravitational Scattering,''
[arXiv:2211.06538 [gr-qc]].

%\cite{Adami:2021nnf}
\bibitem{Adami:2021nnf}
H.~Adami, D.~Grumiller, M.~M.~Sheikh-Jabbari, V.~Taghiloo, H.~Yavartanoo and C.~Zwikel,
``Null boundary phase space: slicings, news \& memory,''
JHEP \textbf{11} (2021), 155
doi:10.1007/JHEP11(2021)155
[arXiv:2110.04218 [hep-th]].

\end{thebibliography}
\end{document}